\RequirePackage{fix-cm}
\documentclass[natbib,smallextended]{svjour3}       
\smartqed  
\usepackage{graphicx}
\usepackage{subfigure}
\usepackage{amssymb}
\usepackage{amsmath}
 \journalname{my journal}

\newcommand{\aj}{AJ}
\newcommand{\apj}{ApJ}
\newcommand{\apjs}{ApJS}
\newcommand{\apjl}{ApJ}
\newcommand{\mnras}{MNRAS}
\newcommand{\aap}{A\&A}

\newcommand{\Mstar}{\ensuremath{M_{\ast}}}
\newcommand{\Rstar}{\ensuremath{R_{\ast}}}
\newcommand{\estar}{\ensuremath{\eta_{\ast}}}
\newcommand{\Mdot}{\ensuremath{\dot{M}}}
\newcommand{\vinf}{\ensuremath{v_{\infty}}}
\newcommand{\Ralf}{\ensuremath{R_{\rm A}}}
\newcommand{\Rkep}{\ensuremath{R_{\rm K}}}
\newcommand{\thooc}{$\theta^{1}$\,Ori\,C}
\newcommand{\sorie}{$\sigma$\,Ori\,E}
\newcommand{\tbrake}{\ensuremath{\tau_{\rm J}}}
\newcommand{\tmdot}{\ensuremath{\tau_{\rm mass}}}
\newcommand{\kelv}{\ensuremath{\rm K}}
\newcommand{\keV}{\ensuremath{\rm keV}}
\newcommand{\tOriC}{$\theta^{1}$~Ori~C}
\newcommand{\etas}{\eta_\ast}
\newcommand{\ra}{R_\mathrm{A}}
\newcommand{\rk}{R_\mathrm{K}}
\newcommand{\rs}{R_\ast}
\newcommand{\teff}{T_\mathrm{eff}}
\newcommand{\beq}{B_\mathrm{eq}}
\newcommand{\bp}{B_\mathrm{p}}
\newcommand{\mdot}{\dot{M}}
\newcommand{\tauj}{\tau_\mathrm{J}}
\newcommand{\tsmax}{t_{\mathrm{s,max}}}
\newcommand{\ts}{t_\mathrm{s}}

\newcommand{\dd}[2][]{\frac{d #1}{d #2}}                  

\newcommand{\e}[1]{\times 10^{#1}}

\begin{document}

\title{
Accretion, Outflows, and Winds of Magnetized Stars}

\titlerunning{Accretion, Outflows \& Winds of Magnetized Stars}
\authorrunning{M.~M.~Romanova \& S.~P.~Owocki}

\author{Marina M. Romanova and Stanley P. Owocki}
\institute{Marina M. Romanova \at Department of Astronomy and CRSR,
Cornell University, Ithaca, NY 14853, USA \\
\email{romanova@astro.cornell.edu} \and Stanley P. Owocki \at
Department of Physics \& Astronomy, University of Delaware,
Newark, DE 19716 USA  \\\email{owocki@udel.edu}}

%
%
\maketitle

\begin{abstract}

Many types of stars have strong magnetic fields that can
dynamically influence the flow of circumstellar matter. In  stars
with accretion disks, the stellar magnetic field can truncate the
inner disk and determine the paths that matter can take to flow
onto the star. These paths are different in stars with different
magnetospheres and periods of rotation. External field lines of
the magnetosphere may inflate and produce favorable conditions for
outflows from the disk-magnetosphere boundary. Outflows can be
particularly strong in the propeller regime, wherein a star rotates
more rapidly than the inner disk. Outflows may also form at the
disk-magnetosphere boundary of slowly rotating stars, if the
magnetosphere is compressed by the accreting matter. In isolated,
strongly magnetized stars, the magnetic field can influence
formation and/or propagation of stellar wind outflows. Winds from
low-mass, solar-type stars may be either thermally or magnetically
driven, while winds from massive, luminous O and B type stars are
radiatively driven. In all of these cases, the magnetic field
influences matter flow from the stars and determines many
observational properties. In this chapter we review recent studies
of accretion, outflows, and winds of magnetized stars with a
focus on three main topics: (1) accretion onto magnetized stars;
(2) outflows from the disk-magnetosphere boundary; and (3) winds
from isolated massive magnetized stars. We show results obtained
from global magnetohydrodynamic simulations and, in a number of
cases compare global simulations with observations.

\end{abstract}

\section{Introduction}
\label{sec:introduction}

Magnetic fields play important and distinct roles in the dynamics
of gas flow around disk-accreting and isolated magnetized stars.
In Sec. \ref{sec:intro disk accreting} and  \ref{sec:intro wind
accreting}, we review different types of disk-accreting magnetized
stars, for which the magnetic field determines the processes of
accretion and outflow. In Sec. \ref{sec:intro winds} we review
isolated magnetized stars, where the magnetic field drives or
shapes the winds from these stars.

\subsection{Accretion onto magnetized stars}
\label{sec:intro disk accreting}

Different types of \textit{disk-accreting} stars have dynamically
important magnetic fields. Some of them are very young and are at
the stage of gravitational contraction; others are very old,
compact stars.
 Young ($\lesssim10^7$ years) low-mass ($0.5-2
M_\odot$) stars have significant magnetic fields of a few
kG \citep{DonatiEtAl2007,Johns-Krull2007}, which is thousands of
times stronger than the globally averaged field of the Sun. Some
of these stars (called \textit{Classical T Tauri Stars, or CTTSs})
are surrounded by accretion disks, and their fields are strong
enough to truncate the disks (see Fig. \ref{sketch-Camenzind}).
The truncation (or magnetospheric) radius $r_m$ is determined by
the balance between the magnetic stress in the magnetosphere and
the matter stress in the disk \footnote{We further discuss the
disk-magnetosphere interaction and the magnetospheric radius in
Sec. \ref{sec:truncation-disk}}. For a number of CTTSs, the
truncation radii are estimated to be of a few stellar radii. The
light curves from such CTTSs show complex patterns of periodic,
quasi-periodic, or irregular variability (see Fig.
\ref{alencar-3}; see also
\citealt{HerbstEtAl1994,StaufferEtAl2014}), which are probably
connected with complex paths of matter flow around the
magnetosphere and rotations of the hot spots that form as a result
of the gravitational impact of the free-falling matter to the
stellar surface (see, e.g., review by \citealt{BouvierEtAl2007a}).

Old, compact stars (i.e., white dwarfs, neutron stars) have even
stronger magnetic fields, originating from the collapse of a main
sequence star under the conservation of magnetic flux. In some
\textit{white dwarfs}, called `polars,' the magnetic field is so
strong, ($B\sim 10^7-10^8$ G) that an accretion disk does not
form, and matter from the secondary star accretes directly to the
large magnetosphere.
In white dwarfs known as `intermediate polars'
(IPs), with strong, but less extreme, fields ($B\sim 10^6$ G)  an
accretion disk forms (e.g., \citealt{Warner1995,Hellier2001}).
This accretion disk is truncated at large distances from the star,
with $r_m \gtrsim 10 R_\star$. Such IPs show periodic variability
associated with the channelling of disk matter onto the magnetic
poles of the star.

Accreting \textit{neutron stars} in binary systems may have very
strong magnetic fields, $B\sim 10^{12}$ G, with accretion disks
that interact with huge magnetospheres, $r_m \gg R_\star$. Due to
their large magnetospheres, neutron stars have relatively long
periods of order seconds. There is also a sub-class of neutron
stars called accreting millisecond X-ray pulsars (AMXPs, e.g.,
\citealt{WijnandsvanderKlis1998}), where the field has decayed to
$B\sim 10^7-10^9$ G  and the neutron star has spun up due to disk
accretion in binary systems (e.g.,
\citealt{Bisnovatyi-KoganKomberg1974,AlparEtAl1982,PatrunoWatts2012}).
In these stars, as in CTTSs, the disk is truncated at a few
stellar radii and the magnetospheric accretion is expected. .
However, due to the much smaller size of the neutron star, the
dynamical time-scale at the truncation radius is only a few
milliseconds. In spite of the large difference in size between
young CTTSs (about $10^{11}$ cm) and millisecond-period
 neutron stars (about $10^6$ cm), the physics of the
disk-magnetosphere interaction is quite similar: the
accretion-induced pulsations are observed in X-ray, which are
associated with the rotating hot spots. In addition,
high-frequency  quasi-periodic oscillations (QPOs) are observed
which carry information about processes in the inner disk (e.g.,
\citealt{vanderKlis2000,PatrunoWatts2012}).
 Section \ref{sec:accretion section} discusses
different aspects of accretion onto magnetized stars.

\subsection{Outflows from the disk-magnetosphere boundary}
\label{sec:intro wind accreting}

\begin{figure*}
  \begin{center}
    \begin{tabular}{cc}
\includegraphics[width=65mm]{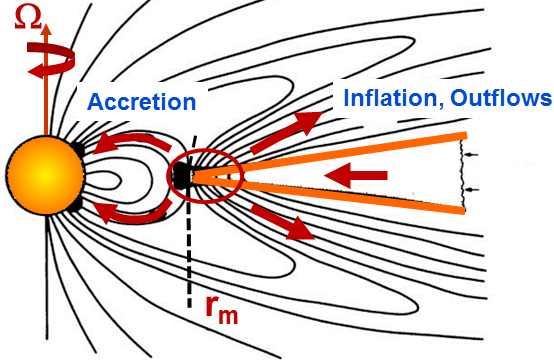}
    \end{tabular}
  \end{center}
  \vspace{-0.4cm}
\caption{Sketch shows interaction of a magnetized star with the
disk. Matter of the disk is stopped by the magnetosphere. Part of
the matter reaches the inner regions of the closed magnetosphere
and accretes onto the star in funnel streams. The field lines of
the external layers of the magnetosphere can inflate and outflows
are possible from the disk-magnetosphere boundary. Based on sketch
from Camenzind (1990).}
 \vspace{-0.4cm}
\label{sketch-Camenzind}
\end{figure*}

\begin{figure*}
  \begin{center}
    \begin{tabular}{cc}
             \includegraphics[width=120mm]{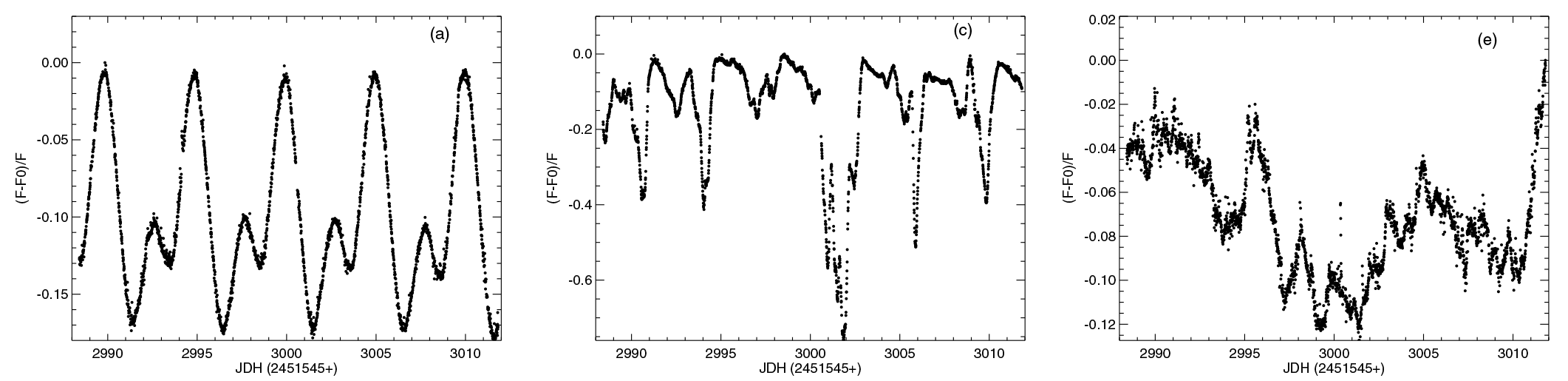}\hspace{0.0cm}
    \end{tabular}
   \end{center}
    \vspace{-0.4cm}
\caption{Different types of light-curves from
    young stars in  NGC 2264 obtained with \textit{CoRoT}: periodic light-curve (left panel),
    AA Tau type light-curve, with characteristic dips (middle
    panel) and irregular light-curve (right panel).
    From \citet{AlencarEtAl2010}.}
 \vspace{-0.4cm}
\label{alencar-3}
\end{figure*}

The field lines of the external parts of the magnetosphere  may
inflate due to the differences in angular velocities of the
foot-points of the magnetic loops connecting the star and the disk
 \citep{AlyKuijpers1990,Camenzind1990,ShuEtAl1994,LovelaceEtAl1995,UzdenskyEtAl2002}.
 Different signs of episodic inflation are observed in CTTSs (see,
 e.g., review by \citealt{BouvierEtAl2007a}).
Inflation creates favorable conditions for outflows from the
disk-magnetosphere boundary, at which the magnetic field of the
star participates in driving outflows (see Fig.
\ref{sketch-Camenzind}). Collimated outflows are observed from a
number of CTTSs (e.g., \citealt{RayEtAl2007}). A significant
number of CTTSs show signs of outflows in spectral lines,
particularly in the near-infrared He~I~$\lambda$10830\AA~line, for
which two distinct components of outflows had been found (e.g.,
\citealt{EdwardsEtAl2006}). Outflows are also observed from
accreting compact stars such as accreting white dwarfs in
symbiotic binaries (e.g., \citealt{SokoloskiEtAl2008}), as well as
from the vicinity of neutron stars (e.g.,
\citealt{Fender2004,HeinzEtAl2007}). The origin of these outflows
is not yet clear. One of attractive possibilities is that they may
originate at the disk-magnetosphere boundary (e.g.,
\citealt{ShuEtAl1994,RomanovaEtAl2009}). In Sec. \ref{sec:outflows
section} we discuss different types of outflows from the
disk-magnetosphere boundary.

\subsection{Winds from isolated magnetized stars }
\label{sec:intro winds}

Many stars are isolated, some of them having significant magnetic
fields. The magnetic field of the Sun is relatively small - a few
Gauss on average. Nonetheless, it plays an important role in the
heating of the corona and in the formation of the wind, which is
driven by gas pressure. Powerful events of inflation and
reconnection of the solar field lines lead to coronal mass
ejections (CMEs), which enhance the wind density and induce storms
in the Earth's magnetosphere (e.g., \citealt{BoikoEtAll2012}).

In young  low-mass stars, the magnetic field is thousands of times
stronger than that of the Sun, and much stronger winds and CMEs
are expected. These winds can be either thermally or magnetically
driven (e.g., \citealt{WeberDavis1967,LovelaceEtAl2008,
VidottoEtAl2009, VidottoEtAl2011}). In magnetically driven winds,
the pressure gradient of the magnetic field determines the
acceleration of matter to the wind. This is possible when,
for gas pressure $P$ and magnetic field $B$,  the
plasma parameter $\beta=8\pi P/B^2 < 1$ at the base of the wind
\citep{LovelaceEtAl2008, VidottoEtAl2011}.
In more massive stars ($M > 5 M_\odot$) of O, B and A types,
powerful winds are driven by the star's radiation pressure. About
10 percent of O and B stars harbor large-scale, organized (often
predominantly dipolar) magnetic fields, ranging in dipolar
strength from a few hundred to tens of thousands Gauss. This field
can influence matter flow in the wind and can determine many
observational properties of these stars. It can channel wind
material or trap the  wind, forming \textit{wind-fed}
magnetospheres that develop from closed magnetic loops (e.g.,
\citealt{ud-DoulaEtAl2013,OwockiEtAl2014}).

Section \ref{sec:winds section} discusses in greater detail the
wind-fed magnetospheres from magnetized massive stars along with
corresponding observational properties associated with the
magnetic field.

\smallskip

The problems of the disk-magnetosphere interaction, outflows and
winds from magnetized stars are multidimensional and require
global axisymmetric or three-dimensional (3D) numerical
simulations. Below, we describe results of recent numerical
simulations of  accretion and outflows from disk-accreting
magnetized stars,
wind-fed magnetospheres of isolated massive stars, and show
examples of comparisons of models with observations.

\section{MAGNETOSPHERIC ACCRETION} \label{sec:accretion section}

Magnetospheric accretion is a complex process, in which the inner
disk matter interacts with the magnetosphere of the star. The
result of such an interaction depends on a number of factors, such
as the period of stellar rotation, the structure of the magnetic
field of the star, the size of the magnetosphere, the diffusivity
at the disk-magnetosphere boundary, properties of the accretion
disk, and  other factors. We start from a brief overview of
studies the magnetospheric accretion.

\subsection{Different studies of the magnetospheric accretion}

The disk-magnetosphere interaction has been studied in a number of
theoretical works (e.g.,
\citealt{PringleRees1972,LambEtAl1973,GhoshLamb1978,SpruitTaam1990,Konigl1991,Campbell1992,KoldobaEtAl2002a}).
It was predicted that the disk should be truncated by the
magnetosphere of the star, and matter should flow above and below
the magnetosphere in the funnel flow. Some authors also predicted
that the magnetic field lines of the external parts of the
magnetosphere may inflate and open due to the differential
rotation of the foot-points of the field lines of the external
magnetosphere (e.g., \citealt{AlyKuijpers1990}), and outflows are
possible along the field lines of the inflated magnetosphere
(e.g., \citealt{Camenzind1990,ShuEtAl1994,LovelaceEtAl1995}).

A number of numerical simulations have been also performed. The
process of the disk-magnetosphere interaction is internally
multi-dimensional, so that two or three-dimensional simulations
are required for investigation of this process.
Early axisymmetric simulations (e.g.,
\citealt{HayashiEtAl1996,MillerStone1997,HiroseEtAl1997}) were
relatively brief, only a few Keplerian rotations at the inner
disk. Rapid, almost free-fall accretion has been observed due to
the initial magnetic braking of the Keplerian disk by the magnetic
field of the non-rotating  magnetosphere. Strong inflation of the
field lines and one-time episode of outflows has been observed.
These simulations did show inflation of the field lines and
outflows. They also  have shown brief events of the magnetospheric
accretion; however, the magnetospheric accretion has not been
studied due to non-stationary nature of the process.
 \citet{GoodsonEtAl1997} has been able to perform longer simulation runs and obtained a few
 episodes of accretion, inflation  and outflows from the disk-magnetosphere boundary.
 Their simulations confirmed the suggestion of \citet{AlyKuijpers1990}
 about the possibility of the cyclic inflation of the field lines of the external magnetosphere
 and formation of outflows along the field lines threading the inner disk \citep{LovelaceEtAl1995}.
   However, these authors investigated only the case of a rapidly
rotating young star  (with period of 1.8 days) in the regime which
is close to the propeller regime (e.g.,
\citealt{IllarionovSunyaev1975,LovelaceEtAl1999}). In subsequent
simulations \citep{GoodsonEtAl1999,MattEtAl2002} only
rapidly-rotating stars were considered. It is not clear whether
these results are general, that is can be applied to any
disk-accreting  magnetized stars.

Development of quasi-equilibrium initial conditions for the
disk-magnetosphere configuration helped to obtain a slow accretion
in the disk, and to investigate accretion through the funnel flows
\citep{RomanovaEtAl2002}.\footnote{In these initial conditions the
corona above the disk rotates with the angular velocity of the
disk, so that the magnetic field lines do not experience
discontinuity at the disk-magnetosphere boundary, and there is no
initial magnetic breaking. There is also an initial balance
between the gravitation, pressure and centrifugal forces in each
point of the simulation region.}. Simulations confirmed many
aspects of the disk-magnetosphere interaction predicted
theoretically, such as the truncation of the accretion disk by the
stellar magnetosphere, formation of the funnel flow, and angular
momentum flow between the disk and the star. However, in these
simulations, the corona above the disk is relatively heavy so that
the inflation and outflows were suppressed in most of simulation
runs. Simulations with the lower-density corona confirmed
accretion through funnel flows and the inflation of the field
lines; however, no outflows were observed \citep{LongEtAl2005}.
Similar simulations were performed by \citet{BessolazEtAl2008} who
also compared the position of the truncation radius of the disk
observed in simulations with magnetospheric radii derived from
different theoretical studies. \citet{LongEtAl2005} studied the
angular momentum flow between the disk, star and corona and found
conditions for the rotational equilibrium state at which a star
neither spins-up, nor spins-down. They found, that in this state,
a star rotates somewhat more slowly than the inner disk, because
part of angular momentum flows from the star to corona along
inflated and partially-inflated field lines. More recently,
\citet{ZanniFerreira2013} observed many cycles of inflation and
reconnection in the external magnetosphere with ejection of matter
along the inflated field lines. Simulations show that outflows are
strongly enhanced in the propeller regime of accretion (e.g.,
\citealt{RomanovaEtAl2005b,UstyugovaEtAl2006}), or in cases where
the accretion disk compresses the magnetic field of the external
magnetosphere \citep{RomanovaEtAl2009}. These axisymmetric
simulations helped to understand many elements of the
disk-magnetosphere interaction. They are particularly valuable for
investigation of the inflation and outflows in the external
magnetosphere, because the coronal density should be low enough
for modeling these processes.

The disk-magnetosphere interaction also depends on the properties
of an accretion disk. In many studies it is suggested that the
outward transport of angular momentum is provided by viscosity,
and the viscosity coefficient is proportional to parameter
$\alpha$, where $\alpha<1$  \citep{ShakuraSunyaev1973}. This
approach has been successfully used in many studies of accretion
disks and also has been used in most of modelings of the
magnetospheric accretion (e.g.,
\citealt{GoodsonEtAl1997,LongEtAl2005,ZanniFerreira2013,RomanovaEtAl2014}).
The $\alpha-$viscosity plays an important role in bringing matter
toward the magnetosphere in steady rate.  On the other hand,
$\alpha-$viscosity mimics some more complex processes in the disk,
such as magnetic turbulence \citep{ShakuraSunyaev1973}.  A
significant progress has been made in understanding turbulent
disks where the turbulence is driven by the magneto-rotational
instability (e.g., \citealt{BalbusHawley1991,BalbusHawley1998}).
 MRI-driven turbulence has been
extensively studied in axisymmetric and local/global 3D MHD
simulations (e.g.,
\citealt{HawleyEtAl1995,BrandenburgEtAl1995,StoneEtAl1996,
Armitage1998,Hawley2000,BeckwithEtAl2009,FlockEtAl2011,SimonEtAl2011,McKinneyEtAl2014}).
In most of simulations the central object is non-magnetized,
usually a black hole. Accretion onto a magnetized star has been
studied in much fewer simulations
\citep{RomanovaEtAl2011a,RomanovaEtAl2012}. Simulations show that
many properties of the magnetospheric accretion are similar in
cases of the turbulent and laminar, $\alpha-$disks, in particular,
in cases of the small-scale turbulence.

To model most realistic situation, the disk-magnetosphere
interaction should be considered in three dimensions.  One of the
main issues is that the inner disk matter is expected to penetrate
through
 through the magnetosphere (the effective
diffusivity) in non-axisymmetric, 3D instabilities, such as the
Rayleigh-Taylor and Kelvin-Helmholtz instabilities (e.g.,
\citealt{AronsLea1976}). Such a penetration will determine the
effective ``diffusivity" at the disk-magnetosphere boundary and
thus the level of interaction between the disk and the star.   On
the other hand, the dipole magnetic moment may be tilted about the
rotational axis of the star, or the magnetic field may be more
complex than the dipole field. Therefore, it is important to study
the disk-magnetosphere interaction in global 3D simulations. A
special ``Cubed Sphere" 3D MHD Godunov-type code has been
developed for this purpose  \citep{KoldobaEtAl2002b} and different
3D simulations were performed for the first time. In this section,
we describe mainly the results of the global 3D simulations of the
disk-magnetosphere interaction. In particular, we show results of
accretion onto a star with a tilted dipole magnetosphere (see Sec.
\ref{sec:accretion-tilted}); accretion through 3D instabilities
(see Sec. \ref{sec:stab-unstab}); accretion onto stars with
relatively large magnetospheres (see Sec.
\ref{sec:larger-magnetosphere}), and accretion onto stars with
complex magnetic fields (see Sec. \ref{sec:complex}). Finally, the
3D MHD models have been developed for CTTS stars with realistic
parameters, and results of simulations were compared with
observations of these stars (see Sec. \ref{sec:BP Tau} and
\ref{sec:V2129}).


\begin{figure}[!ht]
\sidecaption
\includegraphics[width=60mm]{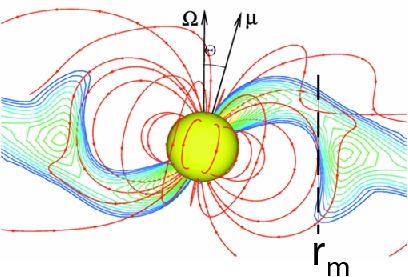}
   \vspace{-0.2cm}
\caption{Slice of the density distribution and selected field
lines resulting from 3D MHD simulations of accretion onto a star
with a tilted dipole magnetic field. The dashed line shows the
position of the magnetospheric radius. From
\citet{RomanovaEtAl2004}.}
 \vspace{-0.4cm}
  \label{funnel-1}
\end{figure}

\subsection{Truncation of the disk by the magnetosphere, and characteristic radii}
\label{sec:truncation-disk}

If the magnetic field of the star is sufficiently strong, then it
truncates the accretion disk at a radius  $r_m$ where the magnetic
stress in the magnetosphere matches the matter stress in the disk.
In the disk, the largest stress is connected with the azimuthal
components of the stress tensor: $T_{\phi\phi}=[p + \rho v_\phi^2]
+ [{B}^2/8\pi-B_\phi^2/4\pi]$, where $\rho$, $p$, $B$,
$B_\phi$ are the local density, gas pressure, total and azimuthal
magnetic field in the disk, respectively.
 At the innermost edge of the disk, $B_\phi \ll B$ and the truncation radius can be found from
 the condition $p + \rho v_\phi^2 = B^2/8\pi$. However, it is
 better to include the total ram pressure which acts  in all three directions and to
 compare the total matter pressure with the total magnetic
 pressure
\begin{equation}
p + \rho v^2 = B^2/8\pi,~~~~{\rm or} ~~~~ \beta_t=8\pi(p + \rho
v^2)/B^2=1 ~, \label{eq:stress balance}
\end{equation}
Here, $\beta_t$ is the generalized plasma parameter, which
includes both \textit{thermal and ram pressure}
\citep{RomanovaEtAl2002}. It is analogous to the standard plasma
parameter $\beta=8\pi p/B^2$, but takes into account the ram
pressure of the matter flow in the disk. Axisymmetric and global
3D MHD simulations show that the condition $\beta_t=1$  is useful
for finding the magnetospheric radius $r_m$ (e.g.,
\citealt{KulkarniRomanova2013}). This radius usually corresponds
to the \textit{innermost} edge of the disk, where the density
drops sharply towards the low-density magnetosphere (in which
magnetic pressure is dominant). This formula can be used to find
$r_m$ in many situations, including those that are non-stationary.
There is also no restriction on the geometry of the magnetic field
of the star (see also Sec. \ref{sec:complex}) \footnote{Sometimes,
the condition $\beta=8\pi p/B^2=1$ is used to find the
magnetospheric radius (e.g., \citealt{BessolazEtAl2008}). This
condition, however, yields somewhat a larger radius at which the
matter flows from the disk to the funnel stream.}.
{Fig.
\ref{funnel-1} shows a snapshot from 3D simulations of a
 laminar, $\alpha-$disk.  The  vertical dashed line shows the
magnetospheric radius  calculated from the stress balance Eq. \ref{eq:stress balance}.}

In earlier theoretical studies  (e.g.,
\citealt{PringleRees1972,LambEtAl1973}) the magnetospheric radius
was estimated from the balance between the largest components of
the stresses, assuming a dipole field from the star and Keplerian
orbital speed in the disk; this gives
\begin{equation}
r_m = k \big[\mu^4/(\dot{M}^2 GM_\star)\big]^{1/7}, ~~~~~k\sim 1~,
\label{eq:alfven}
\end{equation}
\noindent where $\mu=B_\star R_\star^3$ is the magnetic moment of
the star with a surface field $B_\star$, $\dot{M}$ is the disk
accretion rate, and $M_\star$  and $R_\star$ are the mass and
radius of the star, respectively.
 Axisymmetric disk-accretion
simulations by, e.g., \citet{LongEtAl2005}  found that eqs.
\ref{eq:stress balance} and \ref{eq:alfven} give similar values of
$r_m$ if $k\approx 0.5$ \citep{LongEtAl2005}.
\citet{BessolazEtAl2008}  compared different approaches for
estimating the magnetospheric radius, and found that all give
similar values, essentially set by the steep ($B^2\sim r^{-6}$)
decline of the magnetic stress, which dominates over the much more
gradual variation of the disk matter stress.

From a series of 3D MHD simulations, \citet{KulkarniRomanova2013}
fit a scaling for the magnetospheric radius,
\begin{equation}
{r_m}/R_\star \approx 1.06 \large[
{\mu^4}/(\dot{M}^2 GM_\star R_\star^7)\large]^{1/10}
\, ,
\end{equation}
that is somewhat flatter than the $r_m\sim (\mu^2/{\dot M})^{1/7}$
scaling of eqn.\ (\ref{eq:alfven}). This difference can be
attributed to the non-dipole form of the magnetic field, which
results from the compression of the magnetosphere by the accretion
disk (see Fig. \ref{3d-spot-3}, middle panel). Despite this modest
difference in dependencies, the formula (\ref{eq:alfven}) with
coefficient $k\approx 0.5$ still describes the position of the
magnetospheric radius with accuracy sufficient in many
astrophysical situations. But note that this scaling has only been
tested for accretion onto relatively small magnetospheres,
$r_m\lesssim 5 R_\star$; accretion onto stars with larger
magnetospheres has not  yet been studied systematically
\footnote{Simulations of accretion onto stars with larger
magnetosphere require much longer computing time. Test simulations
of such accretion are described in  Sec.
\ref{sec:larger-magnetosphere}. However, to obtain a formula for
magnetospheric radius, multiple simulations are needed.}
\footnote{The compression of the magnetosphere is probably
connected with the ram pressure in the radial direction and so may
depend on the value of the radial velocity (which is proportional
to $\alpha-$parameter of viscosity). The above described
simulations were performed for $\alpha=0.02$. The possible
dependence of the compression on $\alpha-$parameter should be
studied in a separate set of simulations.}.


\begin{figure*}
  \begin{center}
    \begin{tabular}{cc}
\includegraphics[width=90mm]{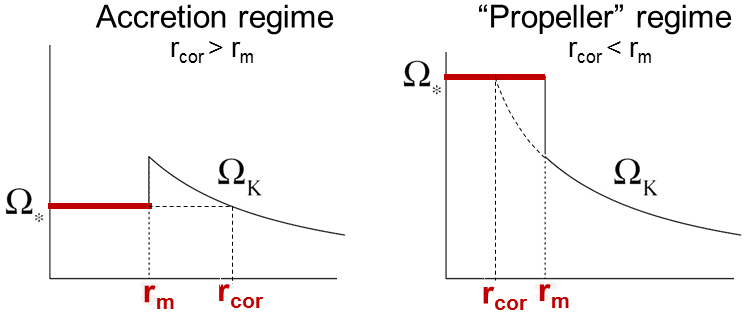}
    \end{tabular}
  \end{center}
   \vspace{-0.4cm}
\caption{\textit{Left panel:} Illustration of
 the ``accretion regime" (left), for which rigid-body
rotation at the stellar rotation frequency $\Omega_\star$ extends
to a magnetic radius $r_m$ less than the corotation radius, $
r_{\rm cor}$,  at which the local Keplerian frequency $\Omega_K =
\Omega_\star$; the {\em sub}-Keplerian rotation of the inner
region facilitates accretion of disk material onto the star.
\textit{Right panel:} The ``propeller regime" (right), wherein
$r_{\rm cor} > r_m$; the {\em super}-Keplerian rotation of the
inner region now causes disk material to be propelled into an
outflow by magneto-centrifugal forces. The condition $r_m=r_{\rm
cor}$ thus represents the boundary between these regimes. From
\citet{UstyugovaEtAl2006}.}
 \vspace{-0.0cm}
\label{sketch-accretion-propeller}
\end{figure*}

Another important radius is the  \textit{corotation radius}, at
which the angular velocity of the star matches the Keplerian
angular velocity of the disk: $\Omega_\star=\sqrt{GM_\star/r_{\rm
cor}^3}$:
\begin{equation}
r_{\rm cor}=[GM_\star/{\Omega_\star}^2]^{1/3} .
\label{eq:corotation}
\end{equation}
As illustrated in fig. \ref{sketch-accretion-propeller},
the condition $r_{\rm cor} = r_m$ characterizes the boundary between ``accretion'' vs. ``propeller'' regimes.
The left panel
demonstrates the case favorable for accretion, in which $r_m <
r_{\rm cor}$, and the inner disk rotates more rapidly than the
magnetosphere of the star; the accreting matter interacts with the
magnetosphere, loses some of its angular momentum, and accretes
onto the star. The right panel of Fig.
\ref{sketch-accretion-propeller} shows the opposite situation, in
which $r_{\rm cor} < r_m$ and the matter of the inner disk gains
angular momentum; this regime is called the ``propeller" regime
(e.g.,
\citealt{IllarionovSunyaev1975,AlparShaham1985,LovelaceEtAl1999}),
in which the inner disk matter can be pushed away by the
rapidly-rotating magnetosphere, while some of the matter can be
ejected as a wind. This situation will be considered in greater
detail in Sec. \ref{sec:propeller}.


\begin{figure*}
  \begin{center}
    \begin{tabular}{cc}
\includegraphics[width=38mm]{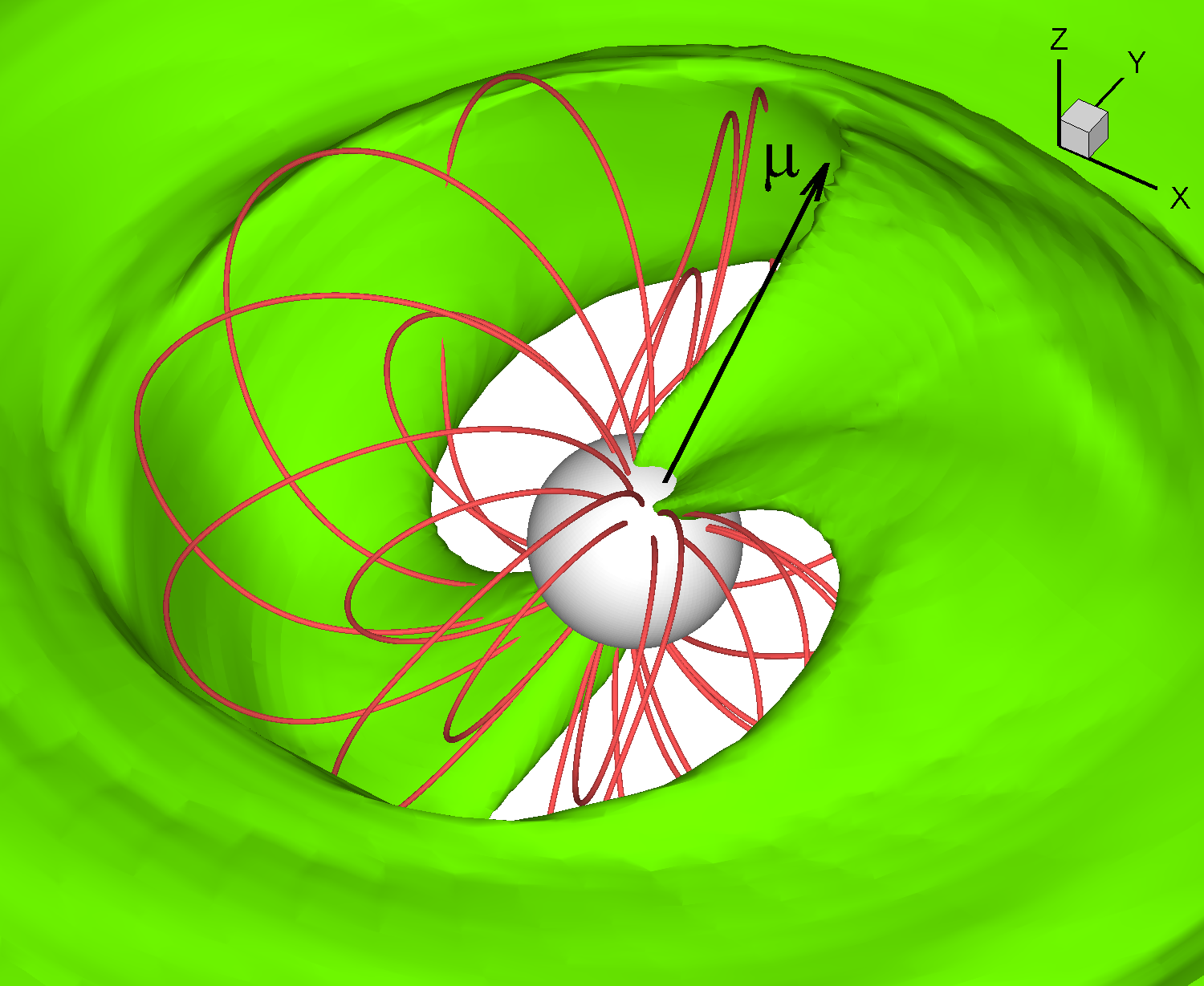}
\includegraphics[width=51mm]{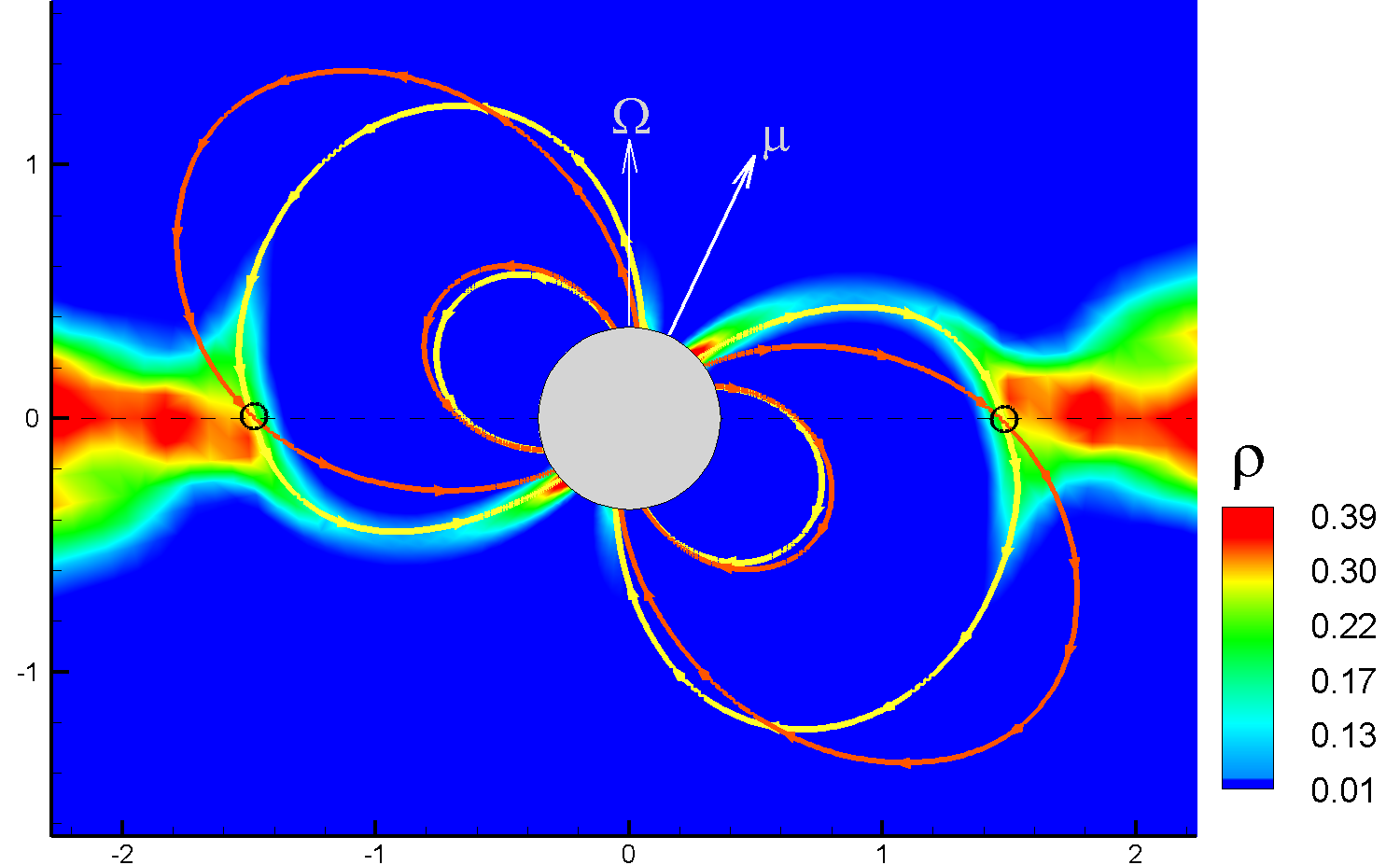}
\includegraphics[width=27mm]{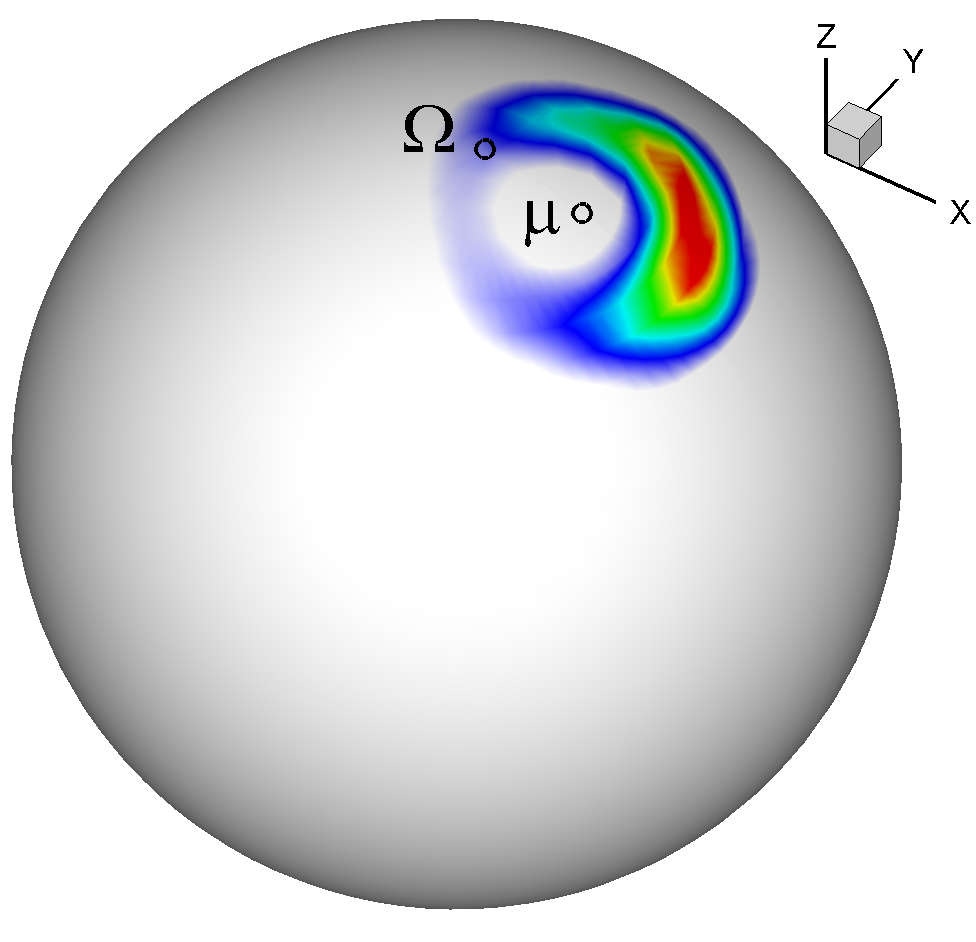}
    \end{tabular}
  \end{center}
   \vspace{-0.4cm}
\caption{\textit{Left panel:} A 3D view of the funnel flow from
the disk to a magnetized star, where the dipole moment $\mu$ is
tilted by $\Theta=20^\circ$ about the rotational axis.  One of the
density levels is shown in green; sample field lines are shown in
red. \textit{Middle panel:} The slice of density distribution
(color background) and magnetic field lines (yellow lines). Red
lines show the dipole field lines at $t=0$. The external field
lines are chosen such that both the yellow and the red lines start
at the inner magnetospheric radius $r_m$, which is marked as a
black circle. \textit{Right panel:} the energy flux distribution
on the surface of the star. Circles show the position of the
magnetic ($\mu$) and rotational ($\Omega$) axes, respectively.
From \citet{KulkarniRomanova2013}.}
 \vspace{-0.4cm}
 \label{3d-spot-3}
\end{figure*}

It is also convenient to characterize the state of the accreting
magnetized star with the fastness parameter, $\omega_s$, which is
defined as the ratio between the angular velocity of the star,
$\Omega_\star$ and Keplerian velocity of the inner disk at
$r=r_m$:
\begin{equation}
\omega_s=\frac{\Omega_\star}{\Omega_K(r_m)} =
\bigg(\frac{r_m}{r_{\rm cor}}\bigg)^{3/2} . \label{eq:fastness}
\end{equation}
This parameter efficiently characterizes different states of the
accreting magnetized star (e.g., \citealt{Ghosh2007}). In the
following sections we show that many processes at the
disk-magnetosphere boundary can be characterized by this
parameter.

\subsection{Accretion onto stars with a tilted dipole magnetic field}
\label{sec:accretion-tilted}

The magnetic axis of the dipole moment, $\mu$,  can be tilted by
an angle, $\Theta$, about the rotational axis, $\Omega_\star$, of
the star. The rotational axis of the star is fixed, and it
coincides with the rotational axis of the disk. In general, the
rotational axis of the star may be also tilted about the
rotational axis of the disk, and this leads to secular warping and
precession of the inner disk (e.g.,
\citealt{LipunovShakura1980,Lai1999}). In this Chapter, however,
we consider only cases in which the rotational axes of the star
and disk coincide.

 Global 3D MHD simulations of accretion onto stars with
different tilt angles of the dipole field, $\Theta$,
 show that matter flows onto the star via two funnel streams
 \citep{RomanovaEtAl2003,RomanovaEtAl2004}. The matter in the funnel streams
 is pulled towards the star by the gravitational force.
Numerical simulations have shown for the first time the structure
and shape of the funnel streams. They have shown that funnel
streams are wide in the meridional
 direction (see left panel of Fig.  \ref{3d-spot-3})
   and     narrow in the vertical direction (see slice of the density distribution in the middle panel of Fig.  \ref{3d-spot-3}).
    They also show that the low-density matter blankets the whole
 magnetosphere, while the funnel streams represent the denser
 parts of the flow (see Fig. \ref{funnel-level-2}).
 The slice in the middle of
 Fig. \ref{3d-spot-3} shows both the dipole field lines, and the field lines
 during the funnel accretion \footnote{We should note that simulations of other groups also show a strong compression of the magnetosphere
 (e.g., \citealt{BessolazEtAl2008,ZanniFerreira2013}).}.
 It is still not clear how important such compression will be in stars with much larger magnetospheres,
 i.e. with $r_m/R_\star > 5$.
 But we can speculate that the field compression might become less significant in such cases,
 because only the external parts of the magnetosphere interact with the disk, while the
  larger part of the magnetosphere is strongly magnetically dominated and so has a nearly dipole structure.


\begin{figure*}
  \begin{center}
    \begin{tabular}{cc}
\includegraphics[width=60mm]{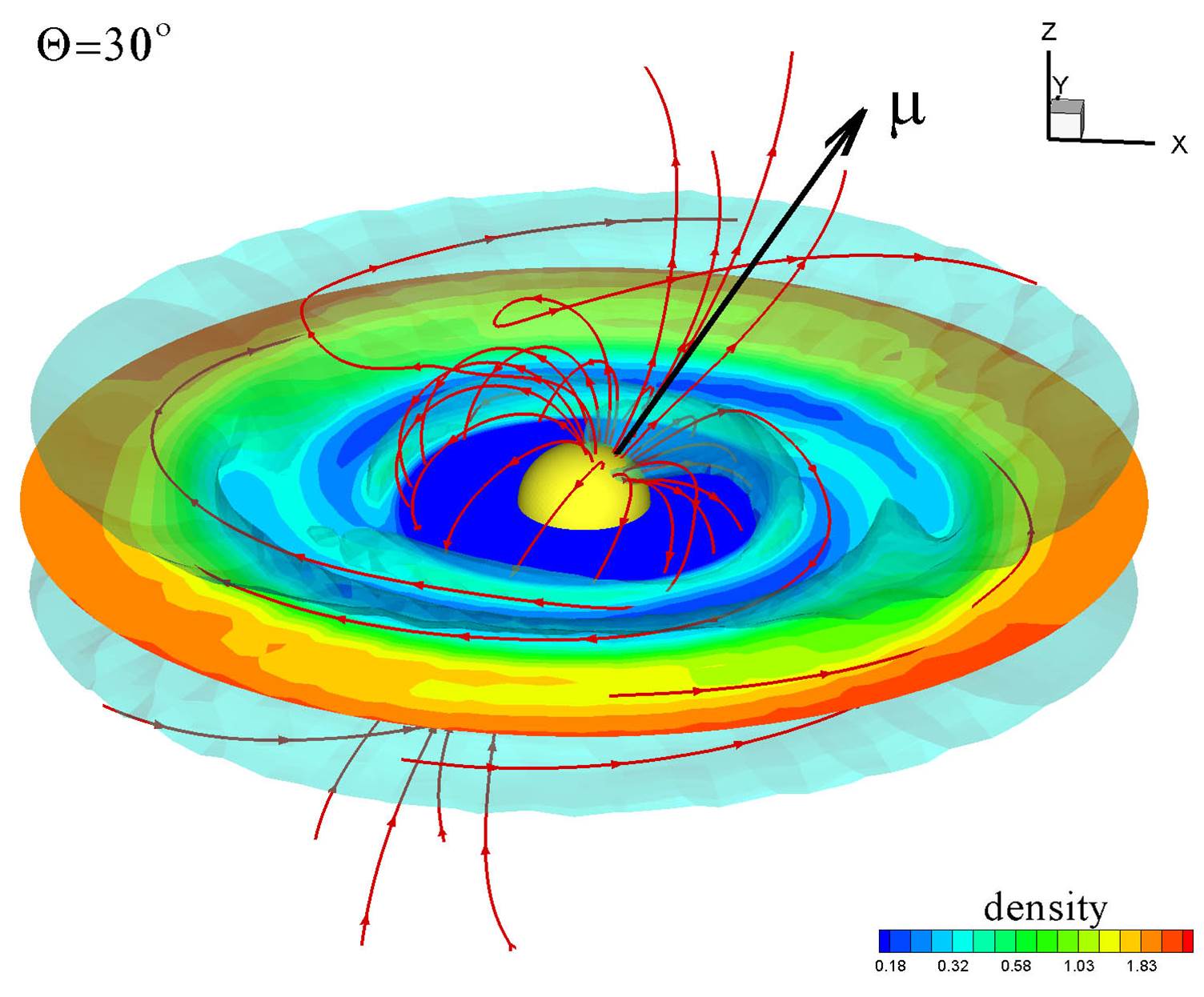}
\includegraphics[width=60mm]{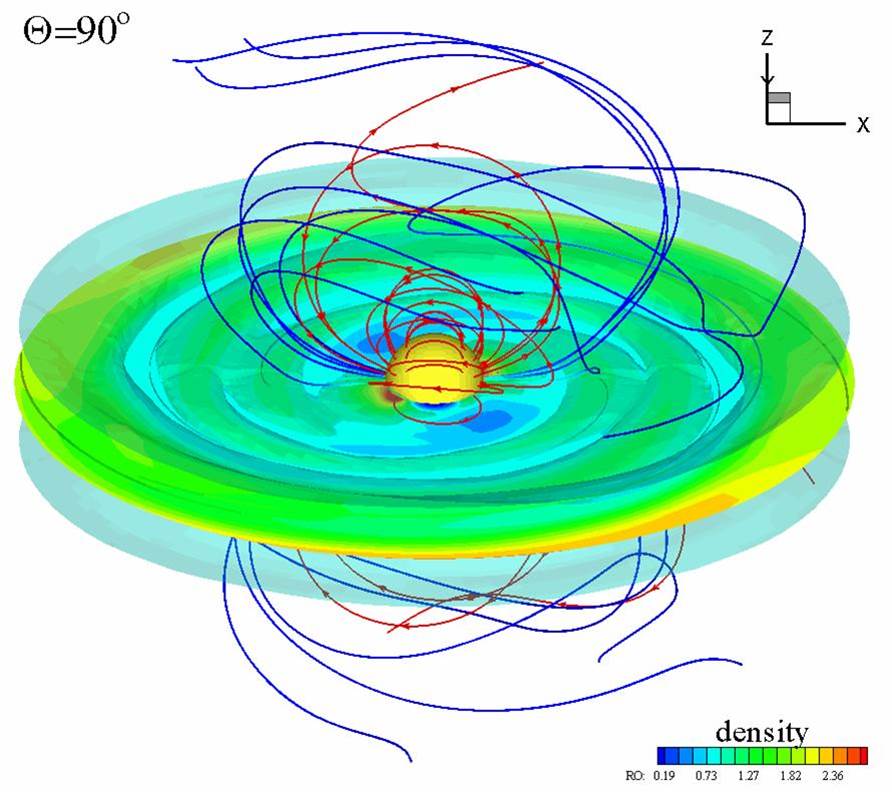}
    \end{tabular}
  \end{center}
   \vspace{-0.4cm}
\caption{\textit{Left panel:} Results of 3D MHD simulations of
accretion onto a star for which the magnetic axis of the dipole is
tilted by $\Theta=30^\circ$ about the rotational axis (where the
rotational axis is aligned with the $z-$axis). \textit{Right
panel:} the dipole axis is tilted by $\Theta=90^\circ$. The color
background shows the density distribution in the equatorial plane.
The blue translucent color shows one of the density levels in 3D.
The lines show sample magnetic field lines. From
\citet{RomanovaEtAl2005a}.}
 \vspace{-0.4cm}
  \label{3d-30-90}
\end{figure*}

\subsubsection{Properties of the funnel streams and hot spots}

The gravitational attraction from the star accelerates the matter
in the funnel streams. This, in turn, causes the formation of two
hot spots where the matter hits the surface of the star. Numerical
simulations helped for the first time to show the likely shape and
structure of the hot spots. Results indicate they tend to be
crescent-shaped, with densities and temperatures highest near the
spot center (see right panel of Fig. \ref{3d-spot-3}). This
numerical finding led to prediction for the wavelength dependence
of the spot radiation, with a different spot size at different
wavelengths. For example, in CTTSs, the innermost small region of
the spot can radiate in X-rays, while an increasingly larger area
radiates in the UV, optical, and IR.
 In fact, the measurements of the spot sizes in several
CTTSs show that the area covered by the hot spots in optical
wavelengths is relatively large, about $10\%$ of the stellar
surface \citep{DonatiEtAl2007,DonatiEtAl2008,DonatiEtAl2011}. On
the other hand, in models of the UV radiation by the accretion
shock, the best match with observations is reached if spots cover
from less than $1\%$ up to a few percents of the stellar surface
(e.g., \citealt{CalvetGullbring1998,GullbringEtAl2000}). More
generally, observations point to  smaller sizes of spots in higher
versus lower energy bands. The inhomogeneous structure of the hot
spots discovered in 3D simulations \citep{RomanovaEtAl2004} can
potentially explain these observational results. Before these
simulations,  hot spots were assumed to have a homogeneous density
and temperature distribution. These simulations thus opened a path
for understanding the properties of hot spots in different
wavebands.

These 3D numerical simulations also have shown for the first time
how the shape of the hot spots depends on the tilt of the magnetic
axis  $\Theta$ \citep{RomanovaEtAl2004}; at small $\Theta$ the
spots have crescent shape, while as $\Theta$ increases to larger
values, the spots acquire a bar shape (see also
\citealt{KulkarniRomanova2005}). The different shapes predict
corresponding differences in the light curves
\citep{RomanovaEtAl2004}, but these also depend on other factors
like observer viewing angle. The differences from light curves
obtained with the homogeneous, round spots of previous, simplified
models is modest, e.g. within $\sim 20\%$ in test runs of
accreting neutron stars (where relativistic effects were taken
into account, \citealt{KulkarniRomanova2005}). While such  simple
round spots thus provide a reasonable first approximation, a more
accurate model of the position and shape of the spot can be
important, for example in studies of accreting neutron stars,
where the mass and radius (or, their ratio) can be found from
relativistically-corrected light-curves (see, e.g., review by
\citealt{PatrunoWatts2012}). For convenience in application, an
analytic formula has been obtained from multiple 3D simulation
runs, with the shape and position of the spots given as a function
of the magnetospheric size $r_m/R_*$ and the the ratio $r_m/r_{\rm
cor}$, which characterizes the level of stellar rotation
\citep{KulkarniRomanova2013}.


\begin{figure}
\sidecaption
\includegraphics[width=80mm]{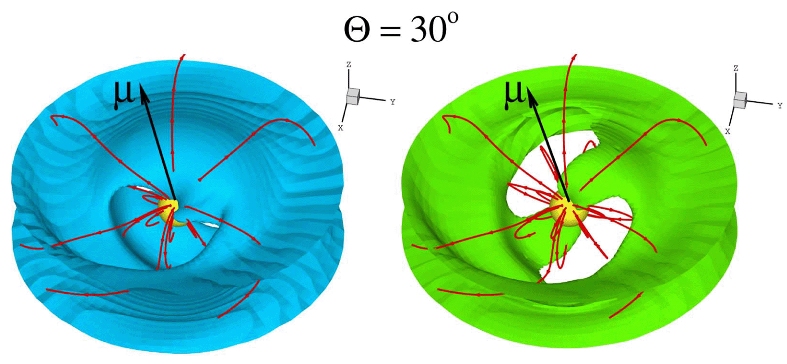}
   \vspace{-0.2cm}
\caption{The two panels compare the configuration of the funnel
streams at two density levels: the density in the left panel is
three times smaller than in the right panel. From
\citet{RomanovaEtAl2005a}.}
 \vspace{-0.4cm}
  \label{funnel-level-2}
\end{figure}

\subsubsection{Moving spots: phase shifts and possible QPOs}

Simulations show that the hot spots on the stellar surface are not
fixed, but instead move about some preferred position. This motion
leads to phase shifts in the light curves
\citep{RomanovaEtAl2003,KulkarniRomanova2013}. In young stars,
this results in the phenomenon of ``drifting period" which is
frequently observed in CTTSs, where the sets of observations are
usually brief (only a few stellar rotations). Recent observations
by the space instrument \textit{MOST} indicate that the CTTS star
TW Hya shows systematic variation of its ``period"
\citep{RucinskiEtAl2008}, which may be connected with the drifting
spot on the stellar surface, as found in earlier 3D simulations.

In millisecond pulsars, moving spots may lead to phase shifts in
light-curves \citep{PapittoEtAl2007,PatrunoEtAl2009} and to timing
noise (e.g., \citealt{PoutanenEtAl2009,LambEtAl2009}).
\citet{PapittoEtAl2007} found a correlation between the pulse
phase shifts and the X-ray flux in millisecond pulsar XTE
J1814-338 (see also \citealt{PatrunoEtAl2009}). The authors argued
that the observed phase shifts were due to movements of the hot
spot in response to variation in accretion rate. This phenomenon
can be explained by the correlation between the hot spot longitude
and the location of the magnetospheric radius as shown by
\citet{KulkarniRomanova2013}.

\begin{figure}[!ht]
\sidecaption
\includegraphics[width=70mm]{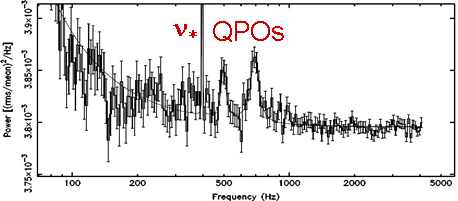}
   \vspace{-0.2cm}
\caption{Power spectral density of the AMXP SAX J1808.4-3658 which
shows the frequency of the star, $\nu_*$ and two high-frequency
QPOs which may carry information about the processes at the
disk-magnetosphere boundary. From \citet{WijnandsEtAl2003}.}
 \vspace{-0.4cm}
  \label{qpo-wijnands}
\end{figure}

In the case of very small tilts of the dipole moment,
$\Theta=(2-5)^\circ$, the funnel streams can be dragged by the
inner disk and the spot may systematically rotate faster or slower
than the star. This can lead to the phenomenon of the
quasi-periodic oscillations (QPOs) in the frequency spectrum and
may possibly explain some of QPOs observed in AMXPs. Usually, one
or two main high-frequency QPO peaks are observed in AMXPs (see an
example in Fig. \ref{qpo-wijnands}). The observed QPO frequencies
show significant temporal variation, but
the difference between peaks often corresponds to either the stellar rotation frequency
 $\nu_*$, or its half-value $\nu_*/2$
\footnote{This phenomenon can be possibly explained by the
beat-frequency model (e.g., \citealt{MillerEtAl1998}).}(e.g.,
\citealt{vanderKlis2000}). In many cases, however, the frequency
difference is nearly fixed at a value, $\sim 300$ Hz, that does
not correlate with the frequency of the star
\citep{BoutloukosEtAl2006,MendezBelloni2007}. The origin of these
QPOs has not yet been understood. One of QPO peaks may originate
at the disk-magnetosphere boundary. The origin of the second peak
is less clear, but an interesting, spot-based model has been
developed by \citet{BachettiEtAl2010}. They performed systematic
3D simulations of accretion onto neutron stars with very small
tilts, $\Theta=2^\circ$, and showed that these rotating spots may
possibly explain one of two QPO frequencies in AMXPs, where the
difference between QPO frequencies does not vary much in time and
is $\sim 300$Hz. \citet{BachettiEtAl2010} found that the lower of
two of QPO frequencies is connected with the rotation of the
funnel stream, while the the higher QPO frequency is connected
with unstable tongues that rotate with the frequency of the inner
disk (see Sec. \ref{sec:stab-unstab}). Simulations show that the
frequency associated with the rotating funnel streams is about
$250-300$Hz lower, and both QPOs  move in parallel, which is in
accord with observations \citep{MendezBelloni2007}. This model,
however, is applicable only to stars with small $\Theta$,
$\Theta\lesssim 5^\circ$.


\subsubsection{Tilted  magnetosphere and waves in the disk}

\begin{figure*}
  \begin{center}
    \begin{tabular}{cc}
             \includegraphics[width=110mm]{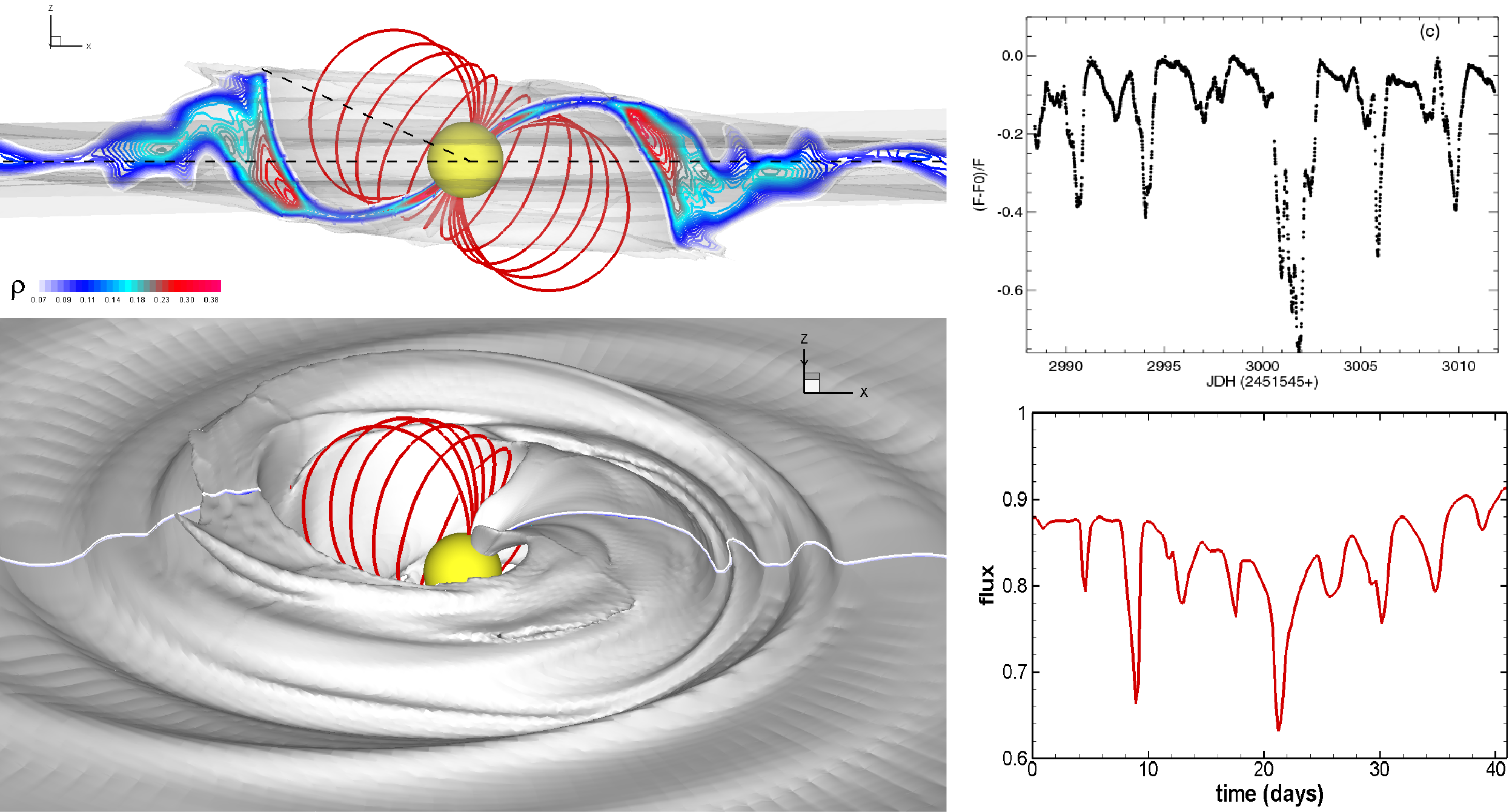}\hspace{0.0cm}
    \end{tabular}
   \end{center}
    \vspace{-0.4cm}
\caption{\textit{Left two panels:} Tilted magnetosphere excites a
warp at the disk-magnetosphere boundary. A slice of density
distribution (top panel) and one of density levels (bottom panel)
are shown. From \citet{RomanovaEtAl2013}. \textit{Right two
panels:} Top panel: typical AA Tau type light-curve observed in
one of stars in young cluster NGC 2264 (from
\citealt{AlencarEtAl2010}). Bottom panel: the light-curve obtained
from 3D MHD simulations.}
 \vspace{-0.4cm}
\label{warp-4}
\end{figure*}

Various types of waves can propagate in accretion disks (e.g,
\citealt{KatoEtAl1998,Kato2004,Kato2007}). A star with a tilted
dipole magnetosphere excites bending and density waves through the
action of magnetic forces on the inner disk (see, e.g.,
theoretical studies by
\citealt{LipunovShakura1980,Lai1999,TerquemPapaloizou2000} and
global 3D simulations by \citealt{RomanovaEtAl2013}). Global
simulations  show that, if the angular velocity of the inner disk
is close to the angular velocity of the magnetosphere (fastness
$\omega_s\sim 1$),
 a strong bending wave forms at the disk-magnetosphere boundary.
Figure \ref{warp-4} shows a snapshot from such a simulation, where
a one-armed bending wave forms at the inner edge of the disk
\footnote{Simulations also show that if a star rotates more slowly
than the inner disk, then bending waves are excited at larger
distances from the star.}. This bending wave may occult or reflect
light from the star.

For example, \citet{BouvierEtAl1999} noticed  that CTTS AA Tau has
dips in the light-curve and, interpreted these as stemming from
occultation of the stellar light by a dusty disk-warp (see also
\citealt{BouvierEtAl2003,BouvierEtAl2007b}). These dips are
quasi-periodic with QPO period of approximately $8.2$ days ,
comparable to the expected rotation period of the star. Doppler
tomography observations of AA Tau show that the dominant component
of the field is a $2-3$ kG dipole field that is tilted at
$20^\circ$ relative to the rotational axis. At this field strength
the magnetospheric radius is close to the corotation radius,
$r_m\approx r_{\rm cor}$ \citep{DonatiEtAl2010}. Simulations show
that in this situation a large amplitude warp forms and rotates
with the frequency of the star, and so could well explain the
observed dips \citep{RomanovaEtAl2013}.

More recently, \citet{AlencarEtAl2010} analyzed the photometric
variability of CTTSs in the young cluster NGC 2264 using data
obtained by the \textit{CoRoT} satellite; they concluded that AA
Tau-like light curves are  fairly common, and are present in at
least ~$30-40\%$ of young stars with inner dusty disks (see also
\citealt{CarpenterEtAl2001,StaufferEtAl2014,StaufferEtAl2015}).
 Fig. \ref{warp-4} compares observed and simulated light curves, for a
case in which  the  star is occulted by the dusty warp. Overall,
results indicate that the warps can produce dips if the dipole
magnetosphere is tilted at sufficiently larger angle,
$\Theta\gtrsim 20^\circ$, relative to the rotational axis.

There are also two persistent density waves which form in the
inner disk and rotate with the frequency which is lower than the
Keplerian frequency of the inner disk.  These waves are good
candidates for explanation of the one of two QPO frequencies
observed in AMXPs. It was frequently suggested that one of QPO
frequencies is associated with some inhomogeneities in the inner
disk, though the nature of inhomogeneities has not been understood
\citep{vanderKlis2006}. Numerical simulations show that these
inhomogeneities may be associated with persistent density waves at
the inner disk \citep{RomanovaEtAl2013}.

\subsubsection{Magnetospheric accretion from turbulent disks}
\label{sec:accretion-mri}

The full dynamical structure of accretion disks is still
uncertain. It is commonly inferred that the disks are turbulent
and the turbulence can be initiated and supported by the MRI
instability. MRI-driven turbulence has been extensively studied in
axisymmetric and local/global 3D MHD simulations (e.g.,
\citealt{HawleyEtAl1995,BrandenburgEtAl1995,StoneEtAl1996,
Armitage1998,Hawley2000,BeckwithEtAl2009,FlockEtAl2011,SimonEtAl2011}).
In most of simulations the central object is non-magnetized,
usually a black hole.

Accretion from MRI-driven turbulent disks onto a magnetized star
has been investigated numerically in both axisymmetric and 3D
simulations \citep{RomanovaEtAl2011b,RomanovaEtAl2012}. These
simulations have shown that the main aspects of  magnetospheric
accretion are similar: the turbulent disk is truncated by the
magnetosphere and matter flows to the star along funnel streams.

It is important to note that the position of the magnetospheric
radius does not depend much on the weak magnetic field associated
with the MRI-driven turbulence. Its magnetic stress is responsible
for the angular momentum transport, but the associated magnetic
pressure is smaller than gas pressure, and the ram pressure $\rho
v^2$ is much larger than gas or magnetic pressure. This is why the
formula in Eq. \ref{eq:stress balance} provides accurately the
position of the truncation radius in turbulent disks as well.

\begin{figure*}
  \begin{center}
    \begin{tabular}{cc}
             \includegraphics[width=110mm]{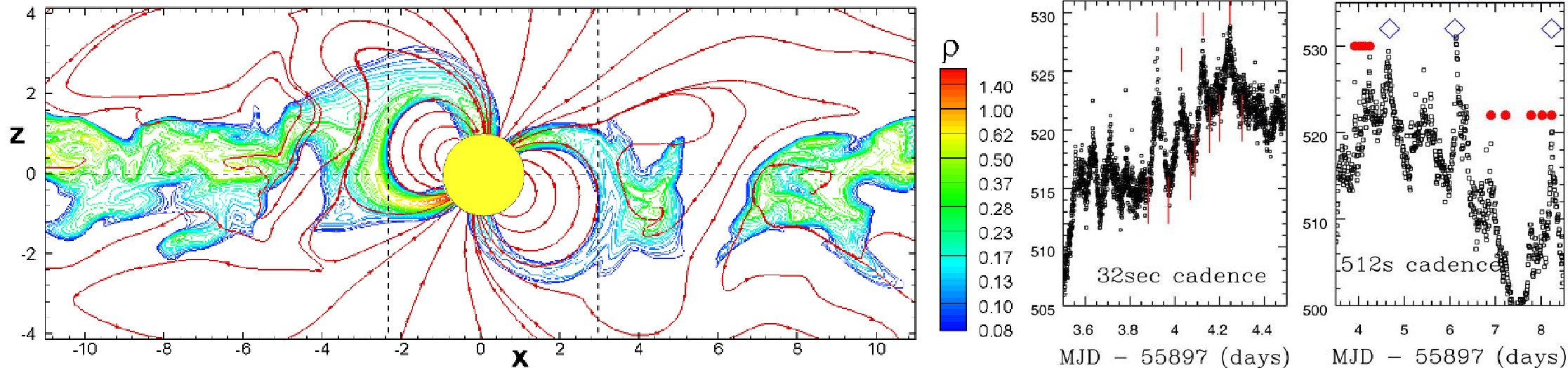}\hspace{0.0cm}
    \end{tabular}
   \end{center}
    \vspace{-0.4cm}
\caption{\textit{Left Panel:} slice of density
    distribution obtained in 3D MHD simulations of matter flow from an
    MRI-driven turbulent disk (from \citealt{RomanovaEtAl2012}).
    \textit{Right two panels:} example of the irregular light curves
 for a star from young cluster NGC 2264
    where variability on the scale
    of 1 day (right panel) and 0.2 days (middle panel) is
    observed. This variability may be caused, e.g.,  by accretion from the
    turbulent disk where the turbulence is present on different
    scales.
    From \citet{StaufferEtAl2014}.}
 \vspace{-0.4cm}
\label{mri-3d-3}
\end{figure*}

The main difference between accretion from the laminar
$\alpha-$disk and turbulent disk is that accretion of the
individual cells may lead to variability in the light-curve.
 The variability time-scale  depends on the
scale of the turbulent cells $L_{\rm turb}$. If it is small
compared with the thickness of the disk, $H$, then the disk acts
as a laminar disk and only small-amplitude variability is
expected, as found in axisymmetric simulations
    \citep{RomanovaEtAl2011b}\footnote{Many axisymmetric simulations also show strong spikes
    which are connected with
    very low (only small numerical) diffusivity in the code, and episodic accumulation of matter at the disk-magnetosphere boundary.}.
 For the largest-scale turbulent cells, $L_{\rm turb}\sim H$,
 much rarer but higher amplitude flares are expected in
 light-curves \citep{RomanovaEtAl2013}.
 Observations of  CTTSs show that many stars have irregular light-curves with
 variability observed at different time-scales (e.g., \citealt{StaufferEtAl2014}).
 Fig. \ref{mri-3d-3} (left panel) shows an example of 3D simulations from the disk
 with a large-scale turbulence,
 while the left two panels show sample light curves at two different time-scales.
 Simulations of turbulent accretion with different scales of the turbulence, together with
 comparisons between modeled and observed  light-curves, may help us understand the
 nature of turbulence in the disk.

\subsubsection{Inflation of the field lines}

Three-dimensional simulations show that the field lines of the
external magnetosphere inflate due to the disk-magnetosphere
interaction.
  Fig. \ref{3d-30-90} (left panel) shows a snapshot from a
simulation in which $\Theta=30^\circ$. One can see that the field
lines of the inner magnetosphere are closed; this is the region in
which magnetic energy dominates.  Some field lines are trapped in
the disk and are azimuthally wrapped by the disk matter. The field
lines connecting the polar regions of the star with the disk
inflate and wrap around the rotational axis forming a magnetic
tower. There is a set of the partially inflated field lines that
thread the inner parts of the disk, and transfer the angular
momentum between the star and the disk and between the star and
corona (see also \citealt{ZanniFerreira2013}). In 3D simulations,
a magnetic tower forms, however,  most simulation runs show little
or no outflow\footnote{This is the main restriction of the current
3D simulations, where the main computing power is used for
resolving the low-density closed magnetosphere and the disk, while
the density in the corona is relatively high, and matter pressure
dominates the magnetic pressure, suppressing magnetic or
magneto-centrifugally driven outflows.}. The inflation of the
field lines is observed at different tilts of the dipole
magnetosphere, including the extreme case of very large tilt of
the magnetosphere, $\Theta=90^\circ$, where the magnetic axis is
located in the equatorial plane of the disk. The right panel of
Fig. \ref{3d-30-90} shows that the magnetic field lines of the
external magnetosphere inflate, and also have a tendency to leave
the disk and to expand into the corona above and below the disk.
Simulations show that the magnetic field lines always wrap around
the rotational axis, independent of of the tilt angle $\Theta$.
However, it is not yet clear whether the outflows represent a
common feature of accreting magnetized stars, or whether they are
more typical for the case of the rapidly-rotating stars  (see Sec.
\ref{sec:propeller}). Outflows were investigated in greater detail
in 2.5D (axisymmetric) simulations, where the density in the
corona is
 lower compared with the 3D simulation cases (see
Sec. \ref{sec:propeller}).

\subsubsection{Stellar spinup/spindown from angular momentum
gain/loss} \label{sec:accretion-angmom}

A star may spin up or spin down, depending on the total angular
momentum flux onto the star. The net angular momentum flux to or
from the surface of the star, $\dot{L}$, is composed of two parts:
the angular momentum flux carried by the matter ($\dot{L}_m$) and
that carried by the magnetic field ($\dot{L}_f$) :

\begin{equation}
\dot{L}=\dot{L}_m+\dot{L}_f,~~~~~
\dot{L}_m=\int\mathrm{d}\mathbf{S}\cdot\rho{r}v_\phi\mathbf{v_p},~~~~~
\dot{L}_f=-\int\mathrm{d}\mathbf{S}\cdot{r}B_\phi\mathbf{B_p}/4\pi,
\end{equation}
where the fluxes are integrated along the surface of the star;
$\mathbf{v_p}$ and $\mathbf{B_p}$ are the poloidal velocity and
the poloidal magnetic field.

Simulations show that the disk and the star exchange angular
momentum mainly through the funnel streams. Near the disk, the
angular momentum in the funnel stream is carried  by the matter,
and the matter then slightly twists the field lines threading the
funnel streams; this twist determines the torque on the star
(e.g., \citealt{RomanovaEtAl2002,LongEtAl2005,ZanniFerreira2013}).
The sign of the term $B_\phi\mathbf{B_p}$ (more precisely, the
sign of the $B_\phi$ component at the stellar surface) determines
whether the star spins up or down. At the stellar surface, the
angular momentum carried by the field, $\dot{L}_f$ is much larger
than that carried by matter, $\dot{L}_f$.

 \citet{LongEtAl2005} performed a set of 2D simulation runs at
different angular velocities of the star, $\Omega_\star$, to find
the critical value of the fastness parameter $\omega_s$,
corresponding to the rotational equilibrium state (i.e., when
$\dot{L}=0$). He found that, in the rotational equilibrium state
$r_{\rm cor}/r_m\approx 1.2-1.3$, that is the fastness parameter
$\omega_s=0.76-0.67$. They found that the inner disk rotates more
rapidly than the magnetosphere, because  the star loses some
angular momentum to the corona along the inflated or
partially-inflated field lines. This negative torque is
compensated by a positive torque from the inner disk. Therefore,
the fastness parameter corresponding to the rotational equilibrium
state, $\omega_s$, strongly depends on the amount of angular
momentum that flows to the inflated or partially-inflated field
lines \footnote{This angular momentum depends on the coronal
density: it can be larger in case of young stars, which can have
strong stellar winds, and smaller in cases of neutron stars.}.

3D simulations also show that a star can  spin up, spins down, or
be in a rotational equilibrium state. It is interesting to note
that the matter fluxes onto the stellar surface, as well as the
torques on the star, are similar across different values of
$\Theta$ (e.g., \citealt{RomanovaEtAl2003}). This is because the
magnetospheric radius is approximately the same for different
$\Theta$.

In case of steady long-lasting accretion, a magnetized star is
expected to be in the rotational equilibrium state. Probably many
CTTSs are in this state (see, e.g., review by
\citealt{BouvierEtAl2007a}). In accreting millisecond pulsars, the
situation is more complex, because accretion occurs during
episodes of accretion outbursts, where the accretion rate
increases a few orders of magnitude, then decreases back to small
values (see reviews by \citealt{vanderKlis2006,PatrunoWatts2012}).
A propeller stage and spinning-down of a neutron star is expected
during the rise and decline of the outburst (e.g.,
\citealt{PatrunoEtAl2009}), while the magnetospheric accretion and
spinning-up is expected during period of the high accretion rate.
These issues are further discussed in Sec. \ref{sec:propeller}.


\begin{figure*}
  \begin{center}
    \begin{tabular}{cc}
             \includegraphics[width=120mm]{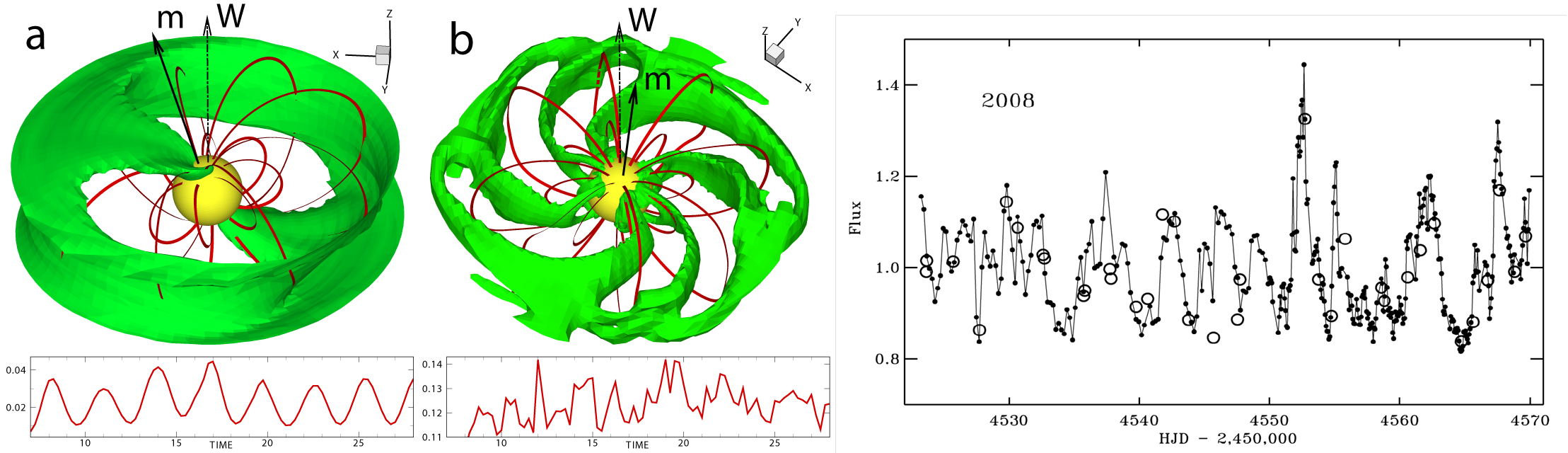}\hspace{0.0cm}
    \end{tabular}
   \end{center}
    \vspace{-0.4cm}
\caption{\textit{Left two panels:} a 3D view of matter flow in the
stable (left) and unstable (right) regime of accretion.
 Bottom panels show the light curves from the hot spots (from
 \citealt{RomanovaEtAl2008}). \textit{Right panel:} The light
curve of the CTTS TW Hya obtained by the \textit{MOST} satellite
  (from \citealt{RucinskiEtAl2008}). }
 \vspace{-0.4cm}
\label{stab-unstab}
\end{figure*}


\begin{figure*}
\centering
         \includegraphics[width=90.mm,clip]{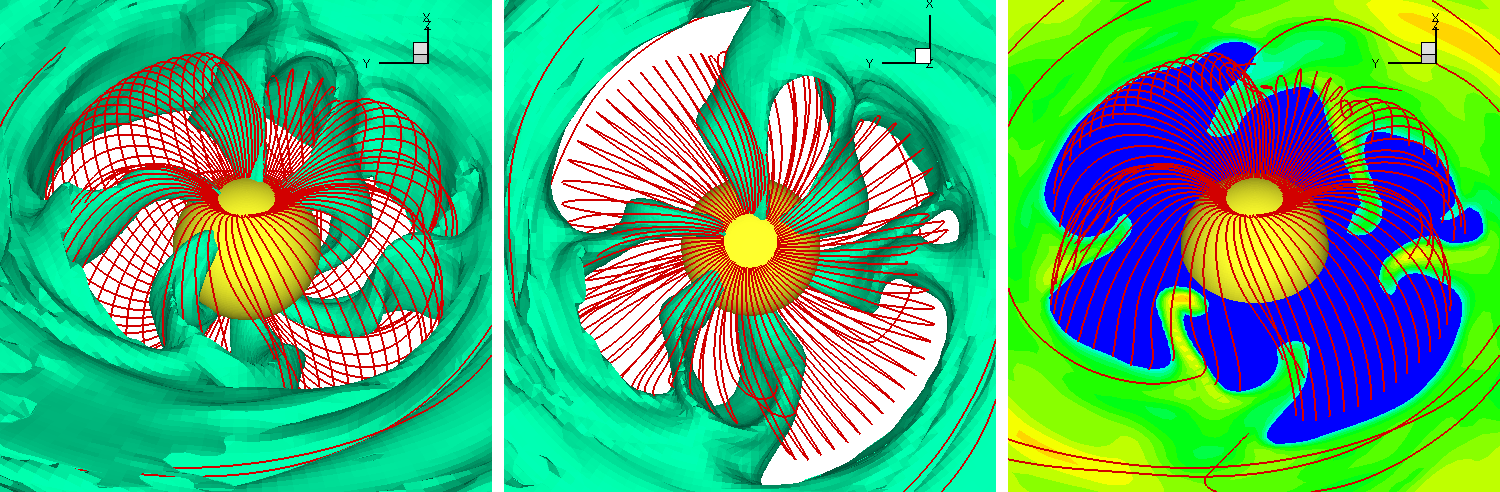}
 \vspace{-0.0cm}
\caption{{\textit{Left panel}: 3D view of matter flow in a case
where chaotic accretion in multiple tongues dominates. One of the
density levels is shown in color, selected magnetic field lines
are shown in red. \textit{Middle panel}: Same but in the face-on
projection. \textit{Right panel:} An equatorial slice of density
distribution is shown in color.} The figure has been created for
the model $\mu1c2.5\Theta5a0.02$ calculated by
\citet{BlinovaEtAl2016}.}
 \vspace{-0.4cm}
\label{tongues-3}
\end{figure*}

\begin{figure*}
\centering
         \includegraphics[width=110.mm,clip]{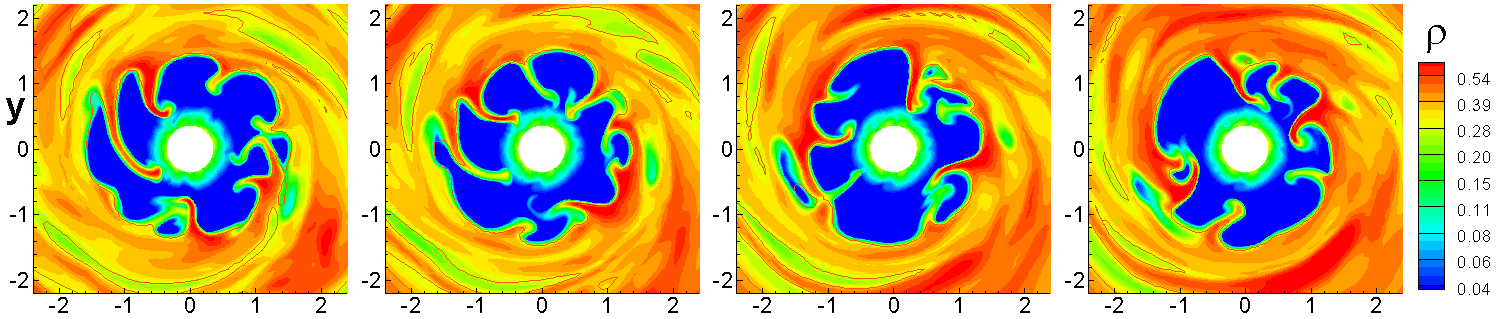}
         \includegraphics[width=110.mm,clip]{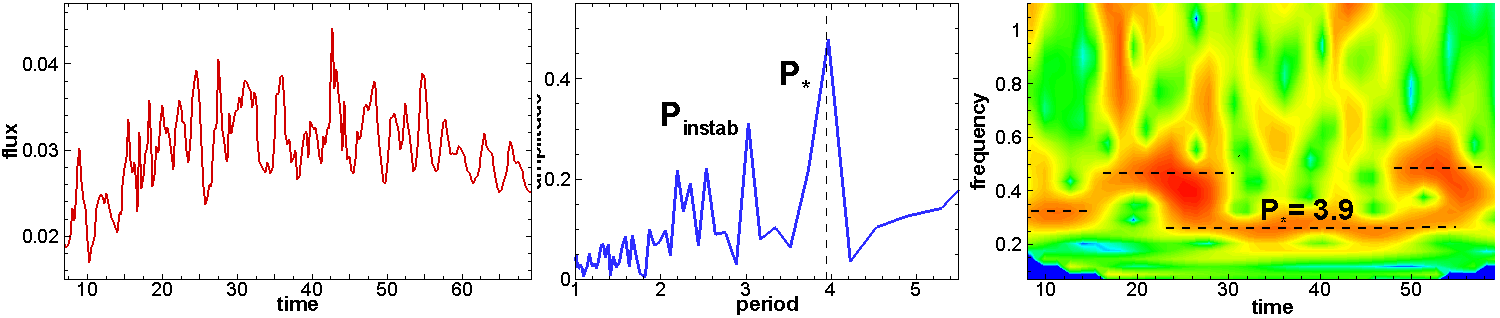}
 \vspace{-0.0cm}
\caption{\textit{Top panels:} Consecutive $xy$-slices  show the
density distribution in the chaotic unstable regime in the model
with $r_{\rm cor}=2.5$ at times $t=16-18$. \textit{Bottom panels:}
The light curve from the rotating hot spots calculated at an
inclination angle of the rotational axis relative to the observer
$i=45^\circ$ (bottom left panel), the Fourier transform obtained
from analysis of the light curve (bottom middle panel), and the
wavelet transform obtained from analysis of the light curve
(bottom right panel). From \citet{BlinovaEtAl2016}.}
 \vspace{-0.4cm}
\label{unstab-chaotic}
\end{figure*}

\subsection{The unstable regime of accretion due to Rayleigh-Taylor instability}
\label{sec:stab-unstab}

A new regime of accretion has been found in 3D simulations of
stars whose dipole moment is only slightly tilted relative to the
rotational axis ($\Theta=5^\circ$). These simulations were
performed in order to better understand AMXPs, for which the tilt
of the dipole magnetosphere is expected to be small. The
simulations show that, in the unstable accretion regime, the
magnetic Rayleigh-Taylor (R- T) instability enables matter to
penetrate the magnetosphere via several chaotically-formed
``tongues''
\citep{KulkarniRomanova2008,RomanovaEtAl2008,BachettiEtAl2010,KurosawaRomanova2013}.
 In earlier studies, it was
suggested that the instabilities at the disk-magnetosphere
boundary may lead to the mixing of plasma with the field in the
external layers of the magnetosphere \citep{AronsLea1976}.
However, global 3D MHD simulations show that the unstable tongues
can deeply penetrate into the magnetosphere. The fact that the
magnetic field is not an obstacle for the Rayleigh-Taylor
instability has been also shown by
\citet{StoneGardiner2007a,StoneGardiner2007b} who performed local
(in the box) 3D simulations wherein the homogeneous magnetic field
is placed at different orientations relative to the boundary
between the light and heavy fluids. These simulations and earlier
ones by
\citet{WangRobertson1984,WangRobertson1985,RastatterSchindler1999}
show that the Rayleigh-Taylor instability leads to the formation
of small-scale waves and filaments that merge to form much larger
filaments, which then deeply penetrate into the
magnetically-dominated, low-density regions.

The matter in the randomly-forming tongues is accelerated by
gravity and is deposited onto the stellar surface as hot spots
with irregular shape and random position. The light curves from
these spots are often irregular, with several peaks per rotational
period. Fig. \ref{stab-unstab} (two left panels) compares 3d views
of matter flow in the stable and unstable regimes, as well as the
light curves from the hot spots.

Observations of CTTSs show that many CTTSs exhibit irregular,
chaotic-looking light curves
\citep{HerbstEtAl1994,AlencarEtAl2010,CodyEtAl2014,StaufferEtAl2014}.
 The right panel of Fig.
\ref{stab-unstab} shows the chaotic-looking light curve of the
CTTS TW Hya obtained with the \textit{MOST} satellite
\citep{RucinskiEtAl2008}. The characteristic time-scale between
flares (a few events per rotational period) corresponds to that
expected in the unstable regime of accretion.

Fig. \ref{tongues-3} shows an example of accretion through
instabilities. The panels show that matter of the disk pushes the
magnetic field lines aside, and penetrates deep onto the
magnetosphere, where it is stopped and channeled into the funnel
streams. These usually do not live long, less than period of
rotation at the inner disk. After depositing matter onto the star,
they become weaker and are destroyed by interaction with the
magnetosphere. Fig.\ \ref{unstab-chaotic} (top panels) shows
several consecutive slices of the density distribution. One can
see that the number and the position of tongues varies in time.
The light-curve calculated from the hot spots (at an observer's
angle $i=45^\circ$) shows somewhat chaotic behavior (see bottom
left panel of the same figure). Fourier analysis of this light
curve shows several frequencies that are associated with chaotic
hot spots, as well as the rotational period of the star. Wavelet
analysis shows at which interval the instabilities are stronger.

\subsubsection{Theoretical background for Rayleigh-Taylor instability}

\begin{figure*}[!ht]
  \begin{center}
    \begin{tabular}{cc}
             \includegraphics[width=80mm]{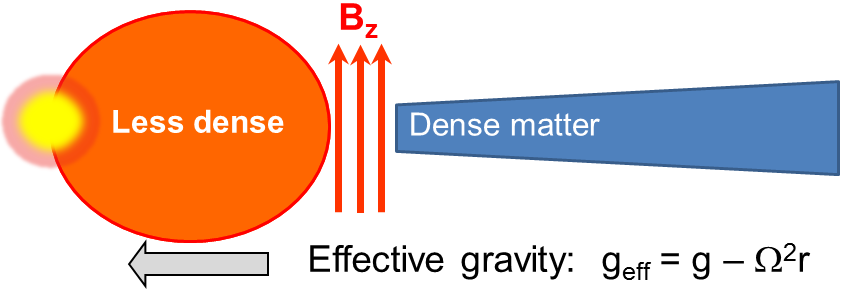}\hspace{0.0cm}
    \end{tabular}
  \end{center}
   \vspace{-0.4cm}
 \caption{A sketch of the situation where accretion
through R-T instability is favorable. The disk is denser than the
magnetosphere, and the effective gravity is directed toward the
star; the poloidal dipole field is not an obstacle for the R-T
instability.}
 \vspace{-0.4cm}
 \label{sketch-stab-unstab}
\end{figure*}

According to theoretical studies (e.g.,
\citealt{Chandrasekhar1961}), a magnetic field that is parallel to
the boundary between the heavy and light fluids is not an obstacle
for the development of R-T unstable modes at the boundary. The
sketch in Fig. \ref{sketch-stab-unstab} demonstrates that the
situation at the disk-magnetosphere boundary is similar to that
considered by \citet{Chandrasekhar1961}.
  In the case of a rotating
disk, the inner disk can be unstable if the effective
gravitational acceleration is negative: $g_{\rm eff}=g+g_c <0$,
where $g=-GM_\star/r^2$ and $g_c=v_\phi^2/r$ are gravitational and
centrifugal accelerations, respectively.

There are a few factors that can suppress the instability,
however. Specifically, the radial shear of the angular velocity,
$2 \left( r \dd[\Omega]{r} \right)^2$, can do so by smearing out
the perturbations. \citet{SpruitEtAl1995} performed a general
analysis of disk stability in the thin disk approximation, taking
the velocity shear into account (see also earlier work by
\citealt{KaisigEtAl1992}). The disk has a surface density $\Sigma$
and is threaded by a magnetic field with a vertical component,
$B_z$. Their analytical criterion for the development of
instability is:
\begin{equation}
 \gamma_{B\Sigma}^2 \equiv (-g_{\rm eff}) \left| \dd{r} \ln
\frac{\Sigma}{B_z} \right| > 2 \left( r \dd[\Omega]{r} \right)^2
\equiv \gamma_\Omega^2~. \label{eq:spruit}
\end{equation}
One can see that the sign and value of the effective gravitational
acceleration $g_{\rm eff}$ are important in this criterion.
Namely, the disk-magnetosphere boundary is unstable if $g_{\rm
eff}$ is negative (i.e., when the effective acceleration is
directed towards the star). The term $\left| \dd{r} \ln
\frac{\Sigma}{B_z} \right|$ characterizes the level of
\textit{compression} of the surface density and magnetic field in
the disk. Instability occurs if the product of $(-g_{eff})$ and
$\left| \dd{r} \ln \frac{\Sigma}{B_z} \right|$ is large enough to
overcome the stabilizing effect of the velocity shear,
$\gamma_\Omega^2$. Simulations show that this criterion, developed
for a disk, also works well for the disk-magnetosphere boundary
(e.g.,
\citealt{KulkarniRomanova2008,KulkarniRomanova2009,BlinovaEtAl2016}).

\begin{figure*}[!ht]
\centering
\includegraphics[width=110mm]{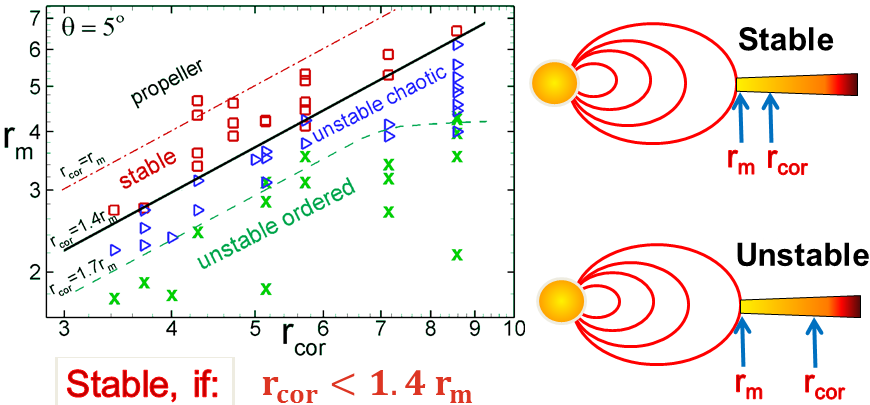}
 \vspace{-0.0cm}
\caption{The boundary between the stable and unstable regimes of
accretion strongly depends on the ratio between the magnetospheric
and corotation radii ($r_m$ and $r_{\rm cor}$, respectively).
\textit{Left panel:} The result of multiple simulation runs.
Squares, triangles, and crosses correspond to the stable,
chaotically unstable, and ordered unstable regimes of accretion,
respectively (from \citealt{BlinovaEtAl2016}). \textit{Right
panels:} The positions of the corotation and magnetospheric radii
in the cases of the stable and unstable regimes of accretion,
respectively. }
 \vspace{-0.4cm}
\label{stab-unstab-diagram}
\end{figure*}

\subsubsection{The boundary between the stable and unstable accretion regimes}

It is important to identify the boundary between the stable and
unstable regimes of accretion. Recent studies show that this
boundary occurs at a critical ratio of magnetospheric to
corotation radii $r_{\rm cor}/ r_m  = k \approx 1.4$,
corresponding to a fastness parameter $\omega_s = (r_m/r_{\rm cor}
)^{3/2} \approx 0.6$ \citep{BlinovaEtAl2016}. Fig.
\ref{stab-unstab-diagram} shows the results of multiple simulation
runs (symbols). The black solid line shows the boundary between
the stable and unstable accretion regimes; accretion is unstable
below the line and stable above the line. The instability becomes
stronger further away from the line, and it is mildly strong near
the line. The red dashed line shows the boundary $r_m=r_{\rm cor}$
($\omega_s=1$), above which the propeller regime is expected.
These simulations were performed for stars with $\Theta=5^\circ$
and small viscosity parameter, $\alpha=0.02$.

Test simulations with larger values of $\alpha-$parameter of
viscosity (i.e., $\alpha=0.1$), show that the instability is
stronger, and the transition to the unstable regime is expected at
the larger values of the fastness parameter, $\omega_s>0.6$ (i.e.,
closer to the propeller line $\omega_s=1)$, compared with the
lower viscosity, $\alpha=0.02$, case. This can be explained by the
fact that, at larger values of $\alpha$, the compression factor
$\left| \dd{r} \ln \frac{\Sigma}{B_z} \right|$ is larger, and thus
the instability starts more readily
\citep{BlinovaEtAl2016}\footnote{The effective $\alpha-$parameter
can be estimated from comparisons of observations with models of
accretion. For example, \citet{Bisnovatyi-KoganEtAl2014} derived
the value $\alpha=0.1-0.3$ for outburst of accretion in X-ray
transient A0535+26/HDE245770.}. In addition, a course grid
resolution suppresses the instability: comparisons of simulations
at different grid resolutions show that the instability starts at
larger values of $\omega_s$ in case of the finer grid, implying
that the courser grid effectively acts to suppress the instability
\citep{BlinovaEtAl2016}. This leads to the question whether the
boundary between regimes may be located at the line $r_{\rm cor}/
r_m \approx 1.0$, if the grid is sufficiently fine ? Addressing
this will require a systematic set of studies at even higher grid
resolution.

An interesting phenomenon has been recently observed in the AMXP
SAX J1808.4-3658, where the rotation frequency of the star is
known ($\nu_\star=401$ Hz), and where the two main QPO frequencies
vary between values lower and higher than this stellar rotation
frequency (see Fig. \ref{qpo-wijnands}).  It has often been
suggested that one of frequencies is associated with the frequency
of the Keplerian rotation at the inner disk, but clear evidence
for this has been lacking. Recent studies
\citep{BultVanDerKlis2015} found that the pulse fraction (which is
associated with the magnetospheric accretion onto the surface of
the neutron star) strongly decreases when the QPO frequency
increases up to the frequency of the star, that is when the inner
disk rotates more rapidly than the magnetosphere of the star. At
the end of the outburst, when the accretion rate decreases, the
QPO frequency decreases, and when it passes the stellar frequency,
the pulse fraction increases back to large value. This phenomenon
provides evidence that this QPO is associated with the inner disk,
and that the transition of the inner disk through the point
$r_m=r_{\rm cor}$ leads to change in the magnetospheric accretion.
One possibility is that accretion becomes unstable and therefore
the pulse fraction decreases when the inner disk reaches the
radius $r_m=r_{\rm cor}$ and moves closer to the star (van der
Klis, privet comm.). If true, then the boundary line between
stable and unstable regimes should be at $r_m\approx r_{\rm cor}$.
To substantiate this hypothesis, more work is needed in both
numerical modeling and observations of different AMXPs.

Modeling of accretion onto stars with larger tilts of the dipole
magnetosphere (i.e., $\Theta=10^\circ, 20^\circ$ and $30^\circ$)
shows that the instability develops, however more matter accretes
in funnel streams above the magnetosphere. The hot spots
associated with these streams provide the sinusoidal component of
the light curve, while the chaotic hot spots provide the irregular
component in the light curve.  The amplitude of this irregular
component is high for $\Theta=10^\circ$, but it decreases for
models with larger $\Theta$ \citep{BlinovaEtAl2016}.

\begin{figure*}[!ht]
\centering
\includegraphics[width=90mm]{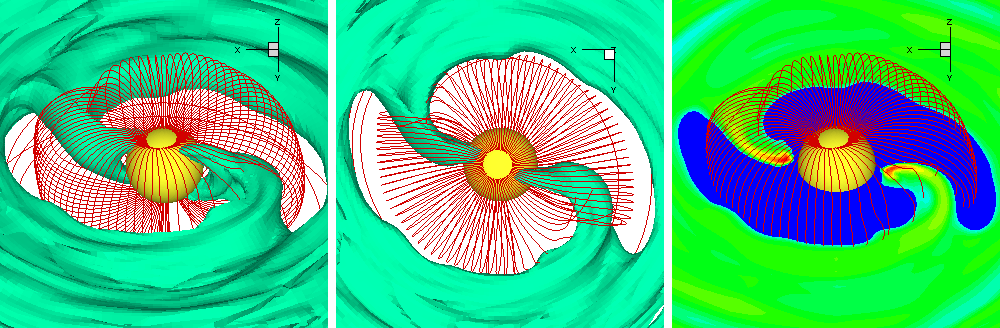}
 \vspace{-0.0cm}
\caption{\textit{Left panel}: 3D view of matter flow in a case
where ordered accretion in two tongues dominates. One of the
density levels is shown in color, selected magnetic field lines
are shown in red. \textit{Middle panel}: Same but in the face-on
projection. \textit{Right panel:} An equatorial slice of density
distribution is shown in color. From \citet{BlinovaEtAl2016}.}
 \vspace{-0.4cm}
 \label{unstab-ordered-3}
\end{figure*}

\begin{figure*}[!ht]
\centering
\includegraphics[width=110mm]{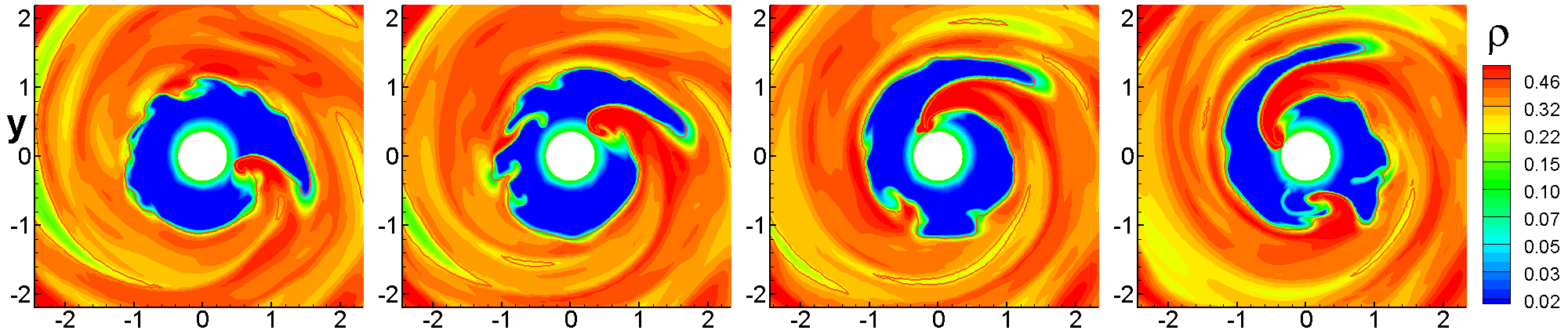}
 \includegraphics[width=110mm]{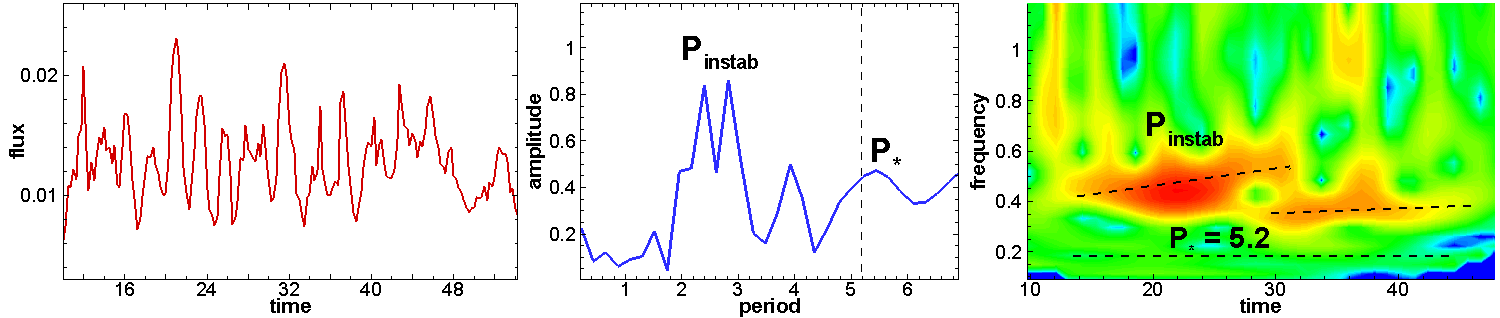}
 \vspace{-0.0cm}
\caption{An example of accretion in the ordered unstable regime.
\textit{Top panels:} Consecutive views of  $xy$ slices. The color
background shows the density distribution, and the lines show
where the kinetic plasma parameter $\beta_t=1$. \textit{Bottom
Panels:} The light curve from the rotating hot spots calculated at
an inclination angle $i=45^\circ$ (bottom left panel), the Fourier
transform obtained from analysis of the light curve (bottom middle
panel), and the wavelet transform obtained from analysis of the
light curve (bottom right panel). From \citet{BlinovaEtAl2016}.}
 \vspace{-0.4cm}
 \label{unstab-ordered-2}
\end{figure*}

\subsubsection{The ordered unstable regime}

Another important regime of accretion has been found in
simulations of accretion onto slowly rotating stars, when the
condition
 $r_{\rm cor} \gtrsim 1.7 r_m$ has been satisfied \citep{BlinovaEtAl2016}.
In this regime, one or two ordered tongues form and rotate with
the angular velocity of the \textit{inner disk}. Fig.
\ref{unstab-ordered-3} shows an example of accretion in this
ordered unstable regime. The top panels of Fig.
\ref{unstab-ordered-2} show equatorial slices, and the bottom
panels show the light curve associated with rotating hot spots, as
well as Fourier and wavelet analysis of this light curve. The
period associated with this instability dominates in both the
Fourier and wavelet spectra. This boundary between chaotic and
this newly found ordered
 unstable regimes is shown in Fig.
 \ref{stab-unstab-diagram}  as the green dashed line.
 It corresponds to the fastness parameter $\omega_s\approx 0.45$.

The frequency of oscillations in the ordered unstable regime
depends on the accretion rate. When the accretion rate increases,
the inner disk moves closer to the star, and the frequency of
oscillations increases (and vice versa). This mechanism may be
important for understanding the high-frequency oscillations in
AMXPs, for which a frequency-luminosity correlation has been
observed (e.g., \citealt{PapittoEtAl2007}). Ordered rotation of
``tongues" probably reflects an ordered rotation of the density
waves in the inner disk; however this issue should be studied
further. The QPO radiation may come from the inner disk waves,
from the ordered tongues, or from the rapidly-rotating hot spots
which glide along the surface of the slower-rotating star.

In the chaotic unstable regime, the whole system of
chaotic-looking tongues  rotates with the angular velocity of the
inner disk. This phenomenon could be detected in AMXPs, where the
observations cover many rotational periods of the neutron star.
This ordered rotation of the system of chaotic spots could induce
a QPO with a frequency $\nu_{QPO}$ that increases when the inner
disk moves inward. During this motion, the unstable tongues become
more and more ordered, and the AMXP switches gradually to the
ordered regime of unstable accretion. During this process, the
quality factor $Q=\nu_{QPO}/\Delta \nu_{QPO}$ increases. In fact
observations do indicate  that the quality factor of QPOs usually
increases with the frequency of QPO  (e.g.,
\citealt{BarretEtAl2007}).

In young stars accretion in the ordered unstable regime is
expected during periods of enhanced accretion rate, that is,
during EXor and FUor stages of evolution (see ,e.g., review by
\citealt{AudardEtAl2014}). Any short-period oscillations may be
connected with  rotation of the hot spots associated with ordered
unstable tongues.
 Recent X-ray observations of the young star
V1647 Ori during two accretion outbursts revealed a $\sim$ 1 day
period that is very short for a young star, and corresponds almost
to the stellar rotation break-up period \citep{HamaguchiEtAl2012}.
Numerical models suggest that this type of variability could be
connected with the ordered unstable regime of accretion. Future
search of the short-term variability in young stars would be
helpful for understanding processes  during their accretion
outburst, like in FUors and EXors.

In the above simulations the unstable regime has been investigated
in case of $\alpha-$disks. In the future, similar studies should
be performed with the MRI-driven turbulent disks.

\begin{figure*}[!ht]
  \begin{center}
    \begin{tabular}{cc}
      \includegraphics[width=80mm]{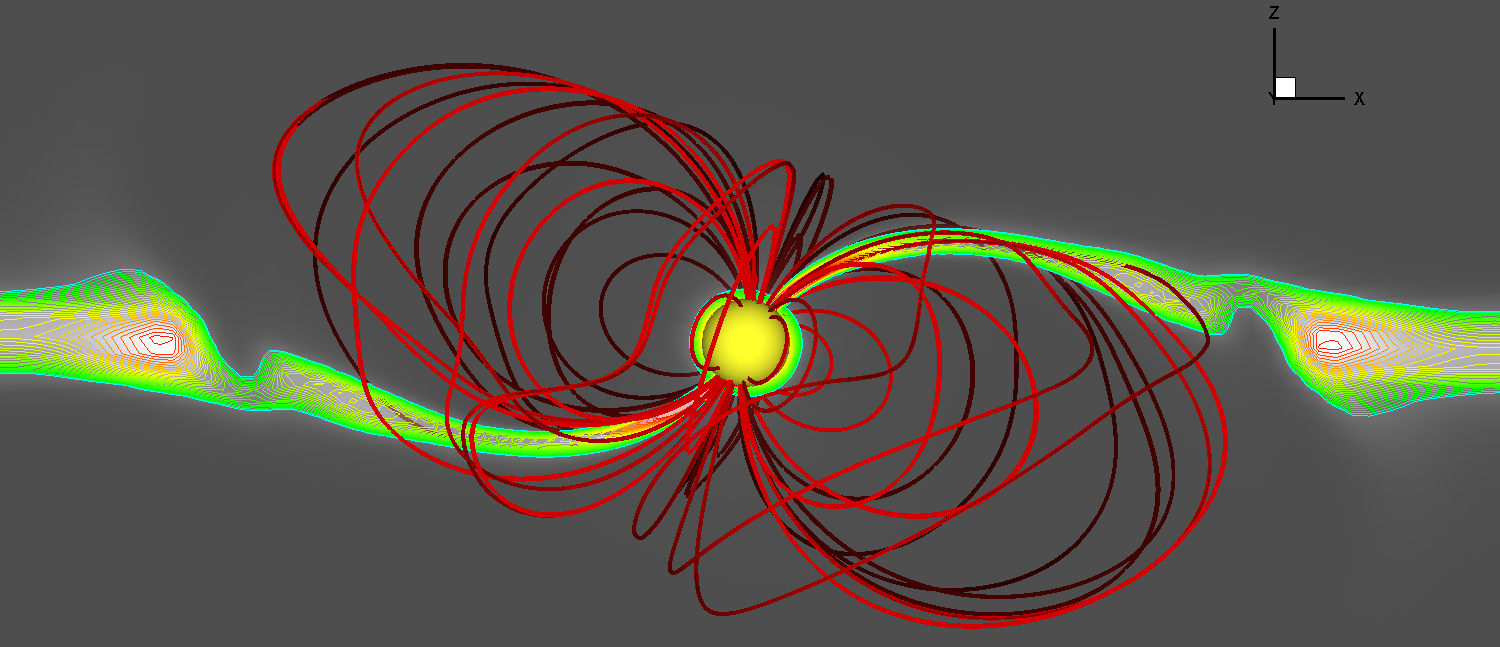}
\end{tabular}
  \end{center}
 \vspace{-0.4cm}
 \caption{An XZ-slice in the density distribution in the case of accretion onto a star
with a large magnetosphere ($r_m\approx 12 R_\star$). The lines
are sample
 magnetic field lines. From \citet{RomanovaEtAl2014}.}
 \vspace{-0.8cm}
  \label{magn-large-slice}
\end{figure*}

\begin{figure*}[!ht]
  \begin{center}
    \begin{tabular}{cc}
             \includegraphics[width=80mm]{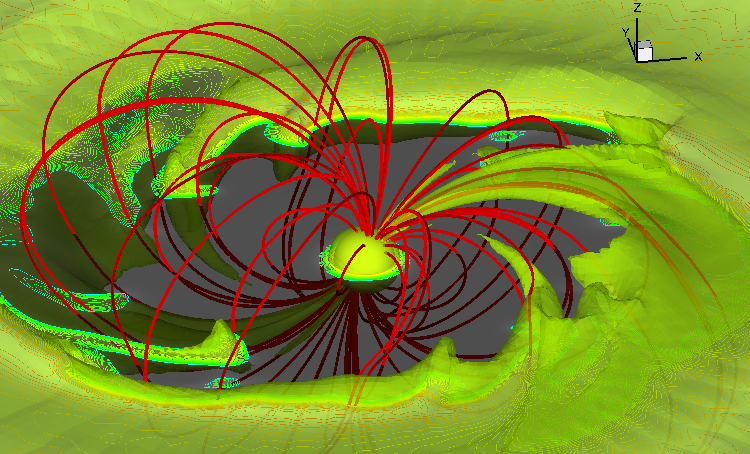}
\end{tabular}
  \end{center}
   \vspace{-0.4cm}
 \caption{A 3D view of matter flow onto a star  with a large magnetosphere. The color background shows
the density distribution in the equatorial slice. The green
translucent layer shows the density distribution in 3D. From
\citet{RomanovaEtAl2014}.} \vspace{-0.4cm} \label{magn-large-3d}
\end{figure*}

\subsection{Accretion onto stars with larger magnetospheres}
\label{sec:larger-magnetosphere}

The simulations described in the above sections are relevant to
stars with relatively small magnetospheres,  such as CTTSs and
AMXPs, where the size of the magnetosphere is only a few times
larger than the radius of the star. However, the magnetospheres
are larger for intermediate polars, $r_m/R_\star \gtrsim 10$, and
much larger  for X-ray pulsars. A special set of simulations has
been performed  to investigate the disk-magnetosphere interaction
in systems with larger stellar magnetospheres, where the
magnetospheric radius $r_m>10 R_\star$ \citep{RomanovaEtAl2014}.
Much larger values of the stellar magnetic moments were taken for
these simulations.

The slice of the density distribution in Fig.
\ref{magn-large-slice} shows that the disk matter stops at the
distance $r_m\approx 12 R_\star$ and flows onto the star in two
funnel streams, which are narrow in the vertical direction.  Fig.
\ref{magn-large-3d}, on the other hand, shows that the funnel
stream is wide in the horizontal direction. The corotation radius
$r_{\rm cor}=20 R_\star$   is large compared to the magnetospheric
radius, $r_m\approx 12 R_\star$, and accretion through R-T
instabilities is expected. Simulations show that the matter in the
inner disk penetrates through the external layers of the
magnetosphere due to the R-T instability, and this instability is
observed at the disk-magnetosphere boundary. However, the unstable
tongues only penetrate the outer parts of the external
magnetosphere, while the matter flows onto the star in two ordered
funnel streams. Each funnel stream may split into several streams,
because matter flows into the funnel stream from higher-density
regions (i.e., from the unstable tongues; see Fig.
\ref{magn-large-3d}).

\subsection{Accretion onto stars with complex magnetic fields}
\label{sec:complex}

Measurements of the surface magnetic fields of CTTSs indicate that
they have a complex structure (e.g.,
\citealt{Johns-KrullEtAl1999,Johns-Krull2007}). Furthermore,
measurements of the magnetic fields of nearby low-mass stars using
the Zeeman-Doppler technique show that their fields are, also,
often complex (e.g.,
\citealt{DonatiCameron1997,DonatiEtAl1999,JardineEtAl2002}). The
observed surface magnetic field is often approximated with a set
of multipoles whose magnetic moments have different misalignment
angles relative to the rotational axis (i.e., different tilts), as
well as different phases in the longitudinal direction (e.g.,
\citealt{JardineEtAl2002}). The surface magnetic field of a star
with a complex magnetic field can be approximated by a
superposition of tilted dipole, quadrupole, octupole, and higher
order multipoles: $\bf B_\star = {\bf B}_{\rm dip} + {\bf B}_{\rm
quad} + {\bf B}_{\rm oct} + ...$.

A star with a quadrupole or octupole magnetic field will also
truncate the disk, albeit at smaller distances from the star
(compared to that of a pure dipole field). The magnetospheric
radius can be derived in analogy with the case of a pure dipole
field (see Eq. \ref{eq:alfven}). For simplicity, we assume that
the primary axis of the multipolar field is aligned with the
rotation axis, and therefore that the $n-$th multipolar component
of the magnetic field can be written in the form
$B_n\sim\mu_n/r^{n+2}$. From the balance of the largest stresses
$B^2/8\pi=\rho v_\phi^2$,  and taking $v_\phi = v_{\rm Kep}$, we
obtain:

\begin{figure*}
  \begin{center}
    \begin{tabular}{cc}
             \includegraphics[width=62mm]{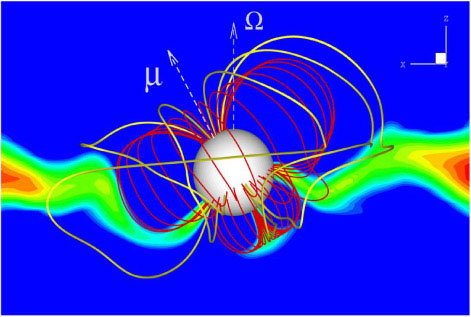}
                \includegraphics[width=56mm]{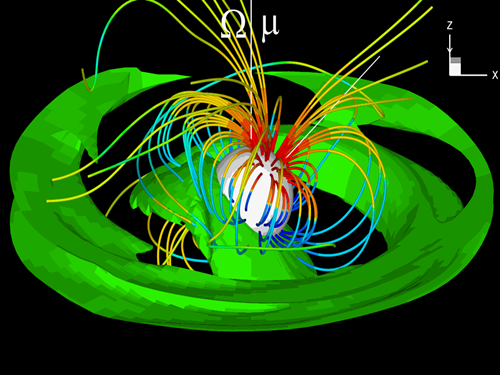}
    \end{tabular}
  \end{center}
 \vspace{-0.4cm}
 \caption{
 \textit{Left panel:} the $XZ-$slice of the density distribution and sample magnetic field lines
 in the case where the dipole and quadrupole magnetic moments are aligned and both are tilted
by an angle $\Theta=30^\circ$ relative to the rotational axis.
\textit{Right panel:} A 3D view of matter flow onto a star for
which the quadrupole moment is tilted about the dipole moment by
$45^\circ$. From \citet{LongEtAl2007,LongEtAl2008}.}
\vspace{-0.4cm} \label{complex-quad-2}
\end{figure*}

\begin{equation}\label{eq:multipole}
r_{m,n}^{(0)}=k_n
\mu_n^{\frac{4}{4n+3}}\dot{M}^{-\frac{2}{4n+3}}(GM_\star)^{-\frac{1}{4n+3}},
\end{equation}
where $k_n\sim 1$ is a coefficient that can be different for
different multipoles.  This formula can be applied in the cases of
aligned multipoles, and it can also be used for estimates of the
magnetospheric radius in the more general case in which the
multipole is not aligned with the rotational axis. We should note
that the $B_z = 0$ in the disk plane for aligned fields of $2n$-th
order multipoles, such as a quadrupole field \citep{LongEtAl2007},
and matter could flow directly onto the star in the disk plane. So
$r_{m,n}$ does not reflect where the inflowing matter stops, but
only the size of the magnetosphere. \cite{Gregory2011} studied
accretion onto stars with the complex magnetic fields
semi-analytically, suggesting a fixed configuration of the field
and showed that complex paths of matter flow are expected.

\begin{figure}
\sidecaption
              \includegraphics[width=60mm]{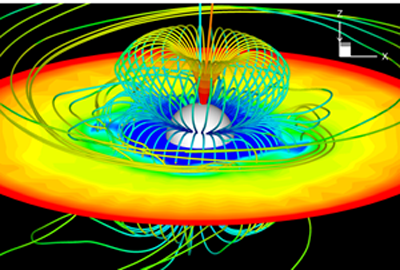}
 \vspace{-0.4cm}
\caption{A 3D view of the magnetic field distribution, as well as
a density slice in the equatorial plane, in the case of accretion
onto a star with a predominantly octupolar field. The colors along
the field lines represent different polarities and strengths of
the field. The thick cyan and orange lines represent the rotation
and octupole moment axes, respectively (from
\citealt{LongEtAl2012}).}
 \vspace{-0.2cm}
\label{octupole-vert}
\end{figure}

A number of global 3D MHD simulations have been performed with the
goal of better understanding accretion onto stars with complex,
non-dipole magnetic fields. Initial simulations were performed to
study the superposition of the dipole and quadrupole fields
\citep{LongEtAl2007, LongEtAl2008}. These simulations show that
matter partially accretes onto the star near the magnetic poles,
and partially to the ring associated with the quadrupole component
of the field.  Fig. \ref{complex-quad-2} shows an example of
simulations where the dipole and quadrupole moments are aligned
(left panel) and misaligned (right panel). In the misaligned case,
the magnetic field looks more complex than in the aligned case.

In the case of a superposition of the dipole and octupole fields,
usually even a small dipole field determines the magnetospheric
accretion, because the octupole field decreases with the distance
rapidly ($B_n\sim r^{-5}$). That is why, to demonstrate accretion
onto a star with octupolar field, the dipole component has been
taken to be very small for demonstration (see Fig.
\ref{octupole-vert}; \citealt{LongEtAl2012}).
In this case, the matter accretion forms two
equatorial belts.

\begin{figure*}[!ht]
  \begin{center}
             \includegraphics[clip,height=0.3\textwidth]{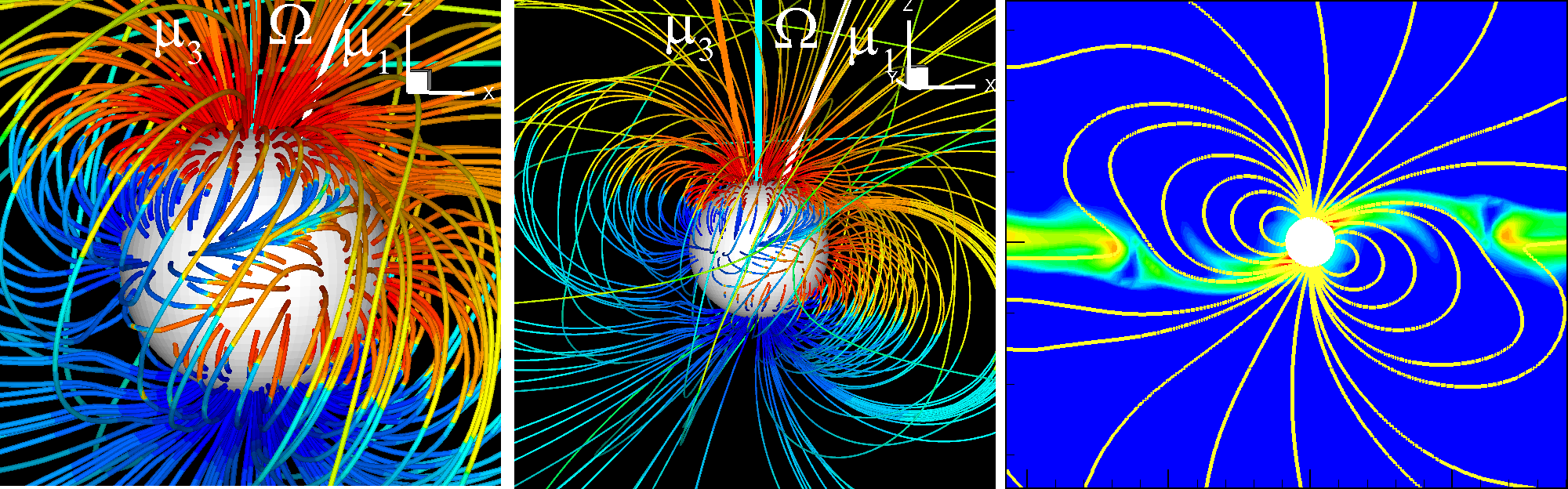}
  \end{center}
   \vspace{-0.4cm}
\caption{\textit{Left two panels:} The magnetic field of BP Tau in
the 3D MHD model. \textit{Right panel:} The density distribution
in the $XZ-$plane. The color of the field lines represents the
polarity and strength of the field. The cyan, white, and orange
lines represent the rotational axis, the dipole moment, and the
octupole moment, respectively.  From \citet{LongEtAl2011}.}
\vspace{-0.4cm} \label{BPtau-3}
\end{figure*}

\begin{figure*}[!ht]
  \begin{center}
    \begin{tabular}{cc}
\includegraphics[width=115mm]{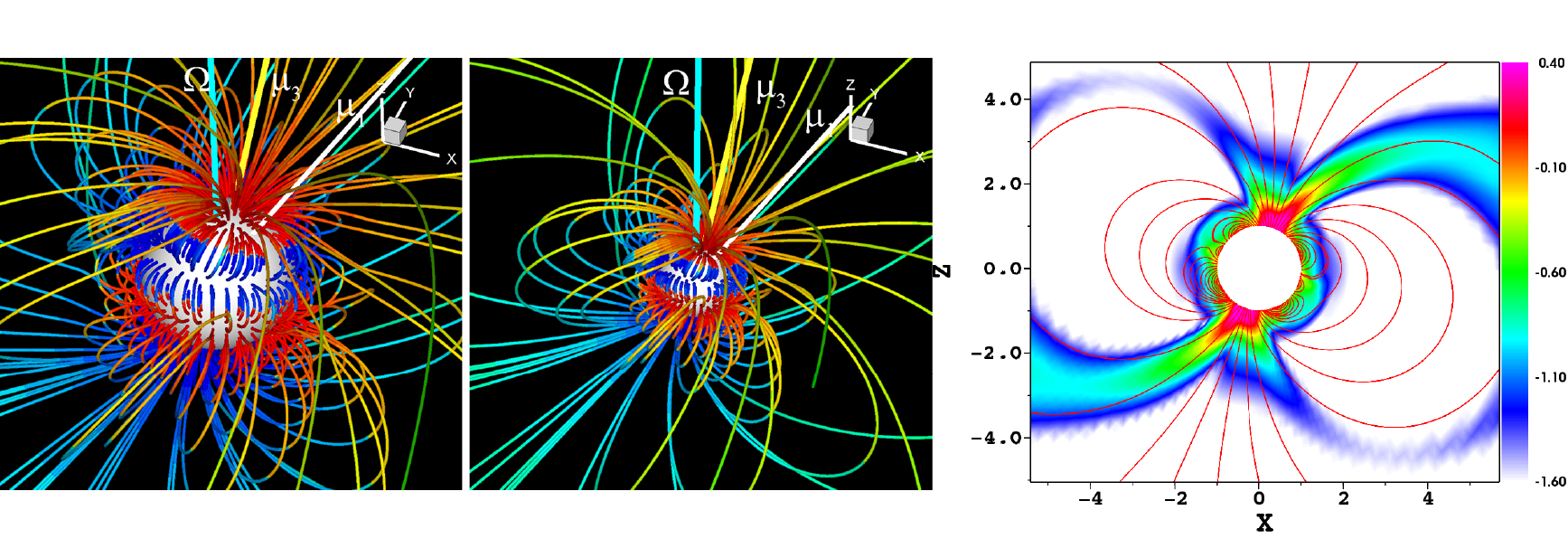}
    \end{tabular}
  \end{center}
   \vspace{-0.4cm}
 \caption{\textit{Left two panels:} The magnetic field of V2129 Oph in the 3D MHD model (from \citet{RomanovaEtAl2011b}).
 \textit{Right panel:} The density distribution in the $XZ-$plane (from \citet{AlencarEtAl2012}).}
\vspace{-0.4cm}
 \label{v2129oph-3}
\end{figure*}

Recent measurements of the surface magnetic field in CTTSs have
shown that, in several stars, the dominant components of the field
are dipole and octupole, while quadrupole and higher-order
components are smaller \citep{DonatiEtAl2007, DonatiEtAl2008}.
Using stellar parameters obtained from these observations, the
modeling of accretion onto the observed stars was somewhat
simplified.
Figures \ref{BPtau-3} and \ref{v2129oph-3} show 3D MHD simulation
results  for two observed CTTSs : BP Tau, where the octupole
component is small \citep{LongEtAl2011}, and V2129 Oph, where the
octupole component is relatively large \citep{RomanovaEtAl2011b}.

\subsection{Modelling accretion onto the CTTS BP Tau}
\label{sec:BP Tau}

The surface distribution of the magnetic field of the CTTS BP Tau
has been measured with the Doppler spectro-polarimetry technique
\citep{DonatiEtAl2008}. The observed surface magnetic field has
been decomposed  into spherical harmonics, and it was found that
the field is mainly poloidal with only $10\%$ of the total
magnetic energy contained within the toroidal field. The poloidal
component can be approximated by dipole ($n=1$) and octupole
($n=3$) moments, which each comprise $50\%$ and $30\%$ of the
magnetic energy, respectively. Other multipoles (up to $n<10$)
have only $10\%$ of the total magnetic energy.
\citet{DonatiEtAl2008} concluded that the magnetic field of BP Tau
is primarily composed of dipole and octupole moments, with $B_{\rm
dip}=1.2$ kG ($\Theta_{\rm dip}\approx 20^\circ$) and $B_{\rm
oct}=1.6$ kG ($\Theta_{\rm oct}\approx 10^\circ$), respectively.

\citet{LongEtAl2011} performed global 3D MHD simulations of
accretion onto a model star with parameters close to those of BP
Tau: $M_\star=0.7 M_\odot$, $R_\star=1.95 R_\odot$, and a
rotational period of $P_\star=7.6$ days (corresponding to a
corotation radius of $r_{\rm cor}\approx7.5 R_\star$). In this 3D
model, the magnetic field has been approximated using superposed
tilted dipole and octupole moments, with polar magnetic field
amplitudes of 1.2 kG and 1.6 kG, respectively. Comparison of the
dipole and octupole fields (using Eqn. \ref{eq:multipole}) shows
that, in BP Tau, the dipole field dominates almost up to the
surface of the star.

Fig. \ref{BPtau-3} shows the distribution of the magnetic field
near the star (left panel) and at larger distances from the star
(middle panel). The right panel shows that matter flows from the
disk onto the star in two funnel streams. However, near the
stellar surface, the octupole component slightly alters the funnel
streams such that matter is deposited closer to the magnetic poles
than those originating from a pure dipole. Moreover, the hot spots
are rounder relative to the crescent-shaped hot spots associated
with the pure dipole case. The shape and position of hot spots
obtained in numerical simulations were in good agreement with
those obtained from observations \citep{LongEtAl2011}.

\subsection{Modelling accretion onto V2129 Oph, and comparisons of spectra}
\label{sec:V2129}

In another example, accretion onto a modeled star with parameters
close to CTTS  V2129 Oph has been investigated
\citep{RomanovaEtAl2011b}. This star has a mass of $M_\star=1.35
M_\odot$, a radius of $R_\star=2.1 R_\odot$, and a rotational
period of $P_\star\approx 6.5$ days. The magnetic field of this
star is dominated by a dipole component of $B_{\rm dip}\approx
0.9$kG and octupole component of $B_{\rm oct}\approx 2.1$kG, both
of which are tilted at small angles about the rotational axis
\citep{DonatiEtAl2011}.

\begin{figure}
  \begin{center}
\includegraphics[clip,height=0.3\textwidth]{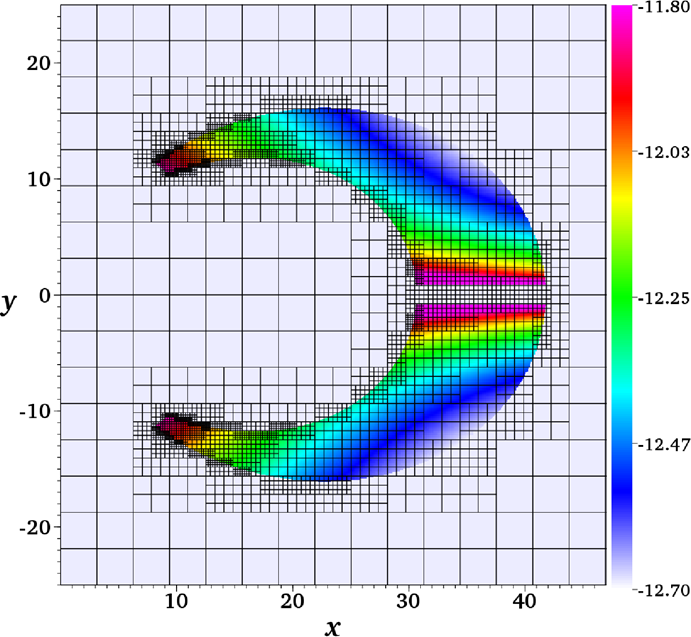}
\includegraphics[clip,height=0.3\textwidth]{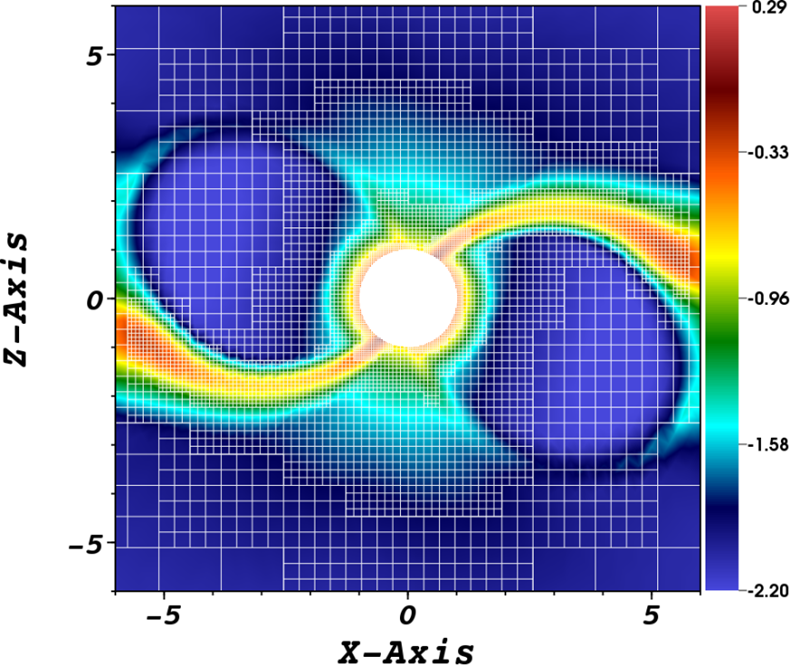}
  \end{center}
\vspace{-0.4cm}\caption{An example of the restructuring grid used
in the radiative transfer code \textit{TORUS} in the case where
the funnel stream density and other parameters are determined by
the analytical formulae from \citet{HartmannEtAl1994} (left panel,
from \citealt{KurosawaEtAl2004}) and from 3D MHD simulations
(right panel).}
 \vspace{-0.2cm}
\label{torus-amr}
\end{figure}

Fig. \ref{v2129oph-3} (left two panels) shows the initial
distribution of the magnetic field. The octupole component of the
field dominates near the star, while the dipole component
dominates at larger distances. The dipole and octupole fields are
equal at approximately $1.6 R_\star$. Simulations show that matter
flows onto the star in two funnel streams above and below the
dipole component of the magnetosphere (see the slice of the
density distribution in the right panel of Fig. \ref{v2129oph-3}).
However, in the region where the octupole field dominates, the
streams are redirected by the octupolar field toward higher
latitudes on the surface of the star

\begin{figure*}[!ht]
  \begin{center}
    \begin{tabular}{cc}
             \includegraphics[width=100mm]{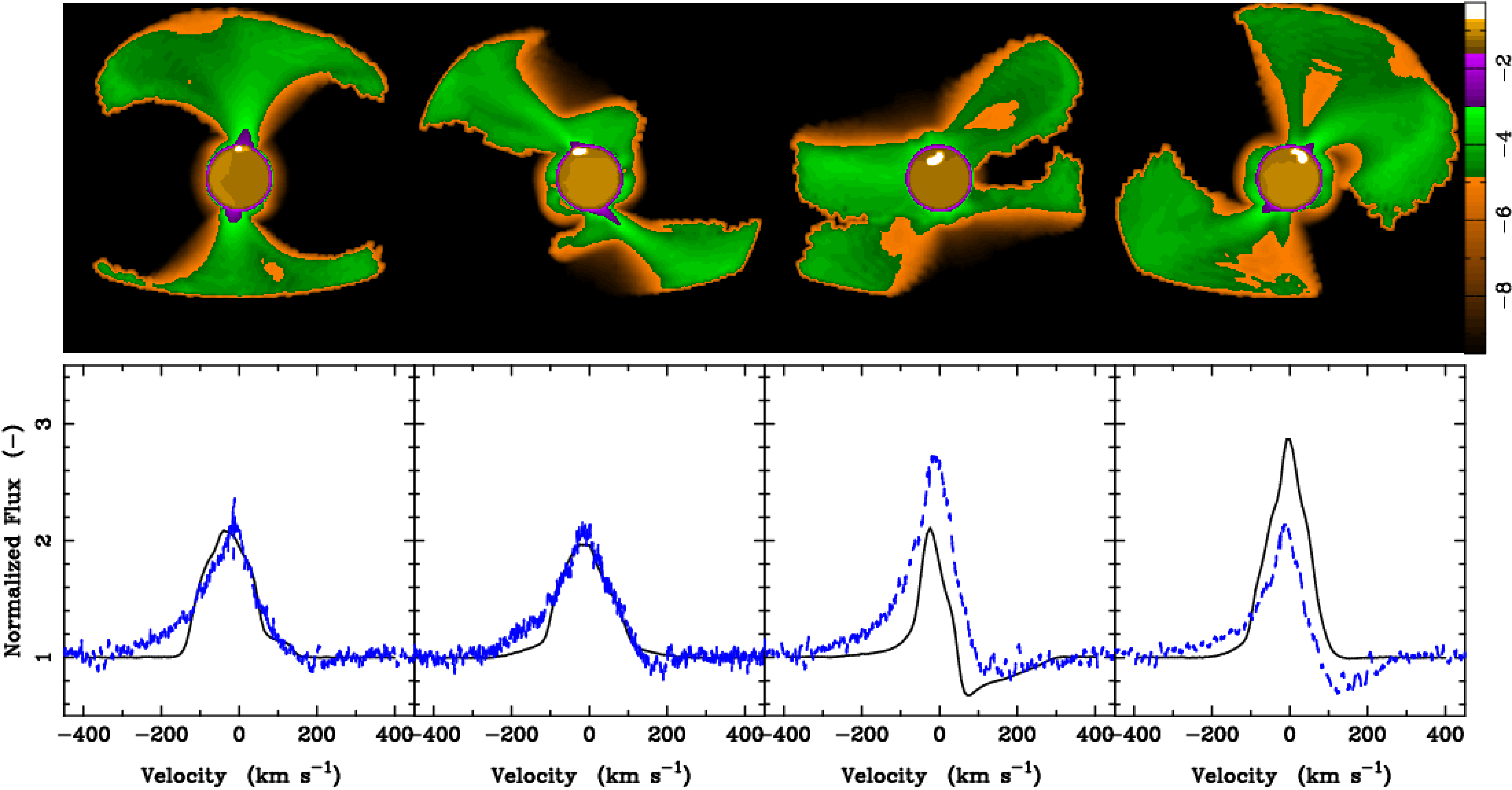}
    \end{tabular}
  \end{center}
   \vspace{-0.4cm}
 \caption{\textit{Top panels:} The emissivity of the funnel flow
calculated in the H$\beta$ spectral line. \textit{Bottom panels:}
A comparison of the observed spectrum in H$\beta$ line (blue line)
with the modeled spectrum (black line). From Alencar et al.
(2012).}
 \vspace{-0.4cm}
 \label{v2129oph-hbeta}
\end{figure*}

Three-dimensional simulations provide us with the dynamical
characteristics of matter flow around magnetized stars (the
distribution of, e.g., density, velocity, temperature). However,
for comparisons with observations, it is important to calculate
the spectrum of the modeled star. To calculate the radiation from
the modeled matter flow and, subsequently, the spectrum in the
Hydrogen lines, the 3D Monte Carlo radiative transfer code
\textit{TORUS}  has been used (e.g.,
\citealt{Harries2000,KurosawaEtAl2004,KurosawaEtAl2008}). The
results of 3D MHD simulations (density, velocity etc.
distribution) were projected onto the adaptive mesh refinement
grid of the \textit{TORUS} code (see Fig. \ref{torus-amr}).
Fig.\ref{v2129oph-hbeta} (top panels) shows the emission of the
funnel streams calculated in the H$\beta$ spectral line shown for
different phases of stellar rotation. The bottom panel of the same
figure compares the modeled and observed spectra of the H$\beta$
line. The plot shows that the observed and modeled spectra are in
good agreement \citep{AlencarEtAl2012}. This is an exciting
example where a 3D MHD model, combined with a 3D stellar radiative
transfer model, with realistic parameters has been compared with
detailed observations and resulted in a good match. This model
shows that global 3D simulations can properly describe realistic
matter flow in CTTSs, and that the 3D radiative transfer code
\textit{TORUS} can provide realistic spectra. These comparisons of
simulations with observations also act as a ``proof of concept''
of the magnetospheric accretion paradigm, suggested earlier
theoretically (e.g., \citealt{Camenzind1990,Konigl1991}). Of
course, this is an example of stable accretion, for which the
magnetospheric and corotation radii are similar in value, which is
typical for the stable accretion case.

\begin{figure*}
  \begin{center}
    \begin{tabular}{cc}
             \includegraphics[width=120mm]{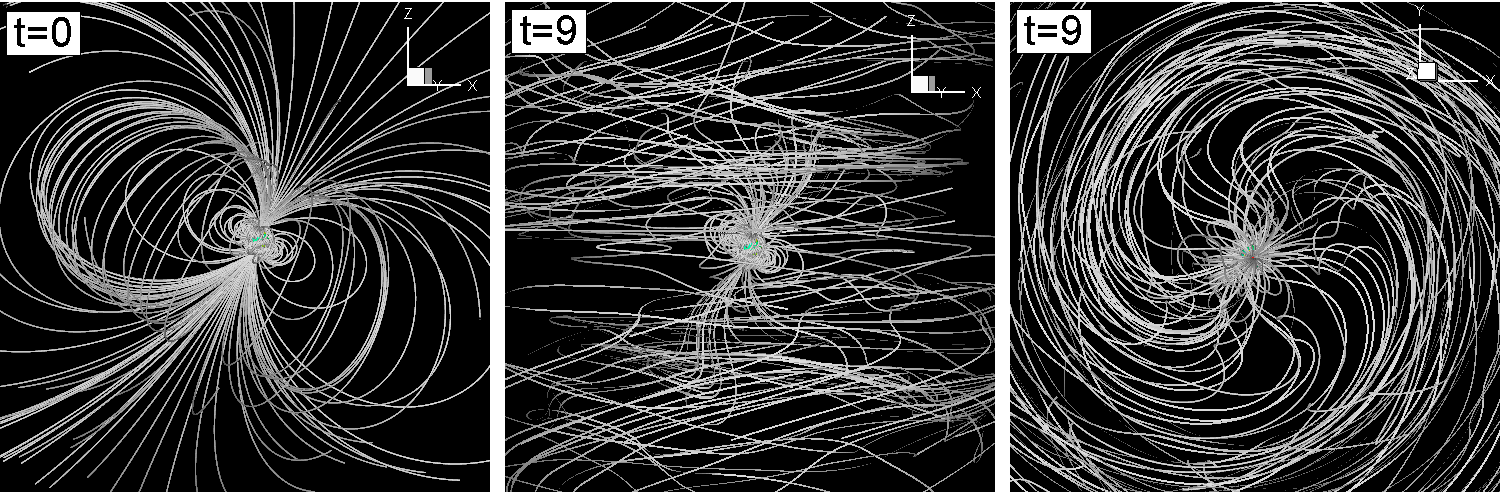}
    \end{tabular}
  \end{center}
   \vspace{-0.4cm}
 \caption{\textit{Left panel:} The initial configuration of  the magnetic field in V2129
 Oph on a large scale. \textit{Middle panel:} The magnetic field configuration at $t=9$.
 \textit{Right panel:}  A top view of the magnetic field configuration. From \citet{RomanovaEtAl2011b}.}
  \vspace{-0.4cm}
\label{v2129-lines-3}
\end{figure*}

Simulations also predict the magnetic configuration of V2129 Oph
on a larger scale (see Fig. \ref{v2129-lines-3}). The
disk-magnetosphere interaction leads to the winding and inflation
of the external magnetic field lines, which thread the disk at
radii  $r\gtrsim r_m$. The figure shows that a strong azimuthal
component of the field is present, and that magnetic tower
structures form about the rotational axis.

\subsection{Summary and future outlook on magnetospheric accretion}

As discussed in the above subsections, 3D simulations have shown,
for the first time, how matter flows onto stars with a tilted
dipole, and onto stars with the complex magnetic field. This has
allowed direct modeling of the  structure of the funnel stream,
and the position and shapes of the hot spots. Specific simulations
were performed for CTT stars with realistic parameters, BP Tau and
V2129 Oph, and the spectrum has been calculated for V2129 Oph
using the 3D radiative transfer code \textit{TORUS}. Comparisons
with observations have shown excellent match, providing a strong
substantiation of the magnetospheric accretion paradigm.

Numerical simulations also led to discovery of two distinct
states, characterized by stable vs.\ unstable regimes of
accretion. The unstable regime may help explain the short-term
variability, with characteristic time of a few events per
rotational period, that is observed in a significant number of
CTTSs.

 The above research concentrated on the
processes at the disk-magnetosphere boundary and the
magnetospheric accretion. Inflation of the field lines is also
modeled, but the high coronal density assumed means it is
matter-dominated, effectively suppressing possible
magneto-centrifugal mechanisms for driving outflows. Future 3D
simulations should concentrate on regimes with density low enough
to allow magnetically driven outflows.  Future simulations can
also aim at modeling of stars with larger magnetospheres, in order
to better understand accretion onto intermediate polars, as well
as X-ray pulsars. Both steps would require more computing power,
because larger parts of the simulation region will be covered by
both a magnetically dominated corona, and a magnetically
dominated, large-scale magnetosphere. Within more computationally
tractable 2.5D (axisymmetric) simulations, the next sections
describe such investigations of outflows from disk-accreting
magnetized stars.

\begin{figure*}[!ht]
  \begin{center}
    \begin{tabular}{cc}
             \includegraphics[width=110mm]{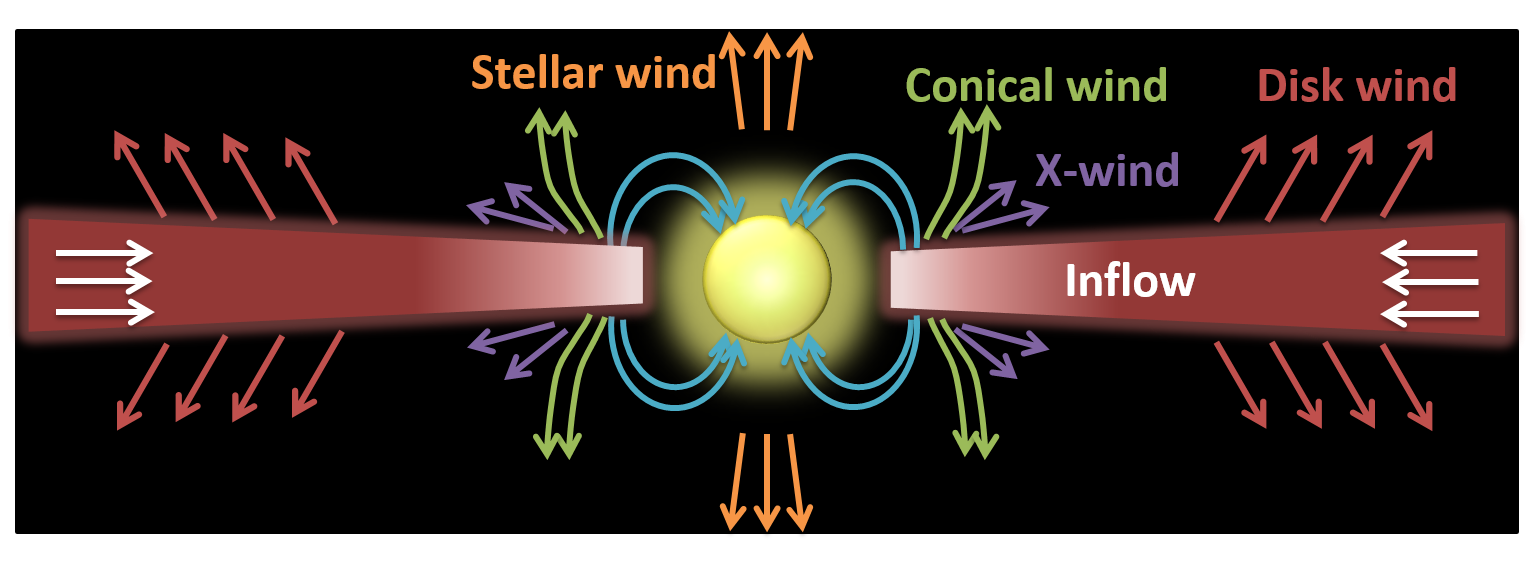}
    \end{tabular}
  \end{center}
   \vspace{-0.4cm}
\caption{A sketch illustrating the physical processes in the
vicinity of accreting magnetized stars. The inner disk matter may
flow onto a star in funnel streams (blue arrows), or it may fly
away from the disk-magnetosphere boundary either in
centrifugally-driven X-winds or magnetically-driven conical winds
(purple and green arrows, respectively). Matter may also flow from
the surface of the star in the form of stellar wind, or from the
disk in the form of the disk wind (orange and red arrows,
respectively). Sketch by M.L Comins.}
 \vspace{-0.4cm}
 \label{sketch-winds}
\end{figure*}

\section{OUTFLOWS FROM THE DISK-MAGNETOSPHERE BOUNDARY}
\label{sec:outflows section}

Different theoretical models have been proposed to explain the
winds and jets from accreting stars  (see review by
\citealt{FerreiraEtAl2006}). Fig. \ref{sketch-winds} demonstrates
a few different possibilities. Outflows can  be accelerated by a
magneto-centrifugal mechanism along the field lines threading the
disk at different distances from the star and tilted by
$>30^\circ$ about the $z-$axis (e.g.,
\citealt{BlandfordPayne1982}); or they can originate at the
disk-magnetosphere boundary \citep{ShuEtAl1994,RomanovaEtAl2009}.
Stellar winds may also contribute to an outflow if part of the
accreting matter is redirected into a stellar wind
\citep{MattPudritz2005,MattPudritz2008}.

In this section, we concentrate on outflows from the
disk-magnetosphere boundary. We show results from two scenarios:
(1) when the star rotates rapidly in the propeller regime and the
rapidly rotating magnetosphere drives outflows; and (2) when the
star rotates slowly but the field lines are bunched up
at the disk-magnetosphere boundary during episodes of enhanced
accretion, in which case the matter flows into a conically shaped
wind (e.g., \citealt{RomanovaEtAl2009}).

\begin{figure*}
  \begin{center}
    \begin{tabular}{cc}
\includegraphics[width=120mm]{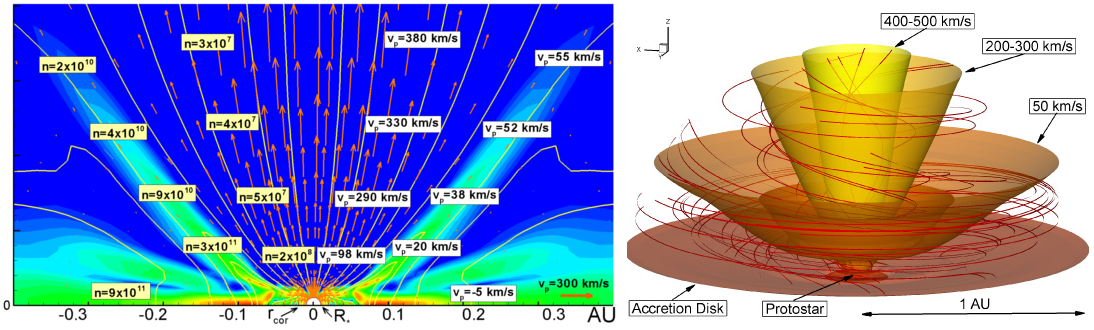}
             \tabularnewline
    \end{tabular}
  \end{center}
   \vspace{-0.4cm}
\caption{\textit{Left panel:}  The density and velocity
distribution in the propeller regime. The numbers correspond to
CTTSs. \citep{RomanovaEtAl2009}. \textit{Right Panel:} A 3D
rendering showing magnetic field lines and density levels
corresponding to different velocities. }
 \vspace{-0.4cm}
\label{propeller-onion-2}
\end{figure*}

\subsection{Propeller-driven outflows}
\label{sec:propeller}

The propeller regime, wherein $r_{\rm cor} < r_m$, was studied in
2.5D MHD simulations in the cases of $\alpha-$disks
\citep{RomanovaEtAl2005b,RomanovaEtAl2009,UstyugovaEtAl2006} and
turbulent disks \citep{LiiEtAl2014}, in which the turbulence is
driven by the magneto-rotational instability (MRI,
\citealt{BalbusHawley1991,BalbusHawley1998}). We consider these
cases in Sec. \ref{sec:propeller-alpha} and Sec.
\ref{propeller-mri}.

\subsubsection{Propeller regime studied with $\alpha-$disks}
\label{sec:propeller-alpha}

The propeller regime has been initially modelled in terms of
$\alpha-$disks,  assuming both axisymmetry and north-south
symmetry, using a code with spherical coordinates. In most
simulation runs, the viscosity parameter is taken to be
$\alpha=0.1-0.3$, which allows a relatively high rate of matter
flow toward the propelling magnetosphere. Magnetic diffusivity is
also incorporated, using a formalism analogous to viscous
diffusion, now proportional to a diffusivity parameter
 $\alpha_d<1$ (e.g.,
\citealt{Bisnovatyi-KoganRuzmaikin1976}).
Such a diffusivity mimics more complex processes, such as 3D instabilities at the
disk-magnetosphere boundary, and is important because it allows
penetration of  inner disk matter onto field lines of the
external magnetosphere.
Most  simulations \citep{RomanovaEtAl2005b,RomanovaEtAl2009,UstyugovaEtAl2006}.
take a
diffusivity coefficient $\alpha_d=0.1$; the dependence on
$\alpha_d$ has been studied in a special set of simulation runs by
\citet{UstyugovaEtAl2006}.

The propeller may be ``strong" -- where the magnetosphere rotates
much more rapidly than the inner disk (fastness parameter
$\omega_s \gg 1$) --, or weak -- where the angular velocities of
the magnetosphere and the disk are comparable ($\omega_s\sim 1$).
For a weak propeller, most of matter accretes onto the star, while
the angular momentum is transferred outward along the disk by
viscosity (e.g., \citealt{SunyaevShakura1977,SpruitTaam1993}). For
a strong propeller, significant amounts of matter can be
redirected from the inner disk to a direct outflow
\citep{IllarionovSunyaev1975,LovelaceEtAl1999}.
For a case of very strong propeller (with  fastness parameter is
$\omega_s\approx 4$, and $r_{\rm cor}=1$;
\citealt{RomanovaEtAl2009}), Fig. \ref{propeller-onion-2} (left
panel) shows a simulation snapshot during an outflow episode.
Analysis of forces shows that the conical part of outflow is
centrifugally driven, while magnetic forces also drive a smaller
mass flux into a magnetically-dominated, better-collimated
Poynting jet \citep{LovelaceEtAl1991, LovelaceEtAl2002}.
Some accretion is also possible, due to the 2D nature of the
propeller regime, wherein matter may flow around the
centrifugally-rotating magnetosphere.

Propeller-driven outflows are expected at the early stages of
stellar evolution, wherein the star still rotates rapidly and
retains a strong magnetic field. A strong propeller may redirect
most of the disk matter into the outflows. It is possible that the
powerful jets from young (Classes 0 and I) stars are connected
with such a propeller stage in young star evolution. These stars
are usually hidden inside dusty envelopes of forming stars, and it
is presently difficult to test this hypothesis. It is also
possible that the outflows observed in class II young stars
(CTTSs) can also be connected with the propeller regime of
accretion. Observations of jets from CTTSs show that they have an
``onion-skin"  velocity distribution, for which the outflow
velocity is higher in the axial regions and decreases away from
the axis (e.g., \citealt{BacciottiEtAl1999,DougadosEtAl2000}).
Similar onion-skin structure is seen in simulations of outflows
from propelling star (see Fig. \ref{propeller-onion-2}, right
panel), and in simulations by \citet{GoodsonEtAl1997}.

\subsubsection{Propeller regime in case of accretion from turbulent disks}
\label{sec:propeller-MRI}

\begin{figure}[!ht]
\sidecaption
             \includegraphics[width=100mm]{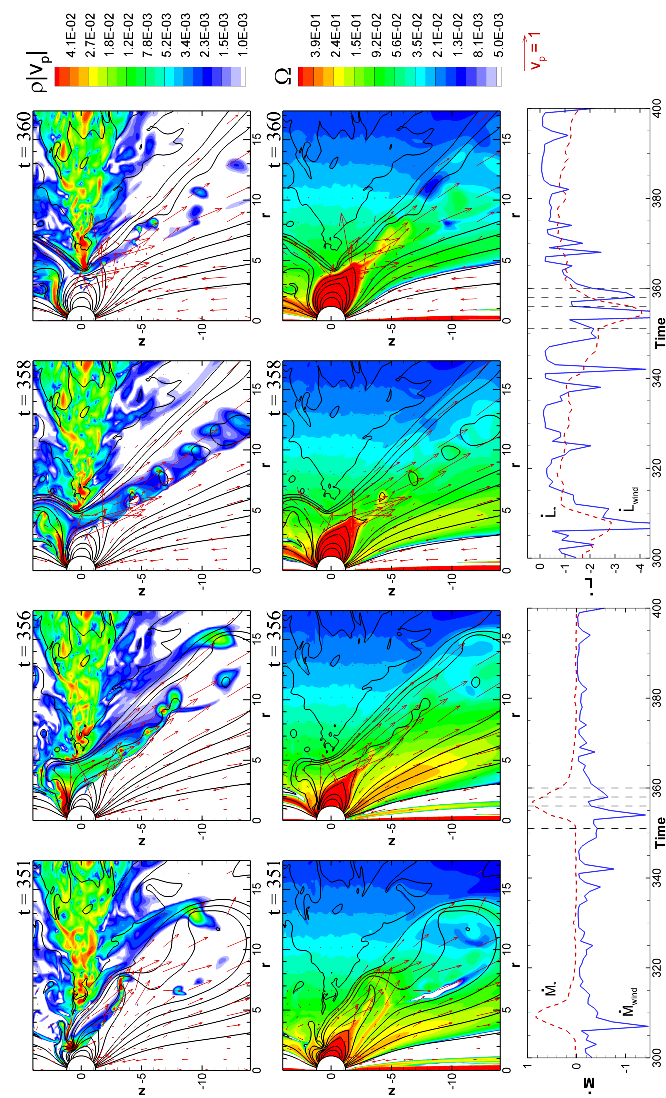}
 \vspace{-0.1cm}
\caption{Axisymmetric simulations of the propeller outflows in the
case of accretion from the turbulent disk ($r_{\rm cor}=1.3$). The
color background shows the matter flux (top row) and angular
velocity (middle row). The bottom row shows the matter fluxes
(left panel) and angular momentum fluxes (right panel) onto the
star and into the wind, respectively. The dimensionless time-scale
can be converted to dimensional one by multiplying the dimensional
time by $P_0$ from Tab. \ref{tab:1}. From \citet{LiiEtAl2014}.}
 \vspace{-0.4cm}
\label{propeller-mri}
\end{figure}

In the case of MRI-driven turbulent disks, the inward disk
accretion can be conveniently provided by this turbulence.
However, the MRI-driven turbulence does not provide the
diffusivity at the disk-magnetosphere boundary. Some matter
penetration is associated with the reconnection events between the
stellar field lines and the field lines of the turbulent cells
\citep{RomanovaEtAl2011b}, but this does not provide significant
diffusivity. Therefore, in axisymmetric simulations an
$\alpha-$type diffusivity is often added, but only to the region
where the disk interacts with the magnetosphere
\citep{LiiEtAl2014}. Actually, most simulation runs of
\citet{LiiEtAl2014} assume no explicit diffusivity ($\alpha_d=0$),
so that only the very small numerical diffusivity provides the
interaction between the disk and magnetosphere. Such small
diffusivity leads to long episodes of the matter accumulation at
the disk-magnetosphere boundary, punctuated by rare events of
accretion onto the star. Fig. \ref{propeller-mri} shows an example
of outflow for  the strong propeller case, with
 $\omega_s\approx 2$ and $r_{\rm cor}=1.3$.
 Along with the strong episodic outflows are evident,  intervals of
matter accumulation and accretion are also evident. Matter
accumulates and accretes when a sufficient amount of matter
reaches the corotation radius.  As  discussed in Sec.
\ref{sec:accretion-mri}, the MRI-driven disk behaves as
matter-dominated disk. In this propeller case, matter of the disk
simply flows above the rapidly rotating magnetosphere of the star.
 In these propeller simuations, top-bottom symmetry was not assumed,
thus allowing matter to flow both above or below the rapidly-rotating magnetosphere,
while pushing the magnetosphere in the opposite direction (see also
 analysis of this phenomenon in \citealt{RomanovaEtAl2011b}). The
 direction of the outflows also changes episodically.


In test simulations with larger diffusivity ($\alpha_d=0.1$) in
the disk-magnetosphere region, the interaction between the disk
and the magnetosphere is stronger, and more matter is ejected to
the outflows, while smaller amount of matter accretes onto the
star \citep{LiiEtAl2014}. In both types of simulations of the
propeller regime, the disk-magnetosphere interaction has a cyclic
character. Initially, the inner disk matter accumulates, then
diffusively penetrates the inner parts of the magnetosphere, where
it acquires angular momentum; some of the matter is then ejected
to a centrifugally-driven wind, while some accretes onto the star.
After this, the magnetosphere expands, and the cycle repeats (see
also \citealt{GoodsonEtAl1997,GoodsonWinglee1999}). Compared with
\citet{GoodsonEtAl1997}, simulations show many more oscillation
cycles. This cyclic oscillations are different from cyclic
accretion discussed for dead disks, where only weak propellers are
considered, and where the excess of angular momentum flows into
the dead disk  (e.g., \citealt{SpruitTaam1993,DangeloSpruit2010}).
There are also similarities: accretion occurs when the inner disk
reaches the corotation radius. However, the propeller-driven
oscillations are expected on much smaller time-scales compared
with the dead disk oscillations, because in the propeller-driven
oscillations only the inner part of the disk oscillates, while in
dead disks much larger parts of the disk are usually involved.


In cases of both turbulent and $\alpha$-type disks, the strength
of the propeller strongly depends on the fastness parameter
$\omega_s$. The estimated value of $\omega_s$ is only approximate,
because the inner disk strongly oscillates, causing the values of
$r_m$ and $\Omega_K(r_m)$ to vary in time. The range of strengths
of the propellers varies depending on the fastness parameter. When
$\omega_s$ is a few times larger than unity, then the propeller is
strong and most of the disk matter can be ejected into outflows.
When $\omega_s\approx 1$, the propeller is weak and most of the
matter accretes onto the star.

The outcome of the propeller regime also strongly depends on the
diffusivity at the disk-magnetosphere boundary. The diffusivity
should be sufficiently high, so that the matter of the inner disk
can interact with the magnetosphere. When the diffusivity is very
low, the rapidly-rotating magnetosphere and slowly-rotating disks
do not exchange angular momentum, and outflows are not possible.
We suggest that, in more realistic 3D simulations, the diffusivity
is not very small, because three-dimensional instabilities such
 as Kelvin-Helmholtz and magnetic interchange instabilities will lead to
mixing of the disk matter with the magnetosphere.

\begin{figure*}
  \begin{center}
    \begin{tabular}{cc}
\includegraphics[width=120mm]{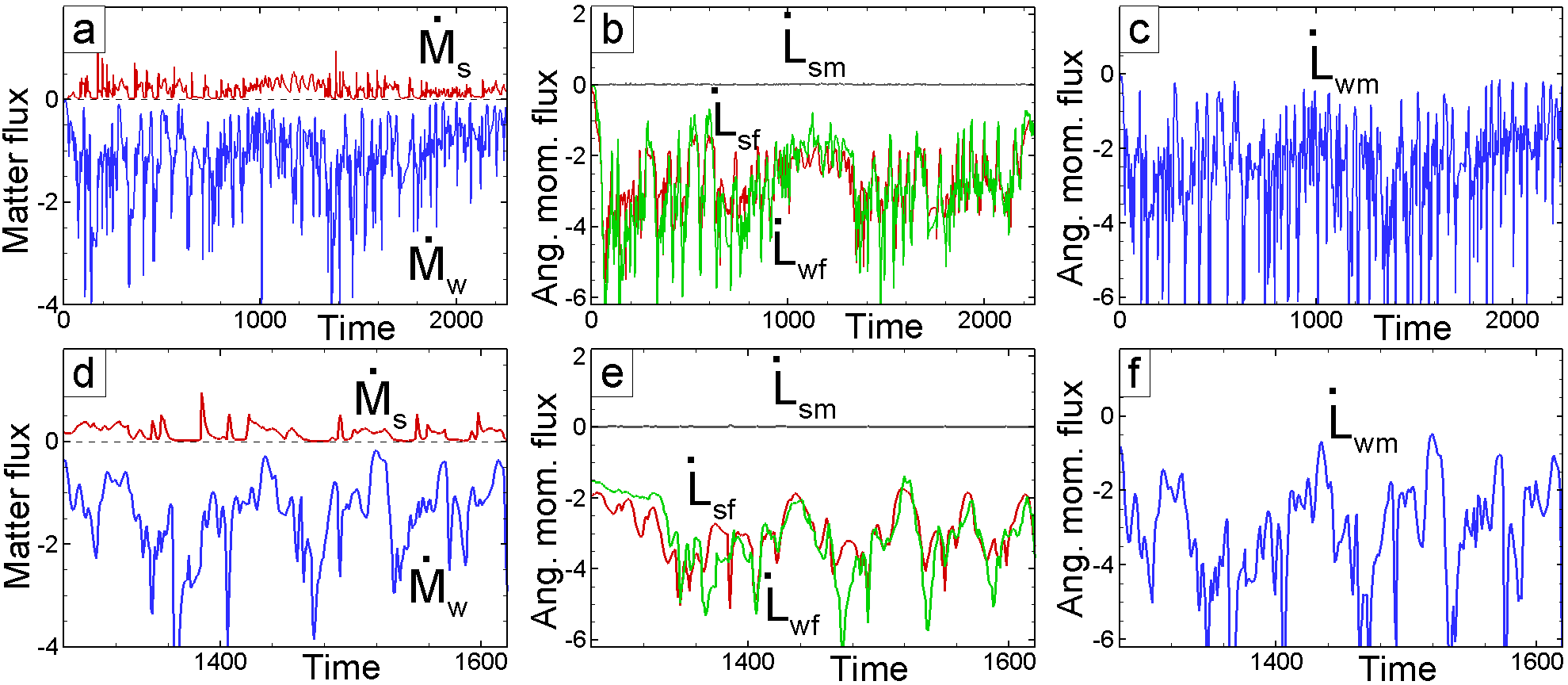}
             \tabularnewline
    \end{tabular}
  \end{center}
   \vspace{-0.4cm}
\caption{\textit{Top left panel:} The matter fluxes onto the star,
$\dot M_s$, and into the wind, $\dot M_w$. \textit{Top middle
panel:}  The angular momentum fluxes carried from the star to the
wind  by the magnetic field, $\dot L_{\rm sf}$, and by matter,
$\dot L_{\rm sm}$. $\dot L_{\rm wf}$ is the angular momentum flux
carried by the field to the wind. \textit{Top right panel:} The
angular momentum flux carried by the matter to the wind. The
bottom panels show the same fluxes as the top panels, but ``zoomed
in'' to focus on a particular period of time in the simulations.
The dimensionless time-scale can be converted to dimensional one
by multiplying the dimensionless time by $P_0$ from \ref{tab:1}.
From \citet{RomanovaEtAl2009}.}
 \vspace{-0.4cm}
 \label{propeller-fluxes-6}
\end{figure*}

\subsubsection{Matter  fluxes and oscillation of the inner disk}
\label{sec:propeller-fluxes}

The amounts of matter flowing onto the star and into the winds
varies according to the strength of the propeller.
 Fig. \ref{propeller-fluxes-6} shows an example of these fluxes in
the case of a strong propeller, calculated in the case of an
$\alpha-$disk. Fig. \ref{propeller-fluxes-6} (panel {\it a}) shows
the  matter fluxes onto the star, $\dot M_s$, and into the winds,
$\dot M_w$, integrated over a surface with radius $r=10$. The
matter flux into the wind is much larger than that onto the star,
$\dot M_w \gg \dot M_s$, implying that almost all of the disk
matter is ejected from the system into the outflows. If the star
rotates more  slowly, then the fraction of the matter flux going
into the wind decreases, and a larger portion of the matter
accretes onto the star (see \citealt{UstyugovaEtAl2006} for
dependences of matter fluxes on $\Omega_\star$, $B_\star$,
$\alpha$, and $\alpha_d$).

Both fluxes strongly vary and show episodic enhancement of
accretion and outflows. The interval between the strongest
outbursts increases when the diffusivity coefficient $\alpha_d$
decreases \citep{UstyugovaEtAl2006}. The bottom panels of Fig.
\ref{propeller-fluxes-6} show the same fluxes as the top panels,
but ``zoomed in'' to focus on a particular time period in the
simulation. The time interval between the strongest outbursts in
the propeller regime is $\Delta t\approx 50-70$ (in dimensionless
units). For protostars and CTTSs ($P_0=1.04$ days), this time
corresponds to $(\Delta t)_{\rm outb}=52-73$ days. In some young
stars (e.g., CTTS HH30 / XZ Tau), the outbursts into the jet occur
at intervals of a few months. This implies that episodic ejections
to the propeller wind may be responsible for some of the
outbursts. In application to accreting neutron stars (AMXPs), we
take into account the reference time-scale  $P_0=1.3$ ms (see Tab.
\ref{tab:1}) and obtain  the dimensional time-scale $(\Delta
t)_{\rm outb}\approx 65-90$ ms. Therefore, rapid variability on
this and smaller time-scales is expected in AMXPs in the propeller
regime. This time-scale is larger, if the diffusivity at the
disk-magnetosphere boundary is smaller, and vice versa. The longer
time scale between bursts is also expected in case of the larger
fastness $\omega_s$ of the propeller.

\begin{table}
\begin{tabular}{l@{\extracolsep{0.2em}}l@{}llll}

\hline
&                                                      & CTTSs         & White dwarfs        & Neutron stars    \\
\hline

\multicolumn{2}{l}{$M(M_\odot)$}                             & 0.8                       & 1                   & 1.4       \\
\multicolumn{2}{l}{$R_*$}                             & $2R_\odot$          & 5000 km             & 10 km     \\
\multicolumn{2}{l}{$R_0$ (cm)}                        & $2.8\e{11}$         & $1.0\e9$            & $2\e6$    \\
\multicolumn{2}{l}{$v_0$ (cm s$^{-1}$)}                & $1.95\e7$              & $3.6\e8$   & $9.7\e{9}$ \\
\multicolumn{2}{l}{$P_0$}                            & $1.04$ days          &  $17.2$ s            & $1.3$ ms   \\
\multicolumn{2}{l}{$B_*$ (G)}                          & $10^{3}$                  & $10^{6}$            & $10^{9}$    \\
\multicolumn{2}{l}{$\rho_0$ (g cm$^{-3}$)}           & $4.1\e{-13}$         & $1.2\e{-9}$         & $1.7\e{-6}$  \\
\multicolumn{2}{l}{$\dot M_0$($M_\odot$yr$^{-1}$)}    & $2.0\e{-8}$         & $1.3\e{-8}$         & $2.0\e{-9}$  \\
\multicolumn{2}{l}{$\dot L_0$ (erg s$^{-1}$)}         & $3.4\e{36}$         & $1.6\e{35}$         & $1.2\e{33}$  \\
\hline
\end{tabular}
\caption{Reference values for different types of stars. We choose
the mass $M$, radius $R_*$ and equatorial magnetic field $B_*$  of
the star. The reference length is $R_0$, the reference velocity is
Keplerian velocity at $R_0$, the reference time-scale $P_0$ is the
period of rotation at $R_0$. The reference density $\rho_0$ is
determined at $R_0$. Reference matter flux $\dot M_0$ and angular
momentum flux $\dot L_0$ are derived from other reference values
(see details in \citealt{RomanovaEtAl2009}). To apply the
simulation results to a particular star one needs to multiply the
dimensionless values from the plots by the reference values from
this table.} \label{tab:1}
\end{table}

\subsubsection{Angular momentum outflow and spinning-down of young  stars}

A star in the propeller regime loses its angular momentum along
the field lines that originate at the surface of the star. Some of
these field lines are strongly inflated and a part of the angular
momentum flows into the magnetically-dominated Poynting flux jet
(e.g., \citealt{UstyugovaEtAl2006}).
 The other set of field lines
originating on the star connect the star to the disk and are only
partially inflated. A fraction of the angular momentum can also
flow out of the star along these field lines (see also
\citealt{ZanniFerreira2013}). The middle panels of Fig.
\ref{propeller-fluxes-6} show the dimensionless values of the
angular momentum fluxes out of the star, calculated at the stellar
surface. The angular momentum is primarily carried by the Poynting
stresses of the magnetic field (see the red curve for $\dot
L_{sw}$ in the middle panels), while the direct angular momentum
carried by the matter is negligibly small (see the black curve for
$\dot L_{sm}$, also in the middle panels). Panels {\it b} and
{\it c} also shows the angular momentum fluxes to the outflows
(calculated at the radius $r=10 R_\star$ from the star)  carried
by the field , $\dot L_{\rm wf}$, and by the matter, $\dot L_{\rm
wm}$. The angular momentum carried by the field to the outflow,
$\dot L_{\rm wm}$, is approximately equal to the angular momentum
out of the surface of the star, $\dot L_{sw}$.

\begin{figure*}
  \begin{center}
    \begin{tabular}{cc}
\includegraphics[width=90mm]{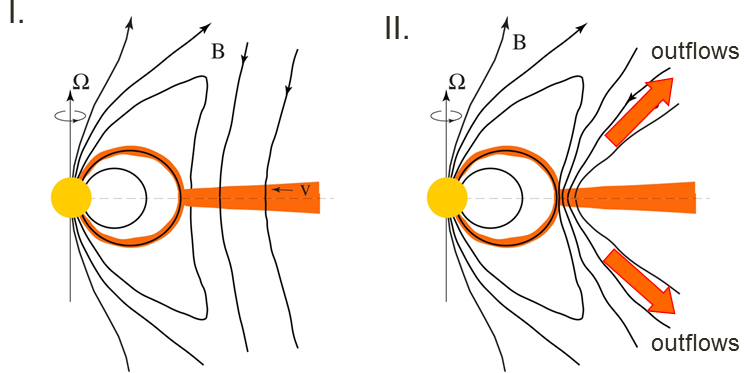}
             \tabularnewline
    \end{tabular}
  \end{center}
 \vspace{-0.4cm}
 \caption{\textit{Left panel:} An example of accretion in which the velocity of the radial
 flow of matter towards the star, $v_{\rm rad}$, is comparable to the velocity of the diffusive penetration
of the disk matter through the field lines of the outer
magnetosphere, $v_{\rm d}$. \textit{Right panel:} An example in
which $v_{\rm rad} \gg v_{\rm d}$, the field lines are bunched by
the accreting matter, and the situation is favorable for
outflows.}
 \vspace{-0.4cm}
\label{conical-sketch}
\end{figure*}

In young solar-type stars, the propeller mechanism may be
responsible for their spinning down from near critical rotation,
which is expected right after their formation, to less than 10\%
of this speed, observed in CTTSs (e.g.,
\citealt{RomanovaEtAl2005b}). If a rapidly rotating protostar has
a magnetic field strength of a few
kG,
then it is in the
strong propeller regime and can rapidly lose angular momentum.
Estimations show that if a protostar rotates initially with period
of $P_\star=1$ day and has a magnetic field of $B=3\times10^3$ G,
then it will lose most of its angular momentum to propeller
outflows  during the time scale of $\tau \approx 3\times 10^5$ ~
{\rm years}. If the magnetic field is $1$ kG, then this time-scale
increases up to $\tau \approx 3\times 10^6$ ~ {\rm years}.  This
is shorter than typical lifetime of CTTSs, which is $\sim 10^7$
years. Therefore, if the magnetic field of protostars is
sufficiently strong, then the propeller mechanism may be
responsible for the fast spinning-down of these protostars to the
slow rotation speeds observed in CTTSs.

\subsubsection{Propeller regime in transitional millisecond pulsars}

In accreting neutron stars the propeller regime is expected at the
end of the accretion outburst, when the accretion rate decreases,
and the truncation radius of the disk can be larger than the
corotation radius.    For example, in AMXP SAX J1808.4-3658, 1Hz
flaring oscillations have been observed at the end of an outburst
\citep{vanStraatenEtAl2005}.
 These oscillations may be caused by the inner disk oscillations in the
 propeller regime
\citep{PatrunoEtAl2009,PatrunoDangelo2013}.
 For the typical
AMXP, the time-scale of spinning down is estimated to be $\tau
\approx 2.5\times 10^7$ years. This time scale is shorter than the
expected lifetime of the accreting millisecond pulsars. However,
these stars spend  relatively little time in the propeller regime.

 Recently discovered  `transitional millisecond pulsars', which
switch between an accretion stage and a radiopulsar stage (e.g.,
\citealt{PapittoEtAl2013,FerrignoEtAl2014,Linares2014}), support
the earlier suggested scenario that millisecond radiopulsars
represent recycled pulsars that are re-spun-up by disk accretion
in the binary system
\citep{AlparEtAl1982,Bisnovatyi-KoganKomberg1974}. Such pulsars
are observed as accreting millisecond X-ray pulsars during periods
of high accretion rate, and as radiopulsars during periods of low
accretion rate. In these stars, the propeller stage is inevitable
and is expected between these two regimes. In fact, strong rapid
oscillations in X-ray have been observed in the transitional
pulsar AMXP IGR J18245-2452, which may be associated with the
propeller regime \citep{FerrignoEtAl2014}. Similar oscillations
have been observed in another transitional pulsar PSR J1023+0038
(e.g., \citealt{PatrunoEtAl2014}). At the same time, enhanced
radiation in the radio band has been observed during this stage,
which is interpreted as possible outflow from the system. Strong
oscillations of the inner disk and outflows were predicted earlier
in theoretical and numerical models (e.g.,
\citealt{LovelaceEtAl1999,RomanovaEtAl2005b,UstyugovaEtAl2006,LiiEtAl2014}).

These new discoveries present an opportunity to compare models
with observations, both to test and constrain the models and open
up a new puzzles. For example, X-ray pulsations were found in PSR
J1023+0038 during its very dim state, when the X-ray luminosity
was about factor 100 lower than during the main accretion state,
when the pulsar is expected to be in the propeller regime (e.g.,
\citealt{ArchibaldEtAl2015,PapittoTorres2015}). Different
explanations were proposed to solve this puzzle (see
\citealt{ArchibaldEtAl2015}). But one key point is that the formal
condition for the propeller regime, $r_m > r_{\rm cor}$ only
applies to an idealized, one-dimensional picture, wherein vertical
centrifugal barrier prevents any accretion onto propelling star.
In an axisymmetric, two-dimensional approach,  the centrifugal
barrier is restricted by the closed magnetosphere of the star. In
other parts of the simulation region, matter of the corona rotates
more slowly, and therefore disk material can flow above/below this
rapidly-rotating magnetosphere and accrete onto the stellar
surface. In addition, a small amount of coronal matter may
precipitate onto the star along the inflated coronal field lines.
Such residual 2D accretion represents an important difference from
the idealized 1D picture the propeller regime: Of course, more
matter accretes in case of the weaker propeller regime, but even
in a strong propeller regime, some matter accretes onto the star
(see Fig. \ref{propeller-fluxes-6}). This may explain X-ray
pulsations of PSR J1023+0038 during very dim state. Detailed
comparisons of axisymmetric simulations with observations may
explain variability and other features of transitional pulsars at
their propeller stage. However, even more advanced,
three-dimensional simulations of the propeller regime are required
to model this properly. On the other hand, current X-ray
telescopes do not yet provide sufficient temporal resolution
(about 10 ms) to detect the time variability found in  such
models.

Similar flaring oscillations have been observed in a few
cataclysmic variables, e.g., in AE Aqr (e.g.,
\citealt{Mauche2006}) which indicated that these stars may also be
in the propeller regime during a part of their life-time. It is
often suggested  that this CV periods of accretion alternate with
periods of ejection (e.g. \citealt{WynnEtAl1997}).

\subsection{Conical winds from the disk-magnetosphere boundary}
\label{sec:conical}

 Recent numerical simulations reveal a new type of wind that can
be important in stars with any rotation rate, if the
magnetic flux of the star is compressed by the disk matter at the
disk-magnetosphere boundary
(see Fig. \ref{conical-sketch}.) This situation is possible, if
the inward radial velocity of the disk matter is larger than
velocity of the diffusive penetration of the disk matter through
the field lines of the magnetosphere
\citep{RomanovaEtAl2009,KoniglEtAl2011,LiiEtAl2012}.
 Simulations show that if the incoming matter compresses
the magnetosphere of the star, the field lines inflate due to the
differential rotation between the disk and the star, and conically
shaped winds flow out of the inner disk. These winds are driven by
the magnetic force, $F_M\propto - \nabla (rB_\phi)^2$ (see Fig.
32), which arises from the wrapping of the field lines above the
disk \citep{LovelaceEtAl1991}. The wind is also gradually
collimated by the magnetic hoop-stress, and it can be strongly
collimated when the accretion rate is large \citep{LiiEtAl2012}.

Conical winds model differ from earlier proposed X-winds model
\citep{ShuEtAl1994} in several respects \footnote{(1) In the
X-wind model, one of the necessary requirements is the condition
$r_m = r_{\rm cor}$, that is the inner disk should rotate with the
angular velocity of the magnetosphere. In conical winds model,
there is no such restriction: a star may rotate much more slowly
than the inner disk, with $r_m \ll r_{\rm cor}$. (2) In the X-wind
model matter is driven by the centrifugal force and overall
situation is closer to the weak propeller regime; in conical winds
model the driving force is mainly the magnetic pressure arising
from the winding of the magnetic field lines above the inner parts
of the disk.}.
\begin{figure*}
\sidecaption
\includegraphics[width=60mm]{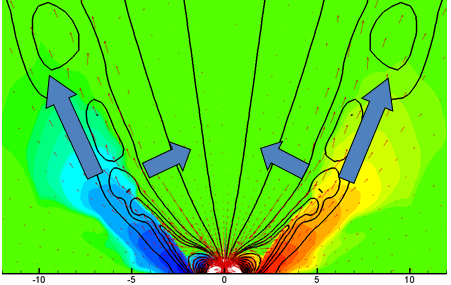}
  \vspace{-0.4cm}
 \caption{The color background shows the poloidal current, which
 results from the large azimuthal component of the magnetic field
 lines above the disk. The magnetic pressure gradient force $F_M\propto - \nabla (rB_\phi)^2$ drives
 matter into cone-shaped outflows. From
 \citet{RomanovaEtAl2009}.}
\label{conical-currents}
\end{figure*}
Conical winds are expected in stars where the accretion rate
strongly increases during the accretion outburst episodes, such as
in AMXPs, and in sub-classes of young stars  EXors  and FUors,
where the accretion rate in the disk increases dramatically (see
review by \citealt{AudardEtAl2014}).
 In fact, in FU Ori itself, H$_\alpha$ shows a strong
blueshifted absorption, providing direct evidence for outflows.
The conical wind model has been applied to FU Ori star and
compared with the empirical model based on the spectral analysis
of the winds in FU Ori \citep{CalvetEtAl1993}. A reasonably good
agreement has been found between these models
\citep{KoniglEtAl2011}. Therefore, the conical winds represent an
attractive model to explain outflows in stars wherein the
accretion rate strongly increases in time. However, this model
should be tested in 3D simulations to check the role of 3D
instabilities at the disk-magnetosphere boundary. The strength of
the winds will probably depend on the ratio between the inward
penetration of matter through instabilities (as discussed in Sec.
\ref{sec:stab-unstab}), and outward flow of matter due to the
winding of the field lines and action of the magnetic pressure
force. Both processes are expected to occur on the time-scale
comparable with the Keplerian time-scale. Global 3D simulation are
required to further investigate conical winds.

\begin{figure*}[!ht]
\includegraphics[width=65mm]{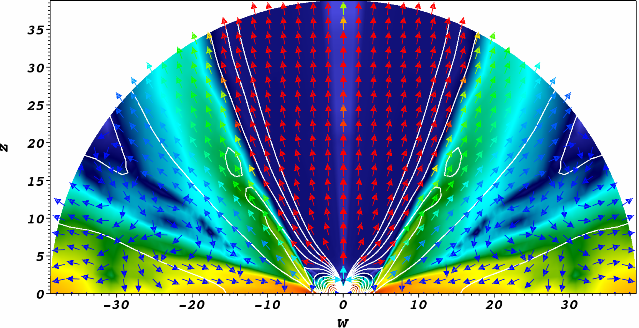}
\includegraphics[width=55mm]{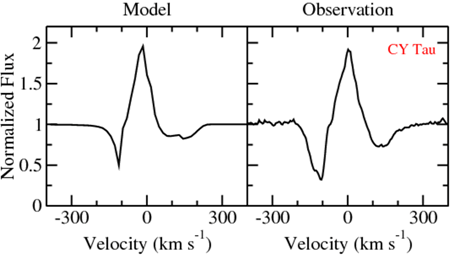}
 \vspace{-0.4cm}
 \caption{\textit{Left panel:} The matter flux distribution, sample field lines and homogeneous velocity vectors in a conical wind.
 \textit{Middle panel:} The spectrum from a conical wind calculated for the He~I~$\lambda$10830\AA~spectral line using the radiative transfer code
 \textit{TORUS}.  \textit{Right panel:} The spectrum of the He~I~$\lambda$10830\AA~line observed in the wind from the CTTS CY Tau. From
\citet{KurosawaRomanova2012}.}
 \vspace{-0.4cm}
\label{conical-spec-4}
\end{figure*}

The observational properties of conical winds, applied to CTTSs,
were investigated by \citet{KurosawaEtAl2011} and
\citet{KurosawaRomanova2012}. In these studies, the spectra of the
H and He lines were calculated using the radiative transfer code
\textit{TORUS}. These simulations show that conical winds produce
a narrow blue absorption component in the spectrum (see Fig.
\ref{conical-spec-4}). Such a blue component is frequently
observed in the spectra of CTTSs (e.g.,
\citealt{EdwardsEtAl2006}).

\begin{figure*}
  \begin{center}
\includegraphics[width=100mm]{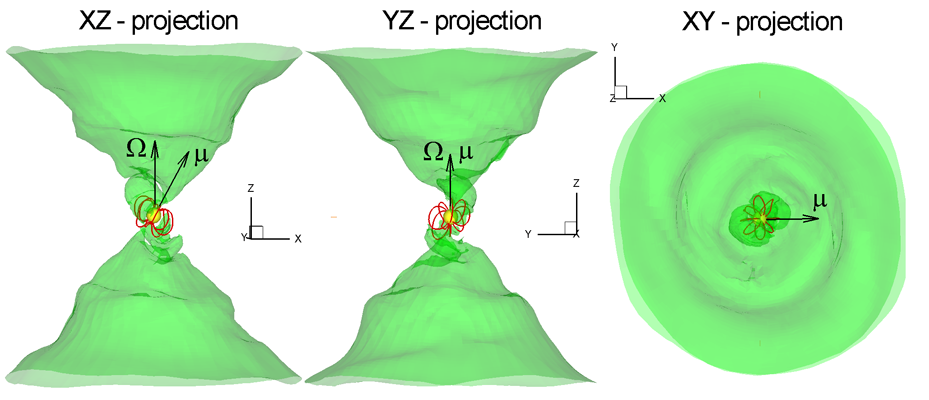}
\end{center}
   \vspace{-0.4cm}
 \caption{3D MHD simulations of conical winds from the disk-magnetosphere boundary
 when the dipole moment of the star is tilted at $\Theta=30^\circ$
show the wavy structure of the inner wall of the conical shell
wind. The green background shows one of the density levels, and
the lines show sample magnetic field lines of the tilted
magnetosphere.}
 \vspace{-0.4cm}
 \label{conical-3d}
\end{figure*}

Conical winds were also studied in test global 3D simulations
where the magnetic axis of the stellar dipolar field has been
tilted by the angle $\Theta=30^\circ$ about the rotational axis
\citep{RomanovaEtAl2009}. These simulations show that, in spite of
the tilted magnetosphere, the winds form a conical structure about
the rotational axis. However, the interaction of the inner disk
with the tilted magnetosphere develops a spiral pattern within the
cone-shaped wind (see Fig. \ref{conical-3d}).

\subsection{Asymmetric and one-sided outflows}
\label{sec:asymmetric-outflows}

\begin{figure*}[!ht]
  \begin{center}
    \begin{tabular}{cc}
\includegraphics[width=120mm]{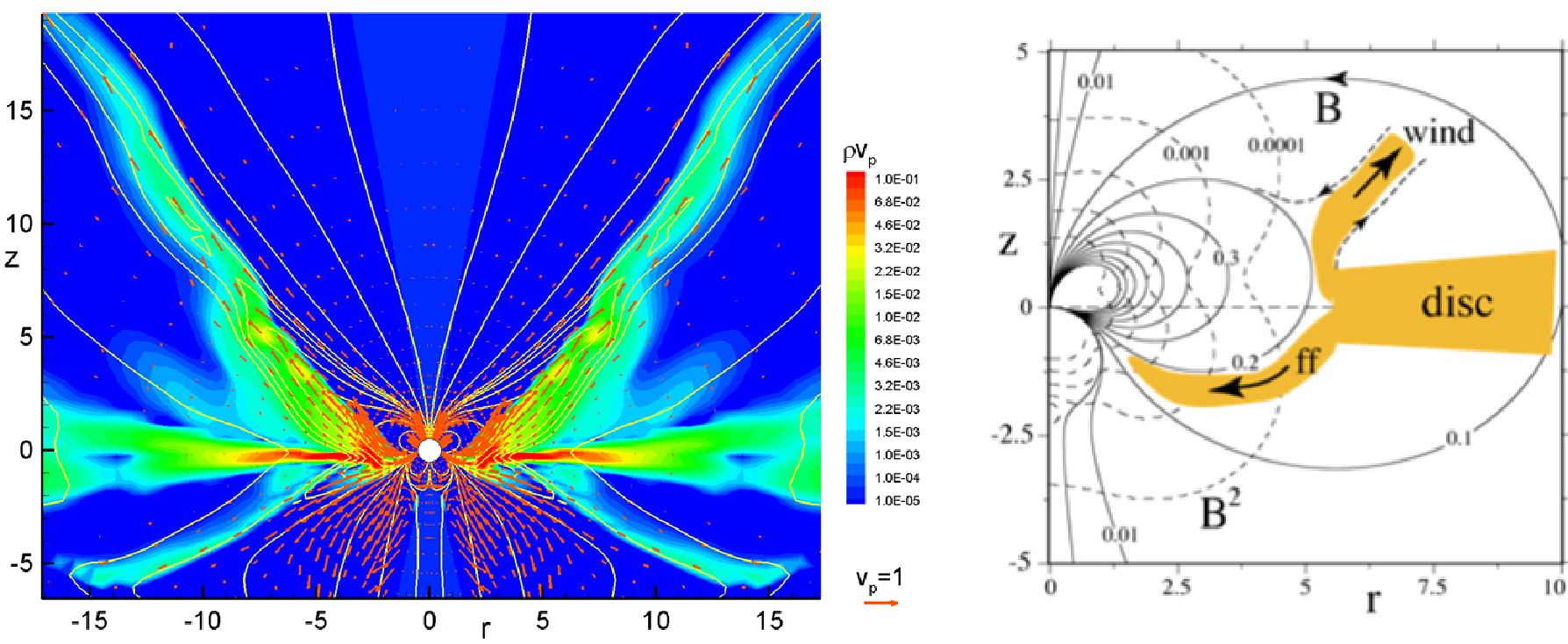}
             \tabularnewline
    \end{tabular}
  \end{center}
 \vspace{-0.4cm}
 \caption{\textit{Left panel:} a density slice and sample field
lines show the result of an axisymmetric simulation of accretion
onto a star with superposed dipole and quadrupole magnetic fields.
\textit{Right panel:} a sketch of accretion and outflow from such
a system (from \citealt{LovelaceEtAl2010}).}
 \vspace{-0.4cm}
\label{asym-2}
\end{figure*}

In stars with a complex magnetic field, outflows may be asymmetric
due to the top-bottom asymmetry of the magnetic field (e.g.,
\citealt{WangEtAl1992}). For example, the superposition of an
axisymmetric dipole field with a quadrupole field leads to a
configuration in which the magnetic flux is larger on one side of
the equatorial plane than the other (see Fig. \ref{asym-2}, right
panel). Axisymmetric simulations of the propeller regime show that
stronger outflows are observed on the side where the magnetic flux
is larger (see left panel of Fig. \ref{asym-2}). In this case, the
matter and energy fluxes will be systematically higher in one
direction and lower in the other direction. One-sided outflows are
observed in a number of young stars (e.g.,
\citealt{BacciottiEtAl1999}).

Axisymmetric simulations of the entire region also show that, even
in the case of a pure dipole field, outflows are usually
one-sided. However, the direction of the outflows switches
frequently, and therefore
the averaged matter and energy fluxes of the outflows above and
below the equatorial plane are expected to be approximately equal
in both directions \citep{LovelaceEtAl2010}. Recent simulations of
the propeller regime in the case of MRI-driven accretion have also
shown that the outflows are one-sided, but that the wind switches
sides much less frequently \citep{LiiEtAl2014}.

\subsection{Future outlook on outflows from the disk-magnetosphere boundary}

In the above subsections, outflows from the disk-magnetosphere
boundary were modeled in two scenarios for which matter flows from
the disk-magnetosphere boundary: rapidly rotating stars in the
propeller regime and slowly-rotating stars (conical winds). Most
of the simulations described were performed using an axisymmetric
approximation. Future simulations should expand to full three
dimensions, so that the disk-magnetosphere interaction through
instabilities can be taken into account. Also, outflows at larger
scales should be further studied.

\section{WIND-FED MAGNETOSPHERES FROM MAGNETIZED MASSIVE STARS}
\label{sec:winds section}

Massive, luminous, hot stars lack the hydrogen recombination
convection zone that induces the magnetic dynamo cycle of cooler,
solar-type stars. Nonetheless, modern spectropolarimetry has
revealed that about 10\% of O, B and A-type stars harbor
 large-scale, organized (often predominantly dipolar) magnetic fields
 ranging in dipolar strength from a few hundred to tens of thousand Gauss.
 \citet{Petit13}
recently compiled  an exhaustive list of 64
confirmed magnetic OB stars with $\teff \gtrsim 16$\,kK, along
with their physical, rotational and magnetic properties; see
figure \ref{fig:etaW} below.

This section summarizes efforts to develop dynamical models for
the effects of such large-scale surface fields on the radiatively
driven mass outflow from such OB stars. The focus is on the
properties and observational signatures (e.g. in X-ray and Balmer
line emission) of the resulting  {\em wind-fed} magnetospheres in
closed loop regions, and on the stellar rotation {\em spindown}
that results from the angular momentum loss associated with
magnetically torqued wind outflow from open field regions.
In this way magnetic fields can have a profound effect on the star's rotational evolution,
giving rotation periods ranging from weeks to even decades, in strong contrast to the
day-timescale periods of non-magnetic massive stars.

The inside-out building of these wind-fed magnetospheres is in
some way complementary to the outside-in nature of the
accretion-fed magnetospheres discussed in previous sections. But
there are also some interesting similarities in the role of the
characteristic magnetospheric and corotation radii, which in this
case are identified below (see equations (\ref{eq:esdef}) and (\ref{eq:rkep})
in Sections \ref{sec:mcp} and \ref{sec:keprad}) as the Alfv\'{e}n radius $\ra$ and the Kepler
radius $\rk$. In particular, the relative sizes of these radii
again plays a key role in determining the magnetospheric
characteristics, in this case leading to dynamical magnetopheres
(DM) when $\ra < \rk$, and centrifugal magnetopheres (CM) when
$\ra > \rk$. The trapping of high-speed wind outflow also leads to
strong shocks and so X-ray emission that can be much harder and
stronger than generally seen for cases of non-degenerate stellar
accretion.

\begin{figure*}
\begin{center}
\includegraphics[width=120mm]{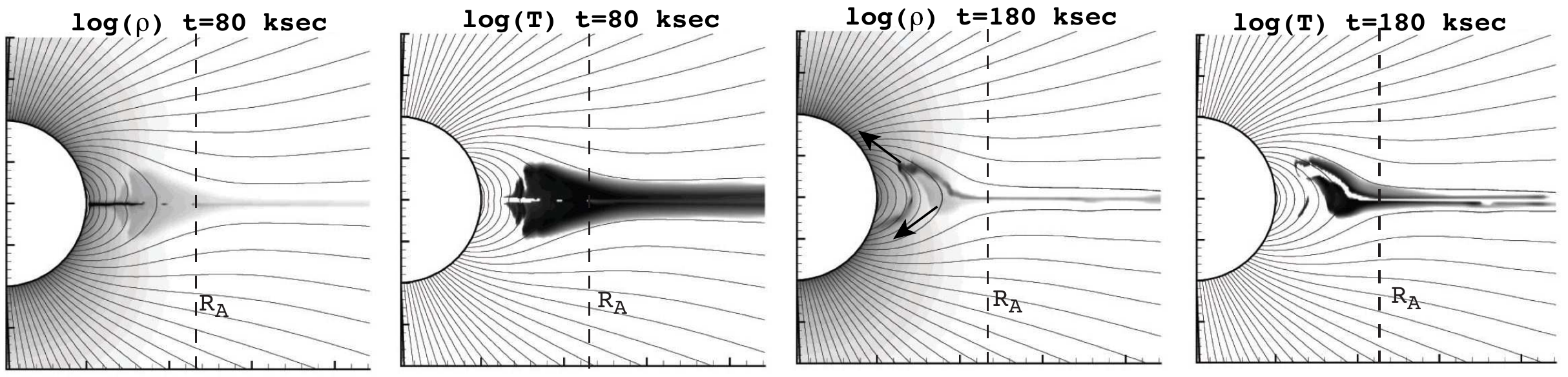}
  \end{center}
\caption{2D MHD simulation for magnetic channeling and confinement
of radiatively driven stellar wind from a non-rotating hot-star
with $\estar=15$ (and so $\ra \approx 2.3 \Rstar$), showing the
logarithm of density $\rho$ and temperature $T$ in a meridional
plane. At a time 80~ksecs after the initial condition, the
magnetic field has channeled wind material into a compressed, hot
region about the magnetic equator, much as envisioned in the
Magnetically Confined Wind Shock (MCWS) paradigm of Babel \&
Montmerle (1997a,b). But by a time of 180~ksecs, the cooled
equatorial material is falling back toward the star along field
lines, in a complex `snake' pattern. The darkest areas of the
temperature plots represent gas at $T\sim10^{7}\,\kelv$, hot
enough to produce relatively hard X-ray emission of a few \keV.
The model reproduces quite well the observed X-ray properties of
$\theta^1$~Ori~C
\citep{Gagne05}.
}
 \label{fig:MCWS}
\end{figure*}

\begin{figure*}
\begin{center}
    \includegraphics[width=120mm]{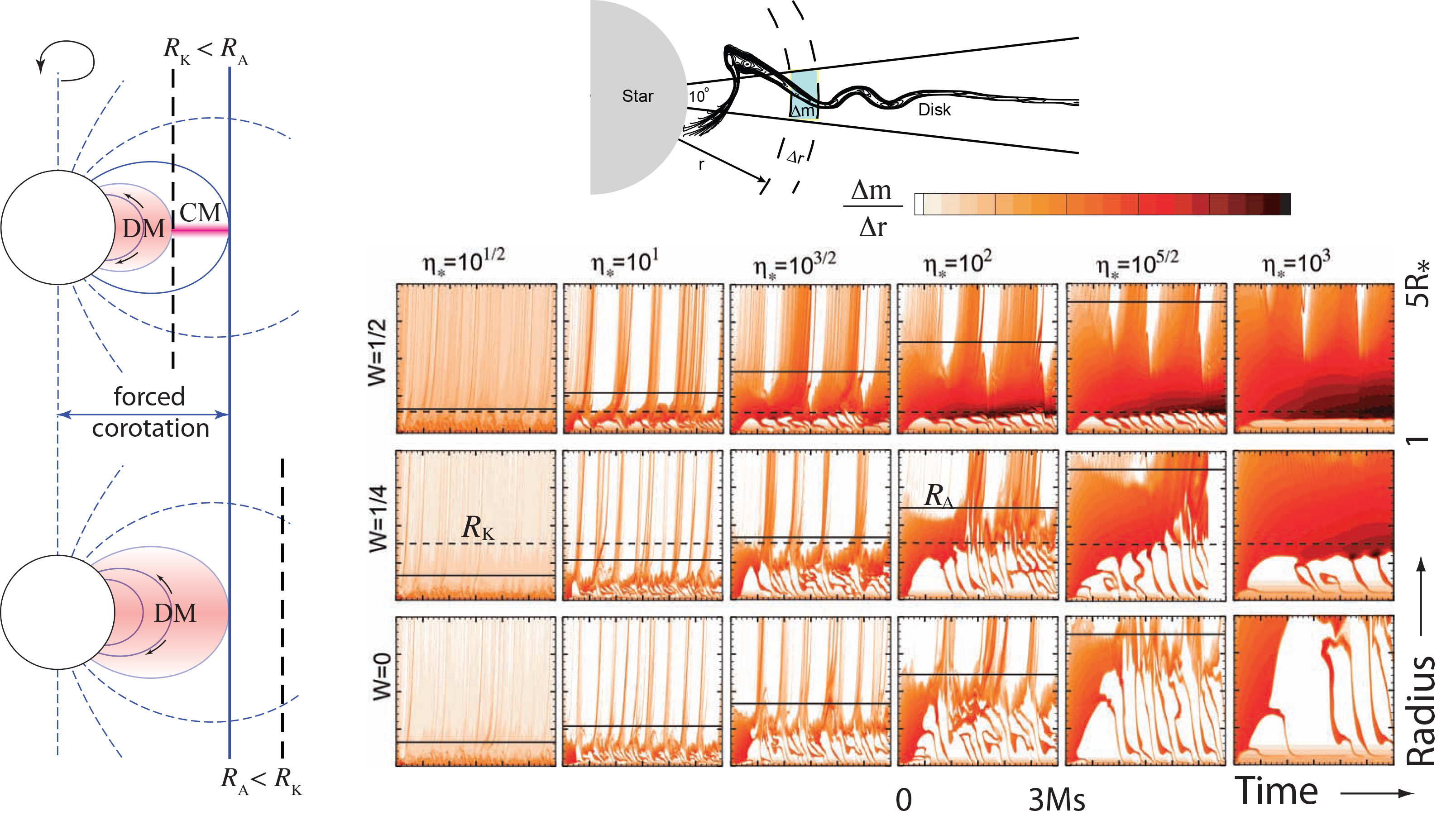}
  \end{center}
\caption{ \small{ {\em Left:} Sketch of the regimes for a
dynamical vs.\ centrifugal magnetosphere (DM vs. CM). The lower
panel illustrates the case of a slowly rotating star with Kepler
radius beyond the Alfv\'{e}n radius ($\rk > \ra$); the lack of
centrifugal support means that trapped material falls back to the
star on a dynamical timescale, forming a DM, with shading
illustrating the time-averaged distribution of density. The upper
panel is for more rapid rotation with $\rk < \ra$, leading then to
a region between these radii where a net outward centrifugal force
against gravity is balanced by the magnetic tension of closed
loops; this allows material to build up to  the much higher
density of CM. {\em Right, Upper:}
 Contour plot for density at arbitrary snapshot of an isothermal 2D MHD simulation with magnetic confinement parameter $\estar=100$ and critical rotation factor $W=1/2$.
 The overlay illustrates the definition of radial mass distribution, $\Delta m/\Delta r$, within $10^\circ$ of the equator.
{\em Right, Lower:} Density plots for  log of $\Delta m/\Delta r$,
plotted versus radius (1-5 \Rstar) and time (0-3~Msec), for a
mosaic of 2D MHD models with a wide range of magnetic confinement
parameters \estar, and 3 orbital rotation fractions $W$. The
horizontal solid lines indicate the Alfv\'en radius \Ralf\ (solid)
and the horizontal dashed lines show  Kepler radius \Rkep\
(dashed). } \label{fig:dmdr} }
\end{figure*}

\subsection{Wind magnetic confinement parameter and Alfv\'en radius }
\label{sec:mcp}

MHD  simulation studies
\citep[e.g.,][]{uDO2002, Uddoula08}
 show that the overall net effect of a large-scale,
dipole magnetic field in diverting such a hot-star wind can be
well characterized by a single \textit{wind magnetic confinement
parameter} and its associated \textit{Alfv\'en radius},
\begin{equation}
\eta_{\ast} \equiv \frac {B_{eq}^2 \, \rs^2} {\mdot \, \vinf} ~~ ;
~~ \frac{\ra}{\rs} \approx 0.3 + \left ( \etas + 0.25
\right)^{1/4} \, ,
\label{eq:esdef}
\end{equation}
where $\beq = \bp/2$ is the field strength at the magnetic
equatorial surface radius $\rs$, and $\mdot$ and $\vinf$ are the
fiducial mass-loss rate and terminal speed that the star
\textit{would have} in the \textit{absence} of any magnetic field.
 This confinement parameter sets the scaling for
the ratio of the magnetic to wind kinetic energy density. For a
dipole field, the $r^{-6}$ radial decline of magnetic energy
density is much steeper than the $r^{-2}$ decline of the wind's
mass and energy density; this means the wind always dominates
beyond the Alfv\'en radius, which scales as $\ra \sim \etas^{1/4}$
in the limit $\etas \gg 1$ of strong confinement.

As shown in figure \ref{fig:MCWS}, magnetic loops extending above
$\ra$ are drawn open by the wind, while those with an apex below
$\ra$ remain closed. Indeed, the trapping of wind upflow from
opposite footpoints of closed magnetic loops leads to strong
collisions that  form X-ray emitting, {\em magnetically confined
wind shocks} (MCWS; Babel \& Montmerle 1997a,b). The post-shock
temperatures $T \approx 20$~MK are sufficient to produce the
moderately hard ($\sim 2$\,keV) X-rays observed in the
prototypical magnetic O-star \tOriC\
\citep{Gagne05}.
%
As illustrated by the downward arrows in the density plot at a
simulation time $t=180$\,ksec, once this material cools back to
near the stellar effective temperature, the high-density trapped
material falls back onto the star over a dynamical timescale.

\subsection{Orbital rotation fraction and Kepler co-rotation radius }
\label{sec:keprad}

The dynamical effects of rotation can be analogously parameterized
\citep{Uddoula08}
 in terms of the {\em orbital rotation
fraction}, and its associated {\em Kepler corotation radius},
\begin{equation}
W \equiv \frac{V_{\rm rot}}{{V_{\rm orb}}} =  \frac{V_{\rm
rot}}{\sqrt{G\Mstar/\Rstar}} ~~ ; ~~ \Rkep = W^{-2/3}\,\Rstar
\label{eq:rkep}
\end{equation}
which depend on the ratio of the star's equatorial rotation speed
to the speed to reach orbit near the equatorial surface radius
$\Rstar$. Insofar as the field within the Alfv\'en radius is
strong enough to maintain {\em rigid-body rotation}, the Kepler
corotation radius $\Rkep$ identifies where the centrifugal force
for rigid-body rotation exactly balances the gravity in the
equatorial plane. If $\Ralf < \Rkep$, then material trapped in
closed loops will again eventually fall back to the surface,
forming a {\em dynamical magnetosphere} (DM). But if $\Ralf >
\Rkep$, then wind material located between \Rkep\ and \Ralf\ can
remain in static equilibrium,
 forming a {\em centrifugal magnetosphere} (CM) that is supported against gravity by the magnetically enforced co-rotation.
As illustrated in the upper left schematic in figure
\ref{fig:dmdr}, the much longer confinement time allows material
in this CM region to build up to a much higher density than in a
DM region.

\begin{figure}[t!]
    \includegraphics[width=120mm]{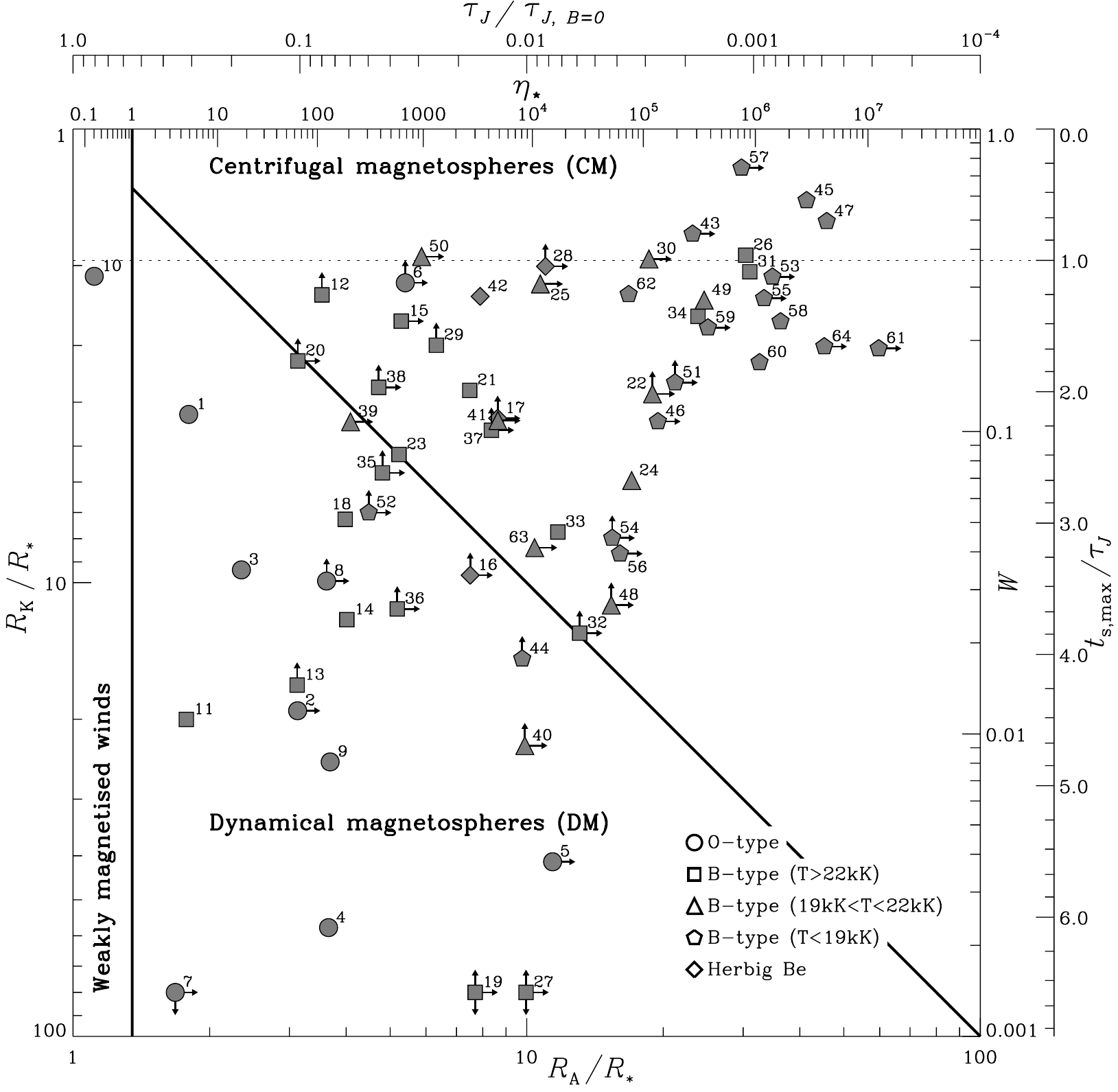}
\caption{Classification of 64 observationally confirmed
magnetic massive stars in terms of  magnetic confinement vs.\
rotation fraction, characterized here by a log-log plot of Kepler
radius $\Rkep$ increasing downward vs.\ Alfv\'{e}n radius $\Ralf$
increasing to the right. The labeled ID numbers are sorted in
order of decreasing effective temperature $\teff$, with stellar identities given in
Table 1 of \citet{Petit13}.
Stars to the left of the vertical solid line have onlyf weakly magnetized winds (with $\etas<1$),
Star below and left of the diagonal solid line have
dynamical  magnetospheres (DM) with $\Ralf<\Rkep$,
while those above and right of this line have centrifugal
magnetospheres (CM) with $\Ralf>\Rkep$.
The additional upper and
right axes give respectively the corresponding spindown timescale
$\tauj$, and maximum spindown age $\tsmax$, as defined in Section
\ref{sec:spindown}. Rapidly rotating stars above the horizontal
dotted line have a maximum spindown age $\tsmax$  that is less
than one spindown time $\tauj$.
 }
 \label{fig:etaW}
\end{figure}

\begin{figure}[t!]
     \includegraphics[width=120mm]{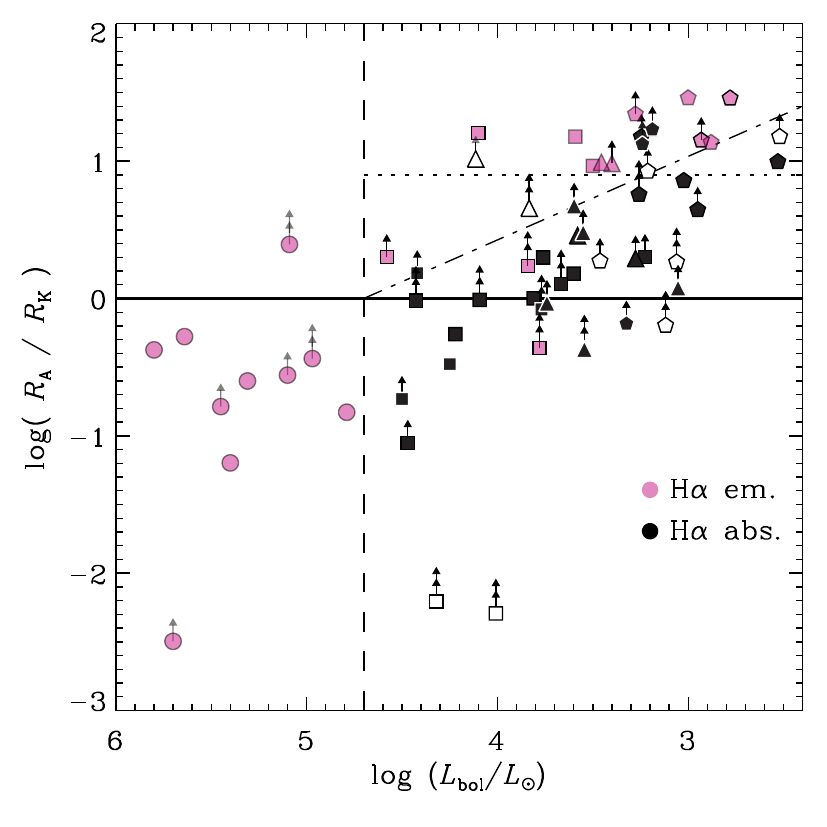}
\caption{
Location of magnetic massive
stars in a $\log$-$\log$ plot of $\ra/\rk$ vs.\ stellar
luminosity. The symbol shadings mark the presence (pink or shaded)
or absence (black) of magnetospheric H$\alpha$ emission, with
empty symbols when no H$\alpha$ information is available. The
vertical dashed line represents the luminosity transition between
O-type and B-type main sequence stars. The horizontal dotted line
and the diagonal dot-dashed line show division of the CM domain
according to potential magnetospheric leakage mechanisms.
 }
 \label{fig:par_ha}
\end{figure}

For full 2D MHD simulations in the axisymmetric case of a
rotation-axis aligned dipole, the mosaic of color plots in figure
\ref{fig:dmdr} shows the time vs. height variation of the
equatorial mass distribution $\Delta m/ \Delta r$ for various
combinations of rotation fraction $W$ and wind confinement
$\estar$ that respectively increase upward and to the right. This
illustrates vividly the DM infall for material trapped below
$\Rkep$ and $\Ralf$, vs.\ the dense accumulation of a CM from
confined material near and above $\Rkep$, but below $\Ralf$.

\subsection{Comparison with Observations of Confirmed Magnetic Hot-stars}

For the 64 observationally confirmed magnetic hot-stars ($\teff
\gtrsim 16$\,kK) compiled by
\citet{Petit13},
figure
\ref{fig:etaW} plots positions in a log-log plane of $\Rkep$ vs.
$\Ralf$. The vertical solid line representing $\estar =1$
separates the domain of non-magnetized or weakly magnetized winds
to left,  from the domain of stellar magnetospheres to the right.
The diagonal line representing $\Rkep = \Ralf$ divides the domain
of centrifugal magnetospheres (CM) to the upper right from that
for dynamical  magnetospheres (DM) to the lower left. Let us now
consider how these distinctions in magnetospheric properties
organize their observational characteristics.

\subsubsection{Balmer-$\alpha$ line emission from DM and CM}
\label{sec:halpha}

Figure \ref{fig:par_ha} plots these observed magnetic stars in a
diagram comparing the {\em ratio} $\Ralf/\Rkep$ vs. stellar
luminosity, with now the symbol coded to mark the presence (light
shading) or absence (black) of magnetospheric H$\alpha$ emission.
The horizontal solid line marks the transition between the CM
domain above and the DM domain below, while the vertical dashed
line marks the divide between O- and B-type  main sequence stars.
Note that {\em all} O-stars show emission, with all but one
(Plaskett's star, which has likely been spun-up by mass exchange
from its close binary companion;
\citep{Grunhut2013}.)
located among the slow rotators with a DM. By contrast,  most B-type stars
only show emission if they are well above the $\Ralf/\Rkep = 1$
horizontal line, implying a relatively fast rotation and strong
confinement that leads to a CM.

The basic explanation for this dichotomy is straightforward. The
stronger winds driven by the higher luminosity O-stars can
accumulate even within a relatively short dynamical timescale to a
sufficient density to give the strong emission in a DM, while the
weaker winds of lower luminosity B-stars require the longer
confinement and buildup of a CM to reach densities for such
emission. This general picture is confirmed by the detailed
dynamical models of DM and CM emission that motivated this
empirical classification.

\begin{figure}[t!]
    \begin{center}
    \includegraphics[width=120mm]{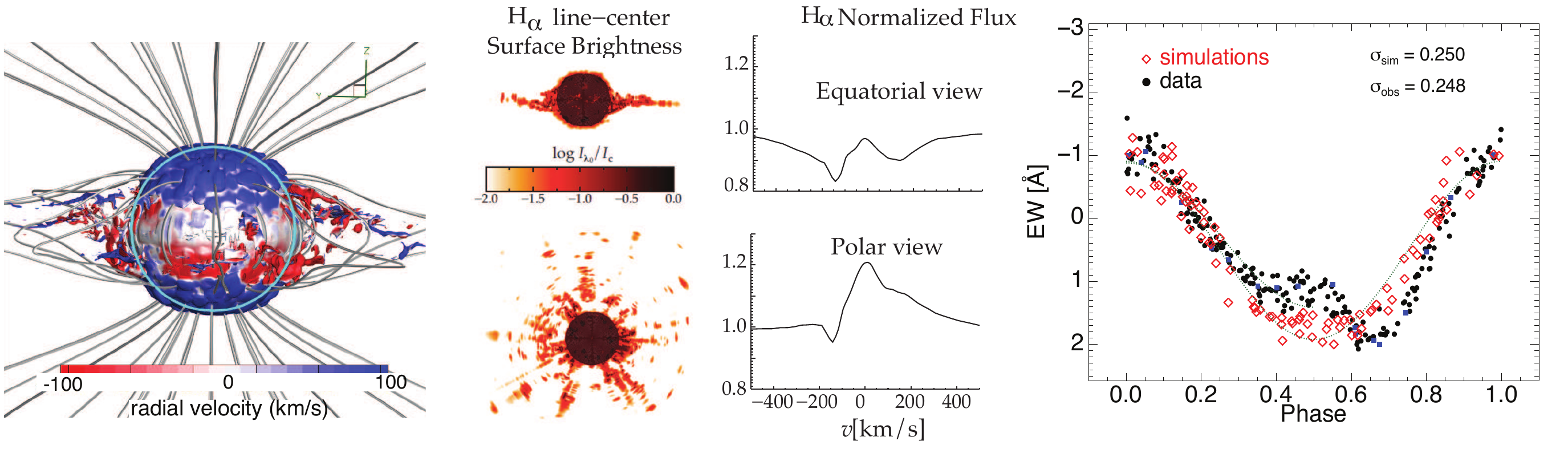}
    \caption{
\label{fig:3DT1OC} 3D MHD model of the dynamical magnetosphere for
the young, slowly rotating (15.4-day period) O7V star \thooc\
(ud-Doula et al.\ 2013). The left panel shows a snapshot of wind
structure drawn as isodensity surface, colored to show radial
component of velocity. The middle panels shows the predicted
equatorial and polar views of H$\alpha$ line-center surface
brightness, along with corresponding line-flux profiles. The right
panel compares the observed rotational modulation of the H$\alpha$
equivalent width (black) with 3D model predictions (red) assuming
a pure-dipole surface field tilted by $\Theta = 45^\circ$ to the
rotation axis, as viewed from the inferred observer inclination of
$i = 45^\circ$.
 }
\end{center}
\end{figure}

For  the slowly rotating  O-stars HD\,191612 and \thooc\ (here
with respective ID numbers 4 and 3), both 2D and 3D MHD
simulations (Sundqvist et al.\ 2012; ud-Doula et al.\ 2013) of the
wind-fed DM reproduce quite well the rotational variation of
H$\alpha$ emission. For the 3D simulations of \thooc, figure
\ref{fig:3DT1OC} shows how wind material trapped in closed loops
over the magnetic equator (left panel) leads to circumstellar
emission that is strongest during rotational phases corresponding
to pole-on views (middle panel). For a pure dipole with the
inferred magnetic tilt  $\Theta=45^\circ$, an observer with  the
inferred inclination $i=45^\circ$ has  perspectives that vary from
magnetic pole to equator, leading in the 3D model to the
rotational phase variations in H$\alpha$ equivalent width shown in
the right panel (shaded circles). This matches quite well both the
modulation and random fluctuation of the observed equivalent width
(black dots), though accounting for the asymmetry about minimum
will require future, more detailed models that include a
secondary, higher-order (non-dipole) component of the inferred
surface field.

\begin{figure}[t!]
\vspace{-0.0in}
\begin{center}
   \includegraphics[width=120mm]{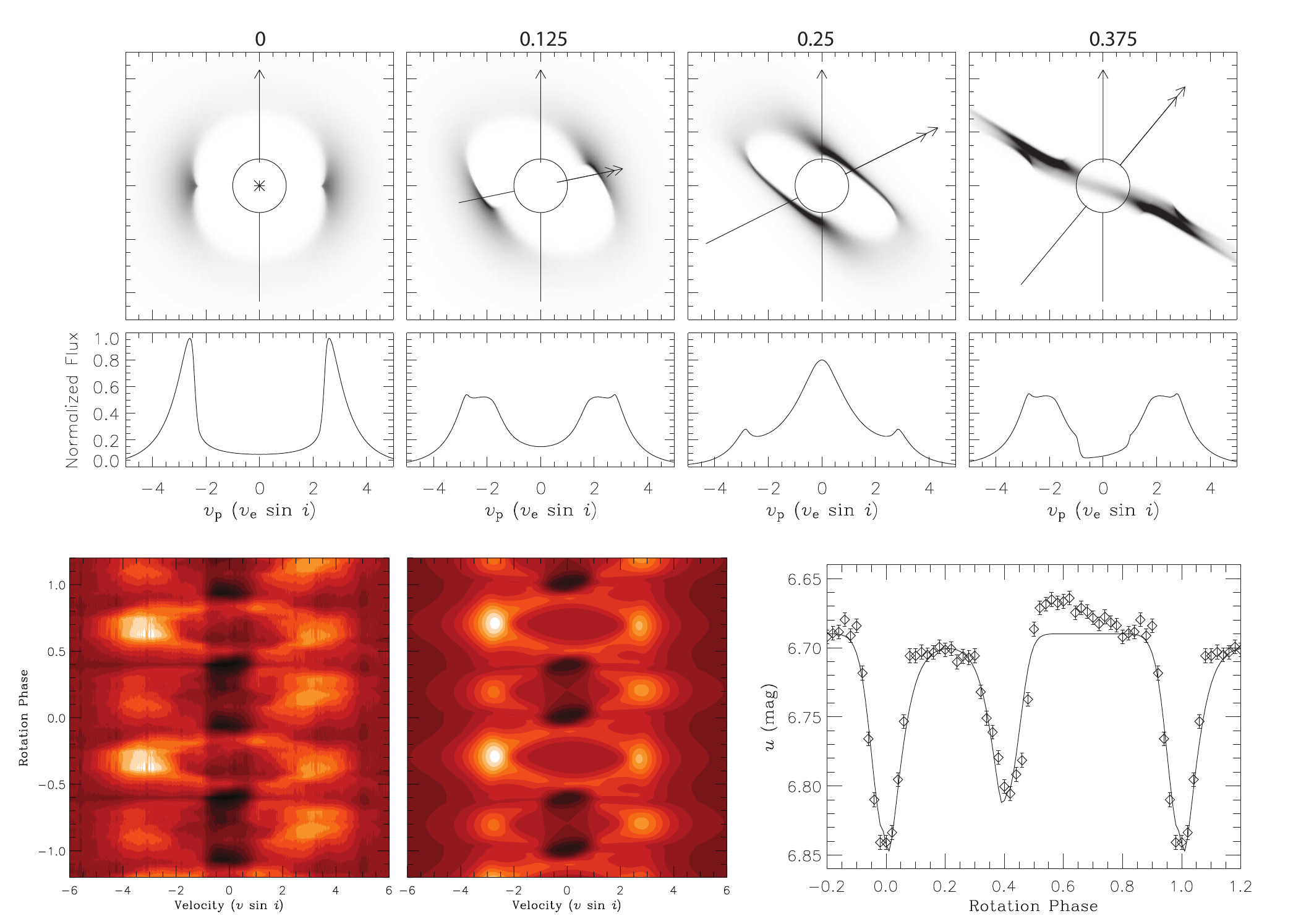}
\end{center}
\vspace{-0.1in} \caption{ Observational signatures of the CM in
the moderately fast rotator (1.2-day period) B2V star \sorie,
compared with results from the RRM model
\citep{TowOwo2005}.
The top row shows surface maps of H$\alpha$ emission and
resulting emission line profiles at the marked rotational phases.
The lower-row density plots are associated dynamic H$\alpha$
spectra, showing the variations relative to the photospheric
profile over two rotation periods of  $\sim 1.2d$; white indicates
emission, and black absorption. The left panel is  based on
echelle observations of the star, while the central bottom panel is the
prediction from the RRM model. The lower-right line plot shows the
Str\"omgren $u$-band light curve of \sorie, revealing the
eclipse-like dimmings that occur when its two magnetospheric
clouds transit in front of the star. The solid line indicates the
predictions of an early RRM model.
  }
  \label{fig:rrm}
\end{figure}

\subsubsection{The Rigidly Rotating Magnetosphere (RRM) model}

In modeling the CM of more rapidly rotating, strongly magnetic
B-stars like \sorie, a key challenge stems from the fact that
their wind magnetic confinement parameters are generally of order
$\estar \sim 10^6$ or more, far beyond the maximum $\estar \approx
10^3$ achieved with direct MHD simulations, which  are limited by
the Courant stability criterion. As an alternative for this  {\em
strong-field limit},
\citet{TowOwo2005} developed a \emph{Rigidly Rotating
Magnetosphere} (RRM) model that uses a semi-analytical
prescription for the 3D magnetospheric plasma distribution, based
on the form and minima of the total gravitationa
l-plus-centrifugal potential along each separate field line.
\citet{Tow2005}
applied this RRM model to synthesize
the emission from material trapped in the associated CM of \sorie.
Figure \ref{fig:rrm} compares the predicted variation of  the
dynamic emission spectrum  over the 1.2 day rotational period with
that obtained from echelle observations of the star. The agreement
is again very good, providing strong general support for this RRM
model for H$\alpha$ emission from the CM of \sorie.

\begin{figure}[t!]
\vspace{-0.0in}
\begin{center}
\includegraphics[width=120mmin]{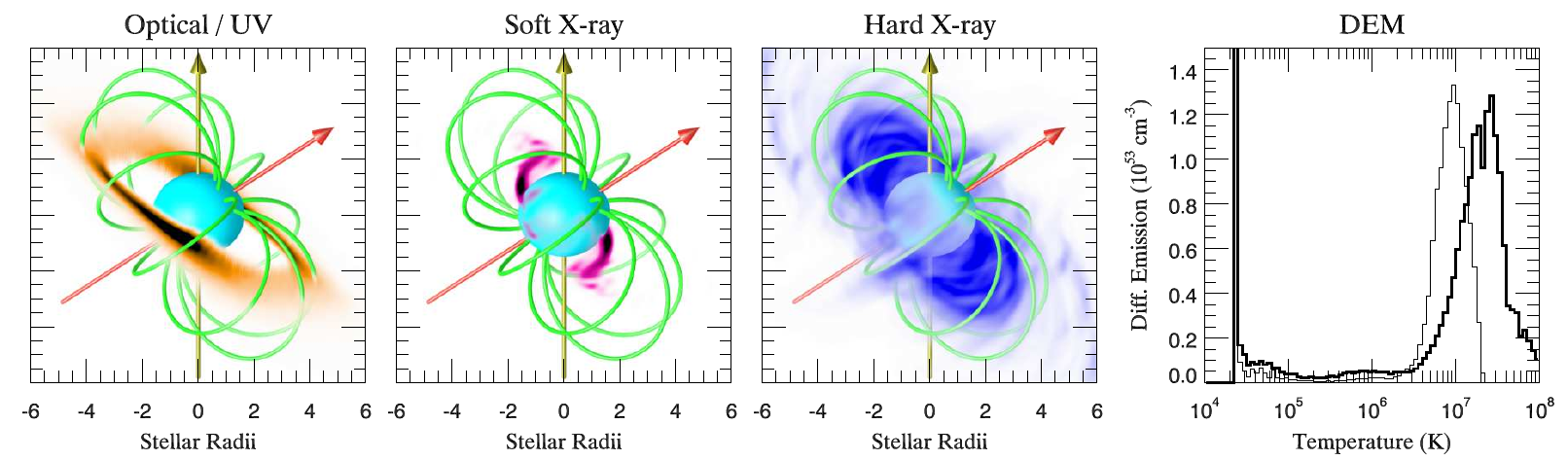}
\end{center}
\vspace{-0.in} \caption{ \small{Snapshots from an RFHD model of
\sorie, showing the
    spatial distribution of magnetospheric emission measure in three
    different temperature bins: optical ($T<10^{6}\,{\rm K}$), soft
    X-ray ($10^{6}\,{\rm K} < T < 10^{7}\,{\rm K}$) and hard X-ray ($T
    > 10^{7}\,{\rm K}$). The plot on the right shows the corresponding
    differential emission measure, for models with (thin) and without
    (thick) thermal conduction.}  }
\label{fig:rfhd}
\end{figure}

The basic RRM concept has been further developed in a successor
\emph{Rigid  Field Hydrodynamics} (RFHD) model (Townsend et al.\
2007), wherein the time-dependent flow along each individual field
line is simulated using a 1D hydro code. By piecing together
independent simulations of many different field lines (typically,
several \emph{thousand}!), a 3D picture of a star's magnetosphere
can be constructed at modest computational cost, leading in turn
to predictions for not only H$\alpha$ but also for  X-ray emission
(and other wind-related observables) of  magnetospheres in the
strong-field limit, as shown figure~\ref{fig:rfhd}. A powerful
aspect of both the RRM and RFHD models is that, within the strong
field limit, they are in principle applicable to arbitrary field
topologies, not just the oblique dipole configurations considered
so far. Thus, for example, they could be used to model the
magnetosphere of HD\,37776, which harbors high-order multipoles
(Kochukhov et al.\ 2011).

\begin{figure}[t!]
\vspace{-0.0in}
\begin{center}
\includegraphics[width=4.8in]{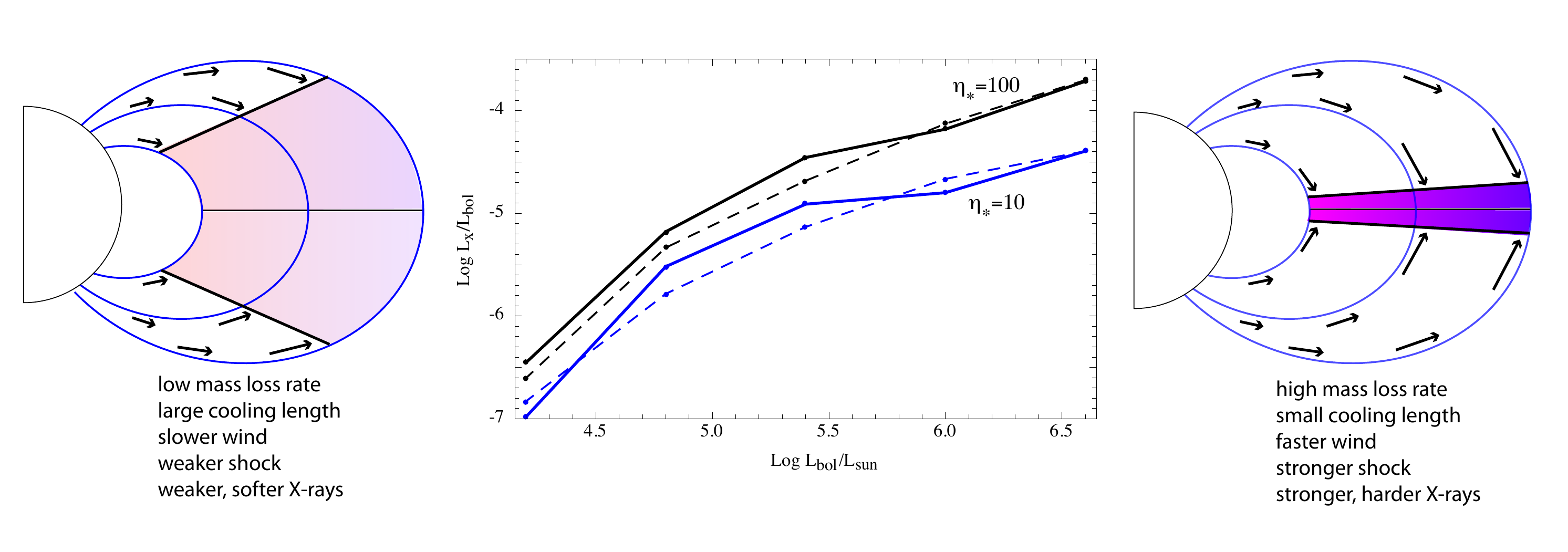}
\end{center}
\vspace{-0.in} \caption{ \small{Scaling of X-ray luminosities
$L_x$ (for energies $E_x > 0.3$\,keV) with stellar bolometric
luminosity $L_{bol}$ (center), for X-ray simulations with (solid)
and without (dashed) inverse Compton cooling (from ud-Doula et
al.\ 2014). The left and right schematics illustrate the effect of
``shock retreat'' in reducing the strength and hardness of X-rays
in lower luminosity stars with lower mass loss rate and thus less
efficient radiative cooling. } } \label{fig:lx}
\end{figure}

\subsubsection{MHD scalings for X-ray luminosity from MCWS}

For the DM cases without dynamically significant rotation,
ud-Doula et al.\ (2014) carried out a MHD simulation parameter
study with a full energy equation to compute the X-ray luminosity
$L_x$ that results from magnetically confined wind shocks (MCWS).
The central panel of figure \ref{fig:lx} plots the ratio
$L_x/L_{bol}$ vs. the bolometric luminosity $L_{bol}$ for models
with magnetic confinement $\etas= 10$ and $100$.  The dashed
curves assume the post-shock cooling is purely by radiative
emission, while the solid curves account also for the effects of
inverse Compton cooling. For the most luminous stars, $L_x$ scales
in proportion to the wind mass loss rate, which for line-driven
winds follows $\Mdot \sim L_{bol}^{1.6}$; but at lower $L_{bol}$,
the lower $\Mdot$ means the radiative cooling becomes inefficient.
As illustrated in the left vs.\ right schematic panels of figure \ref{fig:lx}, the larger
cooling layer  forces a ``shock retreat'' back to lower, slower
wind outflow, leading to weaker shocks, and so lower, softer X-ray
emission.

\begin{figure}[t!]
\vspace{-0.0in}
\begin{center}
\includegraphics[width=4.70in]{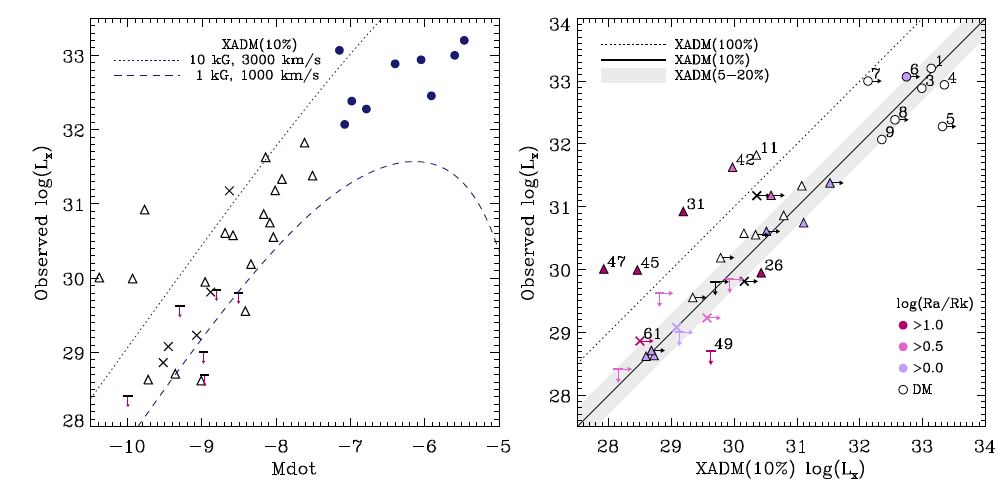}
\end{center}
\vspace{-0.in} \caption{ \small{ {\em Left}: Compilations by
Naz\'e et al.\ (2014) of X-ray luminosity in the 0.5 --10.0 keV
band,
 plotted vs. the log of mass-loss rate, as inferred from stellar parameters using the formula from Vink et al.\  (2000).
 Filled blue dots correspond to O stars, black empty
triangles to B stars, and crosses and downward-pointing arrows to
faint detections and upper limits on the X-ray luminosity,
respectively. The dotted and dashed lines show associated X-ray
luminosities predicted from the X-ray Analytic Dynamical
Magnetosphere (XADM) model of ud-Doula et al.\ (2014), scaled by
10\% efficiency, for the two indicated sets of magnetic field and
wind parameters bracketing the parameters of the sample. {\em
Right}: Direct comparison between the observed X-ray luminosity of
magnetic stars (as in left panel) and the predicted values using
the XADM model, assuming the cited overall efficiency factors. The
agreement is best for slowly rotating DM stars without centrifugal
support (open symbols); rapid rotators, with color indicating
level of centrifugally supported CMs, tend to have higher  $L_x$.
Stars of particular interest are labeled according to their
identification number in Table 1 of Naz\'e et al.\ (2014). } }
\label{fig:xpod-vs-xadm}
\end{figure}

These MHD simulation results of ud-Doula et al.\ (2014) were also
used to calibrate an `X-ray Analytic Dynamical Magnetosphere'
(XADM)  analysis for how the overall X-ray luminosity scales with
stellar magnetic field strength and wind mass loss rate. For the
subset of relatively slowly rotating magnetic OB stars modeled in
these MHD simulations, figure \ref{fig:xpod-vs-xadm} shows that,
when modified by an overall duty cycle efficiency factor to
account for extended intervals of infall of trapped material
without much X-ray emission,  this XADM scaling  gives remarkably
close agreement with empirical trends inferred by Naz\'e et al.\
(2014).

\subsection{Magnetic wind braking, spindown time, and spindown age}
\label{sec:spindown}

Let us now turn to the issue of rotational spindown from magnetic
wind braking. In this regard, the case of \sorie\ provides a key
testbed, because extended photometric monitoring of the timing of
magnetospheric clouds transiting in front of the star (Townsend et
al.\ 2008; see also lower right panel of figure \ref{fig:rrm}) has
allowed a {\em direct} measurement of the change in rotation
period, yielding  a spindown time of 1.34\,Myr (Townsend et al.\
2010). This is remarkably close to the spindown time
\textit{predicted} previously by  ud-Doula et al.\ (2009), based
on the same 2D MHD aligned-dipole parameter study used for figure
\ref{fig:dmdr}.

This MHD study showed that the angular momentum carried out by a
magnetically torqued stellar wind follows the same simple,
split-monopole scaling law derived for the Sun by Weber and Davis
(1967), $\dot J =\frac{2}{3} {\dot M} \, \Omega \, \Ralf^{2}$ --
with, however, the Alfv\'en radius \Ralf\ now given by the {\em
dipole} scaling $\Ralf\sim \estar^{1/4}$, instead the oft-quoted,
stronger scaling ($\Ralf\sim \estar^{1/2}$) for a split monopole.
This leads to an associated general  formula for the rotational
braking timescale,
\begin{equation} \label{eq:tbrake}
     \tbrake \equiv \frac{I \Omega}{{\dot J}}
     = \frac{3}{2} f \tmdot \, \left ( \frac{\Rstar}{\Ralf}\right )^{2}
     \approx 0.15 \frac{\tmdot}{\sqrt{\estar}}.
\end{equation}
Here $\tmdot \equiv M/{\dot M}$ is the stellar mass loss
timescale, and $f \approx 0.1$ is a dimensionless measure of the
star's moment of inertia $I \equiv f M\Rstar^2$.

If we assume for simplicity a fixed radius $\Rstar$ and moment of
inertia factor $f \approx 0.1$, as well as a constant angular
momentum loss rate ${\dot J}$, then the stellar rotation period
$P$ will simply increase exponentially with age $t$ from its
initial value, $P(t) = P_o e^{t/\tauj}$.
This can be used to define a star's \textit{spindown age}, $t_s$,
in terms of the spindown time $\tauj$, and its inferred
present-day critical rotation fraction $W = P_\mathrm{orb}/P$
relative to its initial rotation fraction $W_o$,
$\ts = {\tauj} \ln{W_o/W} $.
Taking the initial rotation to be critical, $W_o =1$, yields a
simple upper limit to the spindown age,
\begin{equation}
\tsmax = \tauj ~  \ln (1/W) \, . \label{tsmax}
\end{equation}
If the initial rotation is subcritical, $W_o < 1$, then the actual
spindown age is shorter by a time $\Delta \ts =  \tauj \, \ln
W_o$.

In figure \ref{fig:etaW} the upper axis gives the spindown
timescale $\tauj$ (normalized by the value in a non-magnetized
wind),
while the right axis gives the maximum spindown age $\tsmax$
(normalized by the spindown time).
Stars above the horizontal dotted line have a maximum spindown age
that is {\em less} than a single spindown time. Together with the
$\Ralf/\Rkep$ vs.\ luminosity plot in figure \ref{fig:par_ha}, we
can identify some important features and trends:
\begin{itemize}
\setlength{\itemsep}{0in} \item All the most rapidly rotating
stars are cooler B-type with weak winds, and thus weak braking,
despite their strong field. The two most extreme examples (ID\, 45
and 47) may be very close to critical rotation, and so provide a
potential link to Be stars, which have {\em not} been found to
have strong ordered fields, but for which rapid rotation is linked
to decretion into an orbiting Keplerian disk.
\item The only rapidly rotating O-star is Plaskett's star (ID\,6),
which has likely been spun up by mass exchange with its close
binary companion
\citep{Grunhut2013}.
Many O-stars have very long rotation period, e.g.\ 538~days for
the field star HD\,191612 (ID\,4), suggesting substantial
main-sequence spindown by wind magnetic braking, with a spindown
age comparable to its estimated main-sequence age. \item In
contrast, the young Orion cluster  star \thooc\ (ID\,3) has a
moderately slow (15.4-day period) rotation,  but is generally
thought to be about 1~Myr old
\citep{Hillenbrand97, Scandariato2012},
much less than its maximum spindown age $\tsmax \approx 3 \tauj
\approx 10$~Myr. Thus its zero-age main-sequence (ZAMS) rotation was likely
already quite
slow, suggesting significant {\em pre-}main-sequence braking,
e.g.\ by pre-main-sequence (PMS) disk-locking, or through a PMS jet and/or wind,
as discussed in Section 3.1.2 above.
\end{itemize}

To reinforce the last point, the recent survey of Herbig Ae/Be
stars by
\citet{Alecian2013a, Alecian2013b}
concludes that magnetic HeAeBe
stars have a slower rotation than those without a detected field.
Among their sample of non-magnetic stars they further find that
those with lower mass evolve toward the ZAMS with a constant
angular momentum, whereas higher mass ($> 5 M_\odot$) stars show
evidence of angular momentum loss during their PMS evolution, most
likely as a result of their stronger, radiatively driven mass
loss.

\subsection{Future Outlook}
The above shows there has been substantial progress in our efforts
to understand the physical and observational properties of
massive-star magnetospheres. But there are still important gaps in
this understanding and key limitations to the physical realism of
the models developed. The following lists some specific areas for
future work:

\begin{itemize}
\setlength{\itemsep}{0in} \item{\em 3D MHD of Non-Axisymmetric
Cases:} Thus far all MHD simulations, whether run in 2D or 3D,
have been restricted to cases with an underlying axial symmetry,
assuming a purely dipole field either without dynamically
significant rotation, or with rotation that is taken to be aligned
with the magnetic dipole axis. Fully 3D simulations are needed for
both the many stars with an oblique dipole, as well as cases with
more complex, higher-order multi-pole fields.

\item{\em Spindown from oblique dipoles or higher-order
multipoles:} An important application of these 3D MHD models will
be to analyze the angular momentum loss from oblique dipole
fields, as well as from  higher-order fields. This will allow
determination of generalized spindown scalings for complex fields,
and provide the basis for interpreting anticipated future direct
measurements of magnetic braking in stars with tilted-dipole or
higher multi-pole fields.

\item{\em Non-Ideal MHD and magnetospheric leakage:} In MHD
simulations of slowly rotating magnetic stars with a DM, the
dynamical infall of material back to the star balances the mass
feeding from the stellar wind, yielding an overall mass and
density that is in quite good  agreement with absorption and
emission diagnostics. By contrast, in CM simulations  the much
longer confinement and mass buildup is limited only by eventual
centrifugal breakout of regions beyond the Kepler radius
\citep{TowOwo2005},
and this now leads to an overall predicted CM
mass and density that significantly exceeds values inferred by
observational diagnostics. To understand better the mass budget of
CM's, it will be necessary to investigate additional plasma
leakage mechanisms, such as the field line interchange transport
that is thought to be key to mass balance of planetary
magnetospheres
\citep{Kivelson2005}.
In addition to
comparison with emission diagnostics of individual stars, this
should aim to derive general scaling laws that can explain the
trends for Balmer emission seen in Figure \ref{fig:par_ha},
particularly the boundary between H$\alpha$ emission and
absorption in B-type stars.

\item{\em Rapid rotation and gravity darkening:} To model the
rapidly rotating magnetic B-stars with $W = V_{\rm rot}/V_{\rm
orb} > 1/2$, there is a need to generalize the lower boundary
condition for both MHD and RFHD models to account for stellar
oblateness, while also including the effect of gravity darkening
for the wind radiative driving. This will also allow a link to Be
stars, to constrain upper limits on the dynamical role of
(undetected) magnetic fields in their quite distinctively
Keplerian (vs. rigid-body) decretion disks. This will also provide
a basis for applying such MHD models to PMS disks of HeAeBe stars.

\end{itemize}

\begin{acknowledgement}

 We thank the organizers of ISSI
Workshop "The Strongest Magnetic Fields in the Universe" for
 excellent meeting and hospitality.
 Resources supporting this work were
provided by the NASA High-End Computing (HEC).  MMR acknowledges
support by NASA grant NNX14AP30G and NSF grant AST-1211318, and
contributions of different collaborators, particularly A.V.
Koldoba,  G.V. Ustyugova, R. Kurosawa, A.A. Blinova and  R.V.E.
Lovelace. MMR thanks M. Comins for editing the manuscript. SPO
acknowledges support by NASA ATP Grants NNX11AC40G and NNX12AC72G,
respectively to University of Delaware and University of
Wisconsin, and extensive contributions of collaborators in the
MiMeS project, particularly D.\ Cohen, V.\ Petit, J.\ Sundqvist,
R.\ Townsend, A.\ ud-Doula, and G.\ Wade.

\end{acknowledgement}

\end{document}